\documentclass[
aps,
pre,
amsmath,amssymb,amsfonts,
reprint,
floatfix,
nofootinbib,
]{revtex4-2}

\usepackage{graphicx}
\usepackage{dcolumn}
\usepackage{bm}

\usepackage{mathptmx}

\usepackage{xcolor}

\usepackage[caption=false]{subfig}
\usepackage{ragged2e}
\DeclareCaptionJustification{justified}{\justifying}

\usepackage{anyfontsize}

\usepackage{amsthm}

\newtheoremstyle{my_theorem_style_remark}{2.5pt}{2.5pt}{}{}{\slshape\bfseries}{.}{  }{\thmname{#1}\thmnumber{ #2}}
\theoremstyle{my_theorem_style_remark}
\newtheorem{remark}{Remark}
\newtheoremstyle{my_theorem_style}{2.5pt}{2.5pt}{\itshape}{}{\bfseries}{.}{  }{}
\theoremstyle{my_theorem_style}

\allowdisplaybreaks

\begin{document}

\title{Many interacting particles in solution. I. Screening-ranged expansions of electrostatic potential~and~energy}

\author{Sergii V. Siryk}
\email{accandar@gmail.com}
\affiliation{CONCEPT Lab, Istituto Italiano di Tecnologia, Via E. Melen 83, 16152, Genova, Italy}
\author{Walter Rocchia}
\email[W. Rocchia \emph{(corresponding co-author)}: ]{walter.rocchia@iit.it}
\affiliation{CONCEPT Lab, Istituto Italiano di Tecnologia, Via E. Melen 83, 16152, Genova, Italy}

\begin{abstract}
\noindent
We present an analytical many-body formalism for systems of spherical particles carrying arbitrary free charge distributions and interacting in a polarizable electrolyte solution, that we model within the linearized Poisson--Boltzmann framework. Building on the detailed spectral analysis of the associated nonstandard Neumann--Poincar\'e-type operators developed in our companion study~\textcolor{red}{\cite{supplem_pre_math}}, we construct exact explicit expansions of the electrostatic potential and energy in ascending orders of Debye screening thereby obtaining systematic ``screening-ranged" series for potentials and energies. These screening-ranged expansions provide a unified and tractable description of many-body electrostatics. We demonstrate the versatility of the approach by showing how it generalizes and improves upon both classical and modern methods, enabling rigorous treatment of heterogeneously charged systems (such as Janus particles) and accurate modeling of higher-order phenomena (such as asymmetric dielectric screening, opposite-charge repulsion, like-charge attraction) as well as yielding many-body generalizations to analytical explicit results previously known only in the two-body setting.
\end{abstract}
\maketitle
\section{Introduction}
\label{section-Intro}
\noindent
Electrostatics is pervasive in the physical realm, especially at nanoscopic and atomistic scales, where it governs a wealth of relevant phenomena and plays a ubiquitous role in biomolecular interactions (e.g.~in protein-ligand binding or in providing specificity to protein-protein association) \cite{SheinermanNorelHonig,PASQUIER2023794}, as well as in colloidal solutions, plasma physics, and atmospheric science \cite{our_jcp,our_jpcb}. Electrostatic interactions can be evaluated at different levels of approximation, depending on the specific application. The Poisson--Boltzmann equation (PBE) and its variants (see \cite{ROHMR,Tabrizi1,Frydel_0,Blossey2023,VMBB_2025,BK2022,BKK2015,SXW2019,HJ2024,GS_2018,MOB_2024,TolFolOnu,Cisneros_ChemRev,BFL_pccp,KLMS_Lang_2022,BR_jma_2023,BR_jcp_2025} and references therein), central to continuum electrostatics models, represent biomolecular systems as low-polarizable media embedded in a highly polarizable solvent and describe ionic species using average, density-based formulations. The electrostatic potential governed by the PBE is a key quantity on which many properties of interest depend~\cite{Blossey2023,BesleyACR,MSP,IAO,our_jcp} (e.g.~solvation and binding energies, reaction field energy, pKa values of titratable sites, electrostatic forces and interaction energies).

While the general PBE is nonlinear, it reduces to the simpler linearized PBE (LPBE)—also known as the Debye--H\"uckel (DH) equation—in the case of low-charged systems that generate sufficiently small potentials (see \cite{our_jcp,our_jpcb,Derb2,Derb4,BMP22,LM1,NakovMat,YDZ,KMMT2018,SLG2024} and references therein). Nonetheless, even for highly charged solutes in nonlinearly responding media, the LPBE can still provide valuable insights—especially at distances larger than the Debye length—provided that the electric field sources are properly renormalized \cite{Alex1,TrizPRL,Triz1,ST1,Boon2015,our_jcp,our_jpcb,Krishnan2017,SchlaichHolm,BritoDenton,BoonJanus2010,Gillespie,AlvTel,Momot2,Doli,Tang1,NXKD,DCRG,EERR_2009,XuRan}. Moreover, works \cite{Fish,JanNetz09,Yu2021} suggest the expanded range of validity and applicability of the DH theory if considering the screening factor ($\kappa$, see below) as an effective system parameter. These features make the study of LPBE-related problems both theoretically and practically important, motivating a steady stream of research from both the theoretical and computational communities (see e.g.~\cite{our_jcp,our_jpcb,Yu3,Yu2021,BritoDenton,Tabrizi1,FilStar_jpcb,BesleyACR,Derb2,Derb4,DerbFil2017,LinQS,Jha1,AddisonSmithCooper2023,Silva1,SCW,BoWangjcp,RMDP,FinBar2016,MAR1,KMRR,BSBBC,FBYJBH,Ether2018,FolOnu,BMP22,Ponce2024,JNQS,VolkXie_PhysE,MajOsh2025,WK1,WK2,CoopJPCB,AM1,Pus_Yam_pccp,BPLPT} and references therein), and many recent contributions address LPBE directly, thereby confirming that the demand for such reductionist models remains strong. LPBE is also an extremely powerful tool in coarse-grained modeling. For instance, in a recent study \cite{CMWA}, the DH description was used to model the bacteriophage HK97 virion electrostatics (with packaging the HK97 genome into the capsid and distributing all capsid charge at the inner surface of a sphere enclosed by a dielectric spherical shell, see \cite{CMWA} for details). Furthermore, very recent work \cite{DGKF_2025} has effectively utilized LPBE for modelling the SARS-CoV-2 electrostatics, especially on studying electrostatic interactions between WildType, Delta and Omicron Spike proteins with charged surfaces, so further extending the fruitful usage of DH theory for the SARS-CoV-2 electrostatics modelling commenced in earlier papers (see \cite[Sec.~1]{DGKF_2025} and references therein for greater details and literature overviews).

For some of the simplest solute geometries, one can solve the LPBE analytically and exactly; for example, in the canonical case of a single isolated charged dielectric sphere (the ``Kirkwood sphere") immersed in a polarizable ionic unbounded solvent \cite{Kirkwood1934}. However, even for \emph{two} dielectric spheres, finding analytical solutions becomes mathematically challenging due to subtle crossovers caused by interaction and mutual polarization effects. The complete solution to this problem has been sought for many years (see \cite{Fish,Derb2,BesleyACR,our_jcp,McClurg,Yu3} and references therein). Recent two-sphere LPBE results for arbitrary free charge distributions and any-order screening components of potentials and energies are reported in \cite{our_jcp}. However, these results cannot be easily extended to general many-sphere systems, which remain analytically elusive.  Indeed, despite extensive studies, the general many-body problem of interacting dielectric spheres—determining potentials, energies, and forces for arbitrary free charge distributions—has never been solved analytically in a fully explicit form. As noted in~\cite{BSL2014}, ``An implicit series expansion is known for the system of two dielectric spheres. For more than two dielectric spheres, numerical treatment is required." A common pitfall is that applying boundary conditions imposes an infinite number of algebraic constraints coupling multipolar expansion coefficients of the potentials in different dielectric regions, or requires the laborious evaluation of surface polarization charges, and as a result leads to a complicated algebraic system of equations (typically with a densely-populated matrix). Common approaches rely on truncation of infinite-dimensional multipole expansions and on restrictive assumptions~\cite{Derb2,our_jcp,Fish}. Furthermore, the widely used DLVO (Derjaguin-Landau-Verwey-Overbeek) theory \cite{Boon2015,FinBar2016,YDZ,NikLeeWas,C5NR04274G,KJSHS,BesleyACR,DJ2010} only captures the leading pairwise single-screened terms, neglecting higher-order screened contributions that govern many-body interactions, ion-exclusion-driven effects like the non-additivity in mixtures \cite{FinBar2016,AL2009}, and polarization/ionic-driven effects, such as counterintuitive phenomena like asymmetric dielectric screening, opposite-charge repulsion or like-charge attraction (see \cite{LCSB,our_jcp,our_jpcb,Yu2021,DY_amjp_2014,DGC_softmat,Wills,Zangi2012,DuanGan2025,DY_2006,LDZLT,LZQZT,ACW_2024,ET_JML_25,WWGWFK,Buyukdagli_PRE_2017,DSVSDIB,VSSAN,OOMTY,BWGSK} and references therein). Higher-order contributions can dominate the interaction energy, especially for dissimilar particles or when higher-order multipoles are present \cite{our_jcp,our_jpcb,Yu2021}. Recent numerical studies \cite{Lindgren_jcis} confirm that many-body induced multipole interactions can contribute over 80\% of the total lattice energy of arrays of charged particles. A more in-depth analysis, as well as relevant references reflecting the development of the theory of electrostatic interactions between charged dielectric spheres, are given in Sec.~\ref{sect:background_progress}.

In this work, we address these challenges by deriving an exact analytical framework for an arbitrary number of interacting dielectric spheres, which obviates conventional approaches that require solving laborious systems of equations for potential coefficients or surface polarization charges. The main achievements of this study can be briefly summarized as follows (an in-depth outline is given in Sec.~\ref{outline_our_results_subs}):~-- We construct an exact explicit solution to the general coupled Poisson-LPBE problem for spheres with arbitrary inhomogeneous free charge distributions, without restrictive assumptions on system parameters.~-- We derive rigorous expansions (and quantify all their terms in explicit form) of the self-consistent potential and total electrostatic energy in ascending Debye screening orders, including higher-order contributions beyond DLVO.~-- Our approach provides a general analytical tool for studying many-body effects, asymmetric dielectric screening, and other phenomena that were previously accessible only numerically or approximately.  

Since this paper is already lengthy, we choose to develop the corresponding screening-ranged expansions for electrostatic forces in a separate joint paper~\textcolor{red}{\cite{supplem_pre_force}}.  

\paragraph*{Structure of the article.} This paper is organized as follows. In Sec.~\ref{sect:background}, we discuss in more detail the current achievements in describing electrostatic interactions between dielectric spheres and related work, and also present the main analytical results of this article. 

In Sec.~\ref{basic_problem_statement0}, we set up the problem of many interacting spherical particles carrying general fixed charge distributions, later expanded in arbitrary-order multipoles. In Sec.~\ref{spherical_Fourier_coeffs_subsec}, we derive the relations that govern the self-consistent potential due to the boundary conditions at the interfaces of the dielectric media. Next, in Sec.~\ref{general_expansions_pot_en}, we develop the general formalism of screening-ranged expansions for potentials and energies. An important special case of fixed charge distributions located at dielectric interfaces is discussed in more detail in Sec.~\ref{nonzero_sigma_free_surfacic}.  

To illustrate the versatility of the proposed general formalism, in Sec.~\ref{application_specific_systems_section} it is instantiated for several important classes of particles. Specifically, particles with centrally-located point charges and charges-dipoles are treated in Secs.~\ref{Appendix_screened_potential_coefficients_point} and \ref{Appendix_screened_potential_coefficients_point_dipoles}, patchy-charged/Janus particles in Sec.~\ref{subsection_janus_particles}, and other distributions, such as particles with clouds of fixed point charges or constant surface charges, in Sec.~\ref{Appendix_screened_potential_coefficients_other}.  

Finally, in Sec.~\ref{section_numeric}, we provide numerical illustrations and convergence tests of the proposed energy expansions, while Sec.~\ref{Discuss_Conclusions_section} summarizes the main results and outlines future prospects. 

Detailed derivations and additional applications are provided in the Appendices. In Appendix~\ref{appendix_2_spheres_case_general}, we apply our novel many-body formalism to the problem of two interacting spheres, a problem that has long been investigated in the literature, and compare it with the two-body analytical formalism of~\cite{our_jcp}. This comparison demonstrates how the novel many-body formalism simplifies calculations and clearly elucidates the behavior of screened components of the mutual interaction energy (the behavior at the limit of zero ionic strength is also discussed).


\section{Background and Related Work}
\label{sect:background}
\subsection{Progress in describing electrostatic interactions between dielectric spheres}
\label{sect:background_progress}
\noindent
For \emph{two} interacting spheres under the Debye-H\"uckel framework, important contributions include early work \cite{Phill1}, which derived simple analytical interaction energy expressions (by fitting them to a numerical solution to LPBE) for equal-sized spheres, later extended to unequal radii in \cite{SF2}. Implicit solutions via multipole expansions under azimuthal symmetry (AS) were constructed in \cite{GlenRus,CC1993} for identical spheres and in \cite{CC1994} for dissimilar spheres. Some extensions to non-AS cases, such as e.g.~arbitrary free charge distributions or nonuniform surface potentials, were developed in \cite{McClurg,SC1999,SadLen,LianMa,Lian2016}. Regarding the two-sphere AS problems, studies \cite{Fish,LLF} first quantified explicitly the leading asymptotic (for well-separated small/weakly-screened particles) double-screened energy corrections (to the single-screened DLVO term) for two centrally-located point charges, while \cite{Derb2,Ether2018,Derb4,FilStar_jpcb} provided analytical expressions for electrostatic forces as well as \cite{Derb2} compared electrostatic versus van der Waals contributions. Additional works \cite{BBKS,LCSB,LCZ} analyzed zero-ionic-strength AS interactions of dissimilar spheres and investigated phenomena such as like-charge attraction. 

Analytical approaches for \emph{many-sphere} systems at zero ionic strength include \cite{DY_2006} (centrally located point charges) and \cite{LinseJCP2008,DOY_2017,Yu2019} which employed arbitrary-order multipole expansions; spatially varying dielectrics were treated in \cite{ODY_2009}. Recent extensions of this type of methods (especially, the Wigner-matrix-free formalism~\cite{Yu2019}) to the LPBE include \cite{Yu3,Yu2021}. Image-charge-based techniques for rapid Green function computations were developed in \cite{Xu2013}. Numerical treatments of many-sphere potentials, energies, and forces are reported in recent~\cite{LinQS,Hassan_Stamm_jctc,Lindgren_jcis,LSB2018,GJLX,HasStammI,Jha1,FBYJBH}. The important \emph{multiple-scattering formalism} (MSF) \cite{Freed2014,QPF,QLLJPF,GXFQ,LQF_2018,Qin2019} essentially reduces the many-sphere problem to local one-body problems iteratively and self-consistently. While it can lower computational cost, convergence issues may arise \cite{BesleyACR}, and no MSF extension exists for the LPBE. Its expansions are in general expressed through iterative integrals, complicating the analysis of higher-order contributions and the derivation of explicit/closed-form formulas (see Remark~\ref{remark_MSF} below for more details).

Despite the literature quoted above, the general many-body problem of interacting dielectric spheres and finding the corresponding potentials, energies and forces, for spheres bearing arbitrary free charge distributions, either within the LPBE setting or at zero ionic strength, has never previously been solved analytically in an exact and explicit/closed form. Indeed, since the resulting algebraic systems, self-consistently coupling the multipole/harmonic expansion coefficients of potentials or surface polarization charges, are intrinsically infinite-dimensional (even in the case of centrally located point free charges  -- see in examples below), as was already noted the common way of handling them is truncation beyond some threshold (sometimes also simultaneously with truncation of the re-expansion coefficients -- see e.g.~recent \cite{Derb2,Derb4} for two-sphere systems) usually alongside with imposing some additional restrictive assumptions on the constitutive system parameters (e.g.~assuming small/weakly-screened particles with large interparticle separations, see \cite{Derb2,our_jcp,Fish}) in order to find approximate analytical solutions.

Another important issue is that the celebrated two-particle DLVO approach (especially, the electrostatic part of DLVO), along with some (polarization-neglecting) variations, despite the widespread and extensive use, is not exact within the underlying LPBE (DH) framework \cite{Boon2015,FinBar2016} but only provides its leading (single-screened, i.e.~$\propto (e^{-\kappa R_{i j}}/R_{i j})^\ell$ at $\ell=1$, with $\kappa^{-1}$ being the Debye screening length, see Sec.~\ref{general_expansions_pot_en} below for details) pairwise terms of electrostatic interactions while neglecting its higher-order ($\ell\ge2$)-screened terms (nevertheless present at the rigorous LPBE level -- such as e.g.~those caused by polarization/finite-sizeness of dielectric particles, ionic effects, etc.) that govern an important part of electrostatic interactions capable to drive important (and sometimes counterintuitive, but still observable in experiments) effects. Such effects include, for example: non-additivity in binary mixtures \cite{FinBar2016}, many-body interactions which cannot be in general factorized in terms of pairwise interactions but start playing a pivotal role in non-diluted/dense solutions of macroions and are important in many colloidal processes (e.g.~crystallization or glass formation) \cite{Boon2015}, repulsion between opposite-charge particles or attraction between like-charge particles (important for e.g.~charge scavenging processes in clouds \cite{LCSB}) at short distances where mutual polarization may dominate, etc. Moreover, those overlooked ($\ell\ge2$)-screened DLVO-beyond contributions, still present at the rigorous LPBE level, can play a critical role in dense solutions and strongly influence the overall interactions as their amplitude may significantly exceed the single-screened/DLVO terms at small interparticle separations, especially when particles are dissimilar (in radii, charges, dielectric constants) and/or when higher-order multipoles are involved \cite{our_jcp,our_jpcb,Yu2021} (let us note that higher-order multipoles can generally play a decisive role in the electrostatic interactions of charge-anisotropic objects such as proteins -- see~\cite{BP2017_BJ}). The recent numerical study \cite{Lindgren_jcis} on lattices/arrays of charged spheres also affirms that many-body induced multipole interactions can create a significant contribution to the total lattice energy and thus many-body electrostatic polarization, rather than pairwise interactions, can play a decisive role in structural stability and finding configurations with global energetic minima. Although such many-body effects in systems of polarizable spheres were also in subject of earlier studies \cite{BarrLui_2014}, there is still no general analytical formalism/approach that is rigorous at the LPBE level or at least in the $\kappa\to0$ limit when the simpler Poisson equation (PE) holds, which allows one to obtain explicit expressions quantifying many-body electrostatic contributions beyond the pairwise DLVO/Coulombic ones. Indeed, despite the great interest to advancing DLVO-beyond contributions in modern literature (see e.g.~recent review \cite{BesleyACR} and also papers \cite{LBD2023,LBD2025} and references therein) and their well agreement with the experimental measurements (see e.g.~\cite{FinBar2016,Derb2,Derb4,BesleyACR}), the corresponding available results quantifying and classifying them are mainly computational in nature, while the analytical explicit/closed-form ones are still very limited and focused mainly on two-sphere systems \cite{our_jcp,our_jpcb} --- e.g.~for the AS case with two dielectric spheres the approximate asymptotic expressions, under some restrictions on the system parameters, for the \emph{double}-screened ($\ell=2$) components of electrostatic forces, were derived in \cite{Derb2, Derb4}. For interaction energy, the corresponding double-screened results were derived earlier in the above-mentioned works \cite{Fish,LLF,McClurg}; in particular, expressions produced by \cite{Fish} are well consistent with experiments \cite{FinBar2016} and played a key role in explaining the non-additivity effects of pair interactions in binary mixtures (see \cite[Sec.~IV]{FinBar2016}). Some explicit results for odd-order screened components of interaction energy in the AS case of two monopolar spheres were obtained in~\cite{Yu3}. General two-sphere LPBE results for arbitrary free charge distributions and any-order ($\ell\ge0$) screening components of potentials and energies were recently obtained in \cite{our_jcp}; however, this approach cannot easily be extended to many-sphere systems (see Appendix~\ref{appendix_2_spheres_case}), which remain analytically elusive. 

\subsection{Outline of the main analytical results}
\label{outline_our_results_subs}
\noindent
Overall, as the analyses above show, previous work provides extensive insights into both two-sphere and many-sphere electrostatics, nevertheless a rigorous, \emph{explicit}, general analytical formalism for the LPBE in many-body systems has to date remained elusive and the general many-body problem (for arbitrary dielectric spheres with arbitrary free charge distributions) has never been solved analytically in a fully explicit form \cite{BSL2014}. Our study aims to fill this gap. The detailed contributions and highlights are summarized below:
\newcounter{my_temp_list_counter0}
\begin{list}{\textbf{\arabic{my_temp_list_counter0}.} }{\usecounter{my_temp_list_counter0} \topsep=0pt \parsep=0pt \itemsep=0pt \partopsep=0pt \parskip=0pt \itemindent=0pt \leftmargin=0pt \labelwidth=0pt \labelsep=0pt}
    \item\label{nov_list_exact_sol}\emph{Exact many-body solution without solving large multipolar systems.} We build an exact explicit analytical solution to the general coupled PE-LPBE problem of an arbitrary number of interacting non-overlapping dielectric spheres with arbitrary inhomogeneous free charge distributions, without restrictive assumptions (such as spherical radii, ionic strength, interparticle separations, existence of specific symmetries, etc.) on the system parameters and without relying on truncation of multipolar expansions or numerical inversion of large, dense matrices. This solution is built constructively using a specific operator series, which thus avoids solving laborious coupled systems and provides exact expansions (ranged in ascending orders of Debye screening and with all terms expressed in an explicit form) for potentials. Convergence of the series is rigorously proven in~\textcolor{red}{\cite{supplem_pre_math}}.
    \item\label{nov_list_energy_force_expansions}\emph{Screening-ranged expansions revealing many-body, nonadditive effects.} Based on the results of Point~\ref{nov_list_exact_sol}, we build exact expansions, in ascending Debye screening orders, of the total electrostatic energy. This generalizes previous approximate/limited-case two-sphere results and provides a rigorous LPBE-level formalism for the exact explicit quantification of many-body contributions of arbitrary screening order. Screening-ranged potential and energy expansions make many-body nonadditivity and the propagation of polarization explicit, enabling systematic inclusion of higher-order screened contributions beyond DLVO-like pairwise interactions.
    \item\label{nov_list_energy_appl}\emph{Unified analytical treatment of diverse charge patterns.} We showcase the versatility of the formalism by applying and instantiating it to various classes of free charge distributions, extending or improving existing results in the literature. Specific examples include:  
    \begin{trivlist}
        \item$\diamond$ Extending explicit two-sphere energy results \cite{Fish,LLF,McClurg,our_jcp,Yu3,DuanGan2025} to many-body systems, reproducing recent zero-ionic-strength polarization energy results \cite{DuanGan2025,RPSA_2024} as limiting cases.
        \item$\diamond$ Improving DLVO-like theory for dipolar or patchy charged particles\footnote{It should be generally noted that there has been an increased interest in constructing anisotropic DLVO-like interaction descriptions for inhomogeneously/patchy charged particles in recent years and this starts playing an increasingly more relevant role in modern colloid and protein studies -- see e.g.~the very recent (2025) work~\cite{GLCBB_2025} and references therein.} \cite{GraafBoon2012} and quantifying previously unknown higher-order screened energy contributions.  
        \item$\diamond$ Providing analytical insight into like-charge attraction / opposite-charge repulsion (including a rigorous treatment of conditions under which these effects can(not) occur in media of given polarizability contrast) and asymmetric dielectric screening~\cite{DY_2006,Yu2019,DY_amjp_2014,Yu2021,DGC_softmat}.
        \item$\diamond$ Finally, as an important benchmark, we also apply the proposed many-body formalism to the classical two-sphere problem. We obtain screened energy components of arbitrary order and analyze their zero-salt limits, also deriving a new method to quantify irregular corrections to Coulomb energy. Our accurate screened components, especially the double-screened term, systematically outperform existing approximations.
    \end{trivlist}
    \item\emph{Direct relevance for simulations and PB solvers.} The explicit expressions derived here are directly usable in Monte Carlo and molecular dynamics simulations, where many-body induced multipoles are known to be important, and they provide an efficient, grid-free electrostatics algorithm in which off-sphere (polarization) contributions are treated entirely analytically. This can make our formalism valuable both as a practical tool for building polarizable coarse-grained models and as a source of accurate analytical benchmarks for testing Poisson-Boltzmann solvers~\cite{DiFlorio2025NextGenPB}.
\end{list}

\section{General construction of exact screening-ranged expansions of potentials and energies}
\label{statements_basics_section}
\subsection{Boundary value problem statement}
\label{basic_problem_statement0}
\noindent
Let us consider a general system consisting of $N$ non-overlapping spherical dielectric particles represented by balls $\Omega_i$ (where $\Omega_i\subset\mathbb R^3$ is mathematically considered as an open set in $\mathbb R^3$) with subscript $i\in\{1,\ldots,N\}$ (or, in a simpler notation and to emphasize the running character of index $i$, $i=\overline{1,\ldots,N}$) immersed into the surrounding medium (electrolytic solvent -- e.g.~water and mobile ions) described by dielectric constant $\varepsilon_\text{sol}$ and Debye screening length $\kappa^{-1}>0$. In this study $\kappa$ is treated as a parameter (of the LPBE \eqref{Lin_eqs_lpb} -- see below), hence, we will not delve into its physical origins and its precise definition \cite{LBD2023,Kjellander_JCP_2016} is therefore not binding to us here. Each particle $\Omega_i$ is centered at $\mathbf x_i\in\mathbb R^3$ and is characterized by its dielectric constant $\varepsilon_i$ and radius~$a_i$. The electrostatic potential $\Phi_{\text{in},i}(\mathbf r)$ in $\Omega_i$ (i.e.~as $r_i < a_i$ where $r_i$ is the radial coordinate of $\mathbf r \in \mathbb{R}^3$ measured from $\mathbf x_i$ so that $r_i = \left\|\mathbf r_i\right\|$, $\mathbf r_i = \mathbf r - \mathbf x_i$) satisfies the PE, while the corresponding potential $\Phi_{\text{out},i}$ in solvent, due to the presence of the $i$-th particle, obeys the LPBE~\cite{our_jcp,our_jpcb}:
\begin{subequations}
\label{Lin_eqs}
\begin{align}
&\Delta\Phi_{\text{in},i}(\mathbf r)= - \rho_i^\text{f}(\mathbf r)/(\varepsilon_i \varepsilon_0), & & \mathbf r\in\Omega_i,\label{Lin_eqs_poisson} \\
&\Delta\Phi_{\text{out},i}(\mathbf r) - \kappa^2 \Phi_{\text{out},i}(\mathbf r) = 0, & & \mathbf r\in\Omega_\text{sol}\label{Lin_eqs_lpb}
\end{align}
\end{subequations}
for all $i=\overline{1,\ldots,N}$, where $\rho_i^\text{f}(\mathbf r)$ is a (free or fixed) charge density distribution supported inside the $i$-th particle, and the solvent domain $\Omega_\text{sol} \mathrel{:=} \mathbb R^3\setminus\bigcup_{i=1}^N\overline{\Omega}_i$ (where $\overline{\Omega}_i$ is the $\mathbb R^3$-closure of $\Omega_i$). Due to the superposition principle adopted in the DH description \cite{Fish,Derb2} the total self-consistent electrostatic potential $\Phi(\mathbf r)$ of the whole system is~then
\begin{equation}
\label{Lin_eq_tot_pot}
\Phi(\mathbf r) = \left[
\begin{aligned}
&\Phi_{\text{in},i}(\mathbf r),\quad \mathbf r\in\Omega_i,\\
&\Phi_{\text{out}}(\mathbf r) \mathrel{:=} \sum\nolimits_{i=1}^N \Phi_{\text{out},i}(\mathbf r),\quad \mathbf r\in\Omega_\text{sol},
\end{aligned}
\right.
\end{equation} 
while at the boundaries of media potential $\Phi(\mathbf r)$ is subject to the following transmission type boundary conditions (BCs):
\begin{subequations}
\label{Lin_eq_standard_bc}
\begin{gather}
\left.\Phi_{\text{in},i}\right|_{r_i \to a_i^-} = \left.\Phi_{\text{out}}\right|_{r_i \to a_i^+}, \label{Lin_eq_standard_bc_1st} \\
\varepsilon_i\left.(\mathbf n_i\cdot\nabla\Phi_{\text{in},i})\right|_{r_i\to a_i^-} - \varepsilon_\text{sol}\left.(\mathbf n_i\cdot\nabla\Phi_{\text{out}})\right|_{r_i\to a_i^+} = \sigma_i^\text{f}/\varepsilon_0 \label{Lin_eq_standard_bc_2nd}
\end{gather}
\end{subequations}
for all $i=\overline{1,\ldots,N}$, where $\mathbf n_i$ is the outer unit normal and $\sigma_i^\text{f}$ is a (free or fixed) charge density (if any) on the boundary $\partial\Omega_i$ ($r_i=a_i$) of $\Omega_i$, and $r_i\to a_i^{\pm}$ denotes approaching $\partial\Omega_i$ from exterior($+$)/interior($-$) of that particle. With no loss of generality of considerations and to make calculations less cumbersome at this stage let us firstly assume all $\sigma_i^\text{f}=0$ in \eqref{Lin_eq_standard_bc_2nd} (inhomogeneously charged surfaces with $\sigma_i^\text{f}\ne0$ play an important role in the electrostatics of patchy colloids and globular proteins \cite{BP2018_jcp} and this case will also be treated a bit later in Sec.~\ref{nonzero_sigma_free_surfacic}), since problem formulations (for the potential) with and without having fixed surface charge are mathematically similar in essence (as we will see below, it affects only right-hand sides of the resulting system of equations governing DH potentials).

For the definiteness, we will focus on the transmission type BCs~\eqref{Lin_eq_standard_bc} which is of primary interest for biomolecular sciences; since the current paper is already lengthy, we would prefer to discuss elsewhere extensions of the proposed analytical formalism to other types of BCs (fixed potentials, linear(ized) charge regulation BCs) \cite{CC1993,CC1994,MAR_EPL_2016,YT_prl_24,Zypman2022,CurLui_2021} and incorporation of the Stern layer into the problem statement. (Let us here note that such extensions may be of considerable interest since e.g.~it was shown in \cite{MSD_2009} that both pair- and many-particle interactions can be well described by the LPBE with constant potential boundary conditions provided that the electrostatic screening length is longer than the interparticle separation, however the corresponding explicit analytical treatments/insights rigorous at the DH-level have still remained elusive, to the best our knowledge.)

Potential $\Phi_{\text{in},i}$ admits \cite{our_jcp,our_jpcb} decomposition 
\begin{equation}
\label{Phi_in_i_decomposition}
\Phi_{\text{in},i} = \Hat\varPhi_{\text{in},i} + \Tilde\Phi_{\text{in},i}, 
\end{equation}
where $\Hat\varPhi_{\text{in},i}(\mathbf r) = \frac{1}{4\pi\varepsilon_0\varepsilon_i}\!\int_{\Omega_i}\!\frac{\rho_i^\text{f}(\mathbf r') \, d \mathbf r'}{\|\mathbf r - \mathbf r' \|}$ is the given particular solution to \eqref{Lin_eqs_poisson} representing the standard Coulombic potential in infinite space produced by $\rho_i^\text{f}$, while $\Tilde\Phi_{\text{in},i}$ fulfills the Laplace equation, $\Delta\Tilde\Phi_{\text{in},i}=0$. We also point out the conventional conditions for potentials satisfying system \eqref{Lin_eqs}, ensuring physical feasibility~\cite{our_jcp,Jack}, namely $|\Tilde\Phi_{\text{in},i}|<\infty$ as $r_i\to0^+$ and $\Phi_{\text{out},i}\to0$ as $r_i\to+\infty$.

\subsection{Representation of potentials through their spherical Fourier coefficients}
\label{spherical_Fourier_coeffs_subsec}
\noindent
We seek the potentials $\Tilde\Phi_{\text{in},i}(\mathbf r)$ and $\Phi_{\text{out},i}(\mathbf r)$ in the form~\textcolor{red}{\cite{supplem_pre_math}}
\begin{subequations}
\label{Lin_eqs_Phi_in_out}
\begin{gather}
\Tilde\Phi_{\text{in},i}(\mathbf r) = \sum\nolimits_{n,m}L_{n m,i} \Tilde r_i^n Y_n^m(\Hat{\mathbf r}_i),\label{Lin_eqs_Phi_in} \\
\Phi_{\text{out},i}(\mathbf r) = \sum\nolimits_{n,m}G_{n m,i}k_n(\Tilde r_i) Y_n^m(\Hat{\mathbf r}_i)\label{Lin_eqs_Phi_out}
\end{gather}
\end{subequations}
with some numerical coefficients $\{L_{n m,i}\}$ and $\{G_{n m,i}\}$; these coefficients are unknown and will be determined from BCs~\eqref{Lin_eq_standard_bc}. Here $i=\overline{1,\ldots,N}$, the scaled dimensionless radial variable $\Tilde r_i \mathrel{:=} \kappa r_i$, $\Tilde a_i \mathrel{:=} \kappa a_i$, the sum $\sum_{n,m} \mathrel{:=} \sum_{0\le|m|\le n} =  \sum_{n=0}^{+\infty} \sum_{m=-n}^n$ (also notation $0\le\left|m\right|\le n$ for indices $n$ and $m$ henceforth means the running $n=\overline{0,\ldots,+\infty}$, $m=\overline{-n,\ldots,n}$), and unit vector $\Hat{\mathbf r}_i \mathrel{:=} \mathbf r_i/r_i$. Spherical angles $\theta_i$ and $\varphi_i$ emanating from $\Hat{\mathbf r}_i$ and then used in (complex-valued) spherical harmonics 
$Y_n^m$ (see definition \eqref{Ynm_definition}), are measured for any $i$ relative to a local coordinate system with center at $\mathbf x_i$ and axes parallel to those of some unique (fixed) global coordinate system. Also $k_n(x) \mathrel{:=} \sqrt{2/\pi}K_{n+1/2}(x)/\sqrt{x}$ and $i_n(x) \mathrel{:=} \sqrt{\pi/2}I_{n+1/2}(x)/\sqrt{x}$ are modified spherical Bessel functions of the 2nd and 1st kind (see Appendix~\ref{appendix_bessel_functions_summary}). Next, for a given $\rho_i^\text{f}$ the corresponding potential $\Hat\varPhi_{\text{in},i}$ can also be expanded around $r_i \to a_i^-$ in multipoles 
\begin{subequations}
\label{varPhi_in_i_multipoles_general}
\begin{gather}
\Hat\varPhi_{\text{in},i} (\mathbf r) = \sum\nolimits_{n,m} \Hat L_{n m,i} \Tilde r_i^{-n-1} Y_n^m(\Hat{\mathbf r}_i) \label{varPhi_in_i_multipoles}
\intertext{with spherical multipole moments (see \cite[Eqs.~(11)-(12)]{our_jcp})}
\Hat L_{n m,i} = \frac{\kappa}{(2 n+1)\varepsilon_i \varepsilon_0}\int_{\Omega_i} \rho_i^\text{f}(\mathbf r_i) \Tilde r_i^n Y_n^m(\Hat{\mathbf r}_i)^\star d\mathbf r_i \label{varPhi_in_i_multipoles_Hat_Lmn}
\end{gather}
\end{subequations}
(superscript $\star$ stands for the complex conjugation). 

Note that the screened spherical harmonics in relations \eqref{Lin_eqs_Phi_out} for different particles refer to the local coordinate frames of the corresponding particles, thus in order for all quantities to be represented through the same spherical harmonics basis set $\{Y_n^m(\Hat{\mathbf r}_i)\}_{0\le\left|m\right|\le n}$ of the $i$-th particle, we first need to express the outside potentials $\Phi_{\text{out},j}$ with $j\ne i$ in this set. The spherical harmonics representation is needed to find relations coupling the spherical expansion coefficients of different quantities on the $i$-th boundary $\partial\Omega_i$ when implementing BCs on it and ensuring the mathematically correct self-consistent treatment of mutual polarization effects in the whole system. We then re-expand the screened harmonics in \eqref{Lin_eqs_Phi_out} emanating from all the other spheres $\Omega_j$ ($j\ne i$) and thereby reduce $\Phi_{\text{out},j}$ to the same spherical basis on $\Omega_i$. This can be accomplished relatively easily when e.g.~all particles' centers $\mathbf x_i$ lie on $z$-axis -- this situation has long been studied in the case of simplified system consisting of only two spheres~\cite{Langbein,GlenRus,Fish,our_jcp,our_jpcb,Derb2,Derb4,FilStar_jpcb,LCSB,BesleyACR}; note that even in such a simplified situation the corresponding re-expansions still contain complicated series/sums over modified Bessel functions (conciser exact re-expansion expressions were derived recently in~\cite{our_jcp} -- see Appendix~\ref{appendix_2_spheres_case}). For arbitrary placement of centers, previous studies have typically used Wigner rotation matrices to first bring vector $\mathbf R_{i j}$ connecting each pair of centers ($\mathbf x_i$, $\mathbf x_j$, $i\ne j$) to $z$-axis in order to reduce this general situation to a sequence of the previous simplified ones (i.e. those with centers along $z$-axis) -- for instance, see \cite{FBYJBH,Head-Gordon2006,Head-Gordon2010,McClurg,DOY_2017,DY_2006} which base their formalisms on similar or analogous procedures; such an approach can be effective when implemented numerically (e.g.~re-expansions of \cite{FBYJBH,Head-Gordon2006}, where PB-AM solver was proposed, now integrated into the well-known APBS software suite \cite{APBS}, are done recursively / via numerical iteration) but makes any rigorous/analytical study of properties/asymptotics of potentials and related quantities extremely challenging. Taking this into account, recent analytical approaches for treating many-sphere systems, underpinned by Wigner-matrix-free fully-analytical re-expansions, were proposed in \cite{Yu2019} (at $\kappa=0$, i.e.~the PE case) and in the important contributions \cite{Yu3,Yu2021} (general LPBE) -- namely, it was advanced in \cite{Yu3,Yu2021} that
\begin{equation}
\label{Yu3_reexp}
k_L(\Tilde r_j) Y_L^M(\Hat{\mathbf r}_j) = \sum\nolimits_{l_1,m_1}\mathcal H_{l_1 m_1}^{L M}(\mathbf{R}_{i j}) i_{l_1}(\Tilde r_i) Y_{l_1}^{m_1}(\Hat{\mathbf r}_i) 
\end{equation}
with re-expansion coefficients originally defined in \cite{Yu3}~as
\begin{equation}
\label{coeffs_HlmLM_definition}
\begin{aligned}
\mathcal H_{l_1 m_1}^{L M}(\mathbf{R}_{i j}) & \mathrel{:=} \sum_{l_2,m_2}\! (-1)^{l_1+l_2} H_{l_1 m_1 l_2 m_2}^{L M} k_{l_2}(\Tilde R_{i j}) Y_{l_2}^{m_2}(\Hat{\mathbf R}_{i j}), \\
H_{l_1 m_1 l_2 m_2}^{L M} & \mathrel{:=} C_{l_1 0 l_2 0}^{L 0} C_{l_1 m_1 l_2 m_2}^{L M} \sqrt{\tfrac{4\pi (2 l_1+1) (2 l_2+1)}{2 L+1}}, 
\end{aligned}
\end{equation}
where $r_i<R_{i j} = \|\mathbf R_{i j}\|$, $\mathbf R_{i j} \mathrel{:=} \mathbf x_j-\mathbf x_i$ points from $\mathbf x_i$ to $\mathbf x_j$, $\Tilde R_{i j} \mathrel{:=} \kappa R_{i j}$, $\Hat{\mathbf R}_{i j} \mathrel{:=} \mathbf R_{i j}/R_{i j}$, and $C_{l_1 m_1 l_2 m_2}^{L M} = \left<l_1 l_2; m_1 m_2 \mid L M\right>$ are Clebsch-Gordan coefficients\footnote{\label{footnote_ClebshGordan_properties}Note that $C_{l_1 0 l_2 0}^{L 0}$ can only be nonzero when $L+l_1+l_2$ is even (thus in fact one can also put $(-1)^L$ instead of $(-1)^{l_1+l_2}$ in the original definition \eqref{coeffs_HlmLM_definition} of $\mathcal H_{l_1 m_1}^{L M}(\mathbf{R}_{i j})$), and $C_{l_1 m_1 l_2 m_2}^{L M}$ can only be nonzero when $M=m_1+m_2$, $|l_1-L|\le l_2\le l_1+L$, which clearly limits the indices $l_2$ and $m_2$ in the sum $\sum_{l_2, m_2}$ in~\eqref{coeffs_HlmLM_definition} to a finite range.}. It is also interesting to note (see \cite[Sec.~IV]{Yu2019}) that, when implemented in numerics, the Wigner-matrix-free re-expansions can speed up calculations by orders of magnitude compared to conventional (Wigner-matrix-based) approaches. In Appendix~\ref{appendix_properties_asympt_H_nmLM} we collect and prove some useful properties and handy values of quantities \eqref{coeffs_HlmLM_definition} that we will need in calculations.

Now using \eqref{Lin_eqs_Phi_in_out}, \eqref{varPhi_in_i_multipoles} and \eqref{Yu3_reexp}, the boundary conditions \eqref{Lin_eq_standard_bc} then result in the following relations connecting the coefficients of the spherical expansions of all potentials involved, on~$\partial\Omega_i$:
\begin{subequations}
\label{implement_bcs}
\begin{align}
&\begin{aligned}
&\Tilde a^n_i L_{n m,i} + \Hat L_{n m,i}/\Tilde a_i^{n+1} - k_n(\Tilde a_i) G_{n m,i} \\ 
&\quad = i_n(\Tilde a_i)\sum\nolimits_{j=1,\, j\ne i}^N\sum\nolimits_{L,M}\mathcal H_{n m}^{L M}(\mathbf R_{i j}) G_{L M,j} \, ,
\end{aligned}\label{implement_bc1}\\
&\begin{aligned}
&\varepsilon_i n \Tilde a^{n-1}_i L_{n m,i} - (n+1)\varepsilon_i\Hat L_{n m,i}/\Tilde a_i^{n+2} - \varepsilon_{\text{sol}} k_n'(\Tilde a_i) G_{n m,i} \\ 
&\quad = \varepsilon_{\text{sol}} i_n'(\Tilde a_i)\sum\nolimits_{j=1,\, j\ne i}^N\sum\nolimits_{L,M}\mathcal H_{n m}^{L M}(\mathbf R_{i j}) G_{L M,j} 
\end{aligned}\label{implement_bc2}
\end{align}
\end{subequations}
for all $i=\overline{1,\ldots,N}$. Now expressing $L_{n m,i}$ from \eqref{implement_bc1} and substituting it into \eqref{implement_bc2}, using Bessel functions derivatives $k_n'(\Tilde a_i)$ and $i_n'(\Tilde a_i)$ given by \eqref{diff_modifiedBessel}, we can then eliminate the coefficients $\{L_{n m,i}\}$ of internal potentials \eqref{Lin_eqs_Phi_in}  and hence obtain relations coupling the coefficients $\{G_{n m,i}\}$ of external potentials  \eqref{Lin_eqs_Phi_out} only:
\begin{equation}
\label{main_eq_G}
\!\!\!\!\!\!
\begin{aligned}
& \bigl(n(\varepsilon_i-\varepsilon_{\text{sol}}) k_n(\Tilde a_i) + \Tilde a_i \varepsilon_{\text{sol}} k_{n+1}(\Tilde a_i)\bigr) G_{n m,i} \\
&\ + \bigl(n(\varepsilon_i-\varepsilon_{\text{sol}}) i_n(\Tilde a_i) - \Tilde a_i \varepsilon_{\text{sol}} i_{n+1}(\Tilde a_i)\bigr)\\
&\ \times \!\sum_{j=1, j\ne i}^N \sum_{L,M} \mathcal H_{n m}^{L M}(\mathbf R_{i j}) G_{L M,j} = (2 n+1)\varepsilon_i \Hat L_{n m,i}/\Tilde a_i^{n+1} .
\end{aligned}    
\end{equation}
Relations \eqref{main_eq_G} can be recast in matrix form~as
\begin{equation}
\label{eqs_G_intermediate_}
\mathsf A_i \Tilde{\mathbf G}_i + \sum\nolimits_{j=1,\, j\ne i}^N \mathsf B_{i j}\Tilde{\mathbf G}_j = \mathbf S_i ,\quad \forall i=\overline{1,\ldots,N},
\end{equation}
where we have introduced (infinite-size) column-vectors $\Tilde{\mathbf G}_i$, $\mathbf S_i$ and matrices $\mathsf A_i$, $\mathsf B_{i j}$, such that 
\begin{equation}
\label{various_matrix_definitions_0}
\begin{aligned}
\Tilde{\mathbf G}_i &\mathrel{:=} \left\{\Tilde G_{n m,i} \right\}_{n m} , \\ 
\mathbf S_i &\mathrel{:=} \left\{(2 n+1)\varepsilon_i\Hat L_{n m,i}/\Tilde a_i^{n+2} \right\}_{n m} ,\\
\mathsf A_i &\mathrel{:=} \operatorname{diagonal}\left\{\alpha_n(\Tilde a_i,\varepsilon_i) \Upsilon_{n,i}\right\}_{n m} , \\
\mathsf B_{i j} &\mathrel{:=} \left\{\beta_{n m, L M}(\Tilde a_i,\varepsilon_i,\mathbf R_{i j})\Upsilon_{L,j}\right\}_{n m, L M} 
\end{aligned}
\end{equation}
(subscripts $n m$ and $L M$ mean running indices $0\le | m |\le n$ and $0\le | M |\le L$ which enumerate rows and columns in the above matrices, respectively) with 
\begin{subequations}
\label{alpha_beta_definitions}
\begin{align}
& \alpha_n(\Tilde a_i,\varepsilon_i) \mathrel{:=} (\varepsilon_i-\varepsilon_\text{sol}) n \frac{k_n(\Tilde a_i)}{\Tilde a_i} + \varepsilon_\text{sol} k_{n+1}(\Tilde a_i), \label{alpha_definition} \\
&\begin{aligned}
\beta_{n m, L M}(\Tilde a_i,\varepsilon_i,\mathbf R_{i j}) \mathrel{:=} & \Bigl(\!\!(\varepsilon_i\!-\!\varepsilon_\text{sol}) n \frac{i_n(\Tilde a_i)}{\Tilde a_i} \!-\! \varepsilon_\text{sol} i_{n+1}(\Tilde a_i)\!\!\Bigr) \\ 
& \times \mathcal H_{n m}^{L M}(\mathbf R_{i j}) .
\end{aligned} \label{beta_definition}
\end{align}
\end{subequations}
In Appendix~\ref{appendix_properties_asympt_alpha_beta} we collate and prove some useful properties and asymptotics of quantities \eqref{alpha_beta_definitions} that we will need throughout the text. In \eqref{eqs_G_intermediate_} and \eqref{various_matrix_definitions_0}, we also introduced the scaling of coefficients of \eqref{Lin_eqs_Phi_out}, namely we put 
\begin{equation}
\label{G_nm_scaling}
G_{n m,i} = \Tilde G_{n m,i} \Upsilon_{n,i} \ \ \;\text{with}\ \ \;\Upsilon_{n,i} \mathrel{:=} \left((2 n+1) k_n(\Tilde a_i) a_i\right)^{-1} ;
\end{equation}
this will help us to conveniently consider $\{\Tilde G_{n m,i}\}$ as spherical Fourier coefficients of $\left.\Phi_{\text{out},i}\right|_{\partial\Omega_i}$ in the appropriate basis for the convenience of rigorous mathematical analysis (see details in the joint paper \textcolor{red}{\cite{supplem_pre_math}}). Note that diagonal matrix $\mathbb A$ has always strictly positive elements on its main diagonal (see Appendix~\ref{appendix_properties_asympt_alpha_beta} for useful properties and asymptotics of the quantities introduced in~\eqref{alpha_beta_definitions}). Matrix equations \eqref{eqs_G_intermediate_} can be assembled into the following global block system whose blocks are (infinite-size themselves) components~\eqref{various_matrix_definitions_0}:
\begin{equation*}
\!\!\!\!\!
\left(\!
\vphantom
{
\begin{pmatrix}
0 & 0 & \ldots & 0 \\
 0 & 0 & \ldots & 0 \\
 \vdots & \vdots & \ddots & \vdots \\
 0 & 0 & \ldots & 0
\end{pmatrix}
}
\right.
\text{\vspace{-1cm}}
\underbrace{\!\!\!\!\!
\begin{pmatrix}
\mathsf A_1 \!\!&\!\! 0 \!\!&\!\! \ldots \!\!&\!\! 0 \!\!\\
\! 0 \!\!&\!\! \mathsf A_2 \!\!&\!\! \ldots \!\!&\!\! 0 \!\! \\
\! \vdots \!\!&\!\! \vdots \!\!&\!\! \ddots \!\!&\!\! \vdots \!\! \\
\! 0 \!\!&\!\! 0 \!\!&\!\! \ldots \!\!&\!\! \mathsf A_N \!
\end{pmatrix}
\!\!\!
}_{\mathbb A}
\;
+
\;
\underbrace{
\!\!\!
\begin{pmatrix}
\!\!0 \!\!&\!\! \mathsf B_{1 2} \!\!&\!\! \ldots \!\!&\!\! \mathsf B_{1 N} \!\\
\!\!\mathsf B_{2 1} \!\!&\!\! 0 \!\!&\!\! \ldots \!\!&\!\! \mathsf B_{2 N} \!\\
\!\!\vdots \!\!&\!\! \vdots \!\!&\!\! \ddots \!\!&\!\! \vdots \!\!\\
\mathsf B_{N 1} \!\!&\!\! \mathsf B_{N 2} \!\!&\!\! \ldots \!\!&\!\! 0\!\!
\end{pmatrix}
\!\!\!\!\!
}_{\mathbb B}
\left.
\vphantom
{
\begin{pmatrix}
0 & 0 & \ldots & 0 \\
 0 & 0 & \ldots & 0 \\
 \vdots & \vdots & \ddots & \vdots \\
 0 & 0 & \ldots & 0
\end{pmatrix}
}
\right)
\,
\underbrace{
\!\!\!\!
\begin{pmatrix}
\!\Tilde{\mathbf G}_1\!\! \\
\!\Tilde{\mathbf G}_2\!\! \\
\!\vdots\!\! \\
\!\Tilde{\mathbf G}_N\!\!
\end{pmatrix}
\!\!\!\!
}_{\Vec{\Tilde{\mathbb G}}}
\;
=
\;
\underbrace{
\!\!\!\!
\begin{pmatrix}
\,\mathbf S_1\! \\
\!\mathbf S_2\! \\
\!\vdots\! \\
\,\mathbf S_N\!
\end{pmatrix}
\!\!\!\!
}_{\Vec{\mathbb S}}
\end{equation*} 
with block matrices $\mathbb A \mathrel{:=} \operatorname{diagonal}\{\mathsf A_i\}_{i=1}^N$, $\mathbb B \mathrel{:=}\{\mathsf B_{i j}\}_{i,j=1;\; i\ne j}^N$, and block column-vectors $\Vec{\Tilde{\mathbb G}} \mathrel{:=} \{\Tilde{\mathbf G}_i\}_{i=1}^N$, $\Vec{\mathbb S} \mathrel{:=} \{\mathbf S_i\}_{i=1}^N$. Taking into account that diagonal matrix $\mathbb A$ has strictly positive elements on its main diagonal (see \eqref{alpha_is_positive_ineq}, \eqref{in_kn_are_positive_}) the system can also be recast~as
\begin{equation}
\label{global_lin_sys1}
(\mathbb I + \mathbb K) \Vec{\Tilde{\mathbb G}} = \mathbb A^{-1} \Vec{\mathbb S}, 
\end{equation}
where $\mathbb K \mathrel{:=} \mathbb A^{-1} \mathbb B = \{\mathsf K_{i j}\}_{i,j=1;\; i\ne j}^N$ with $\mathsf K_{i j} \mathrel{:=} \mathsf A_i^{-1}\mathsf B_{i j}$, and $\mathbb I$ is the identity matrix. These $\mathbb K$ and $\mathbb A^{-1} \Vec{\mathbb S}$  apparently take the following expanded blockwise form which we are recording here for convenience:
\begin{gather*}
\mathbb K =\!\!
\begin{pmatrix}
\!0 \!\!&\!\! \mathsf A_1^{-1} \mathsf B_{1 2} \!\!&\!\! \ldots \!\!&\!\! \mathsf A_1^{-1} \mathsf B_{1 N} \!\!\\
\!\mathsf A_2^{-1} \mathsf B_{2 1} \!\!&\!\! 0 \!\!&\!\! \ldots \!\!&\!\! \mathsf A_2^{-1} \mathsf B_{2 N} \!\!\\
\!\vdots \!\!&\!\! \vdots \!\!&\!\! \ddots \!\!&\!\! \vdots \!\!\\
\!\mathsf A_N^{-1} \mathsf B_{N 1} \!\!&\!\! \mathsf A_N^{-1}\mathsf B_{N 2} \!\!&\!\! \ldots \!\!&\!\! 0\!\!
\end{pmatrix}\!\!, 
\ \ \   
\mathbb A^{-1} \Vec{\mathbb S} =\!\!
\begin{pmatrix}
\!\!\mathsf A_1^{-1} \mathbf S_1\!\! \\
\!\!\mathsf A_2^{-1} \mathbf S_2\!\! \\
\!\!\vdots\!\! \\
\!\mathsf A_N^{-1} \mathbf S_N\!\!
\end{pmatrix}\!\! .
\end{gather*}

\subsection{General construction of expansions (in ascending orders of Debye screening factors) for potentials and energies}
\label{general_expansions_pot_en}
\noindent
Here we derive exact screening-ranged expansions for potential coefficients and energy; regarding those of forces, see the joint paper~\textcolor{red}{\cite{supplem_pre_force}}.
\subsubsection{Expansion of potential coefficients}
\label{general_expansions_pot_}
\noindent
We construct screening-ranged expansions 
\begin{subequations}
\label{GL_componentwise0}
\begin{align}
\Tilde{\mathbf G}_i & = \sum\nolimits_{\ell=0}^{+\infty}\Tilde{\mathbf G}_i^{(\ell)} , \label{G_componentwise0_} \\ 
\mathbf L_i & = \sum\nolimits_{\ell=0}^{+\infty}\mathbf L_i^{(\ell)} \label{L_componentwise0_}
\end{align}
\end{subequations}
for the coefficients $\Tilde{\mathbf G}_i$ and $\mathbf L_i$ ($\mathbf L_i \mathrel{:=} \{L_{n m,i}\}_{n m}$) of potentials~\eqref{Lin_eqs_Phi_in_out}, where superscript $\ell$ indicates the order of screening as detailed below. To this end, in the current complex study (see also our joint work \textcolor{red}{\cite{supplem_prl}} for a plain presentation) we propose and leverage the idea of explicitly constructing general screening-ranged expansions (of electrostatic quantities of interest, such as potentials, energy, and forces) that block representation \eqref{global_lin_sys1} entails owing to the featured spectral properties of operator~$\mathbb K$ established in the joint study \textcolor{red}{\cite{supplem_pre_math}} -- namely that many-body block operator $\mathbb K$ (when considered in the appropriate infinite-dimensional Hilbert space constructed from square-summable sequences of spherical Fourier potential coefficients) has ``discrete" spectrum, is compact, and has a spectral radius less that unity. This enables us to formally bypass the need to solve system \eqref{global_lin_sys1} for the potential coefficients by explicitly expressing its inverse operator $(\mathbb I+\mathbb K)^{-1}$ in the form of an absolutely convergent operator series of Neumann type\footnote{Albeit classical Neumann operator series are well-known tools in mathematics, we are not aware of any corresponding explicit exact solution to the considered many-body problem or respective exactly-quantified Debye screening expansions of electrostatic quantities, similar to those proposed here. Existing studies (see Sec.~\ref{sect:background}) have primarily utilized various intermediate approximations/simplifications and developed methods tailored to particular problems (such as for two-sphere systems, by involving the truncation/simplification of systems governing multipolar coefficients or by introducing specific generating functions for electrostatic quantities, etc.); these approaches have yielded valuable, though perhaps limited, results that are not readily generalizable to more complex/many-body cases. The main mathematical challenge when using Neumann-type operator series is proving their convergence, which can be done by analyzing the related spectral properties; see \textcolor{red}{\cite{supplem_pre_math}}, where we perform the required analysis assuming that the spheres do not overlap (without imposing additional conditions on the particle radii, interparticle distances, or their ratios).}: $(\mathbb I+\mathbb K)^{-1} = \sum\nolimits_{\ell=0}^{+\infty}(-1)^\ell \mathbb K^\ell$, hereby
\begin{equation}
\label{G_neumann_series0}
\Vec{\Tilde{\mathbb G}} = \left(\sum\nolimits_{\ell=0}^{+\infty}(-1)^\ell \mathbb K^\ell \right) \mathbb A^{-1} \Vec{\mathbb S} \, \mathrel{=:} \, \sum\nolimits_{\ell=0}^{+\infty}\Vec{\Tilde{\mathbb G}}^{(\ell)} 
\end{equation}
($\mathbb K^\ell$ is the $\ell$-th power of $\mathbb K$; also, by convention, $\mathbb K^\ell$ at $\ell=0$ is equal to~$\mathbb I$). Superscript $\ell$ in $\Vec{\Tilde{\mathbb G}}^{(\ell)}$ indicates the order of screening -- indeed, since all non-zero elements of $\mathsf B_{i j}$ are $\propto e^{-\kappa R_{i j}}/R_{i j}$ and $\mathbb A$ is completely $R_{i j}$-free, then $\Vec{\Tilde{\mathbb G}}^{(\ell)}$ terms are proportional to homogeneous screening polynomials (consisting of individual Debye screening factors $e^{-\kappa R_{i j}}/R_{i j}$ with indices $i$, $j$, $i\ne j$) of degree~$\ell$. Taking into account that $\Vec{\Tilde{\mathbb G}}$ was defined as $\Vec{\Tilde{\mathbb G}} = \{\Tilde{\mathbf G}_i\}_{i=1}^N$ and thanks to the block-wise structure of~\eqref{global_lin_sys1}, these naturally induce the partitioning of addends $\Vec{\Tilde{\mathbb G}}^{(\ell)}$ of \eqref{G_neumann_series0} in the corresponding (still infinite-sized) subblocks $\Vec{\Tilde{\mathbb G}}^{(\ell)} = (\Tilde{\mathbf G}_i^{(\ell)})_{i=1}^N$. Operating with block-wise matrix multiplications\footnote{These mimic the usual numeric matrix multiplications, except that the order of operations must now be watched with care due to the general non-commutativity of matrix blocks in $\mathbb K$ (note also that the corresponding blocks themselves have infinite size). See the joint paper \textcolor{red}{\cite{supplem_pre_math}} where the justification for the mathematical correctness of the corresponding operations in the current context is discussed.} on $\mathbb K$ in \eqref{G_neumann_series0}, we arrive at relations $\Tilde{\mathbf G}_i^{(\ell)} = -\sum\nolimits_{j=1,\, j\ne i}^N \mathsf A_i^{-1}\mathsf B_{i j} \Tilde{\mathbf G}_j^{(\ell-1)} = (-1)^\ell\sum_{j=1}^N (\mathbb K^\ell)_{i j} \mathsf A_j^{-1}\mathbf S_j$ (matrix $(\mathbb K^\ell)_{i j}$ is the $(i,j)$-th block of block matrix $\mathbb K^\ell$) for addends of \eqref{G_componentwise0_} -- or, in expanded form:
\begin{equation}
\label{G_componentwise0}
\!\!\!\!
\begin{aligned}
& \Tilde{\mathbf G}_i^{(0)} = \mathsf A_i^{-1}\mathbf S_i , \qquad \Tilde{\mathbf G}_i^{(1)} = -\mathsf A_i^{-1}\sum\nolimits_{j=1,\, j\ne i}^N\mathsf B_{i j} \mathsf A_j^{-1}\mathbf S_j , \\
& \Tilde{\mathbf G}_i^{(2)} = \mathsf A_i^{-1}\sum\nolimits_{j=1}^N\sum\nolimits_{k=1,\, k\ne i,\, k\ne j}^N\mathsf B_{i k} \mathsf A_k^{-1}\mathsf B_{k j} \mathsf A_j^{-1} \mathbf S_j , \ \ \ldots \ ,\\
& \Tilde{\mathbf G}_i^{(\ell)} = \sum_{j=1}^N \sum^N_{k_{\ell-1}=1 ,\, k_{\ell-1}\ne i} \mathsf K_{i k_{\ell-1}} \sum_{k_{\ell-2}=1 ,\, k_{\ell-2}\ne k_{\ell-1}}^N \mathsf K_{k_{\ell-1} k_{\ell-2}} \\ 
&\qquad\quad\times\cdots \times \sum_{k_1=1,\, k_1\ne k_2,\, k_1\ne j}^N (-1)^\ell \mathsf K_{k_2 k_1} \mathsf K_{k_1 j} \mathsf A_j^{-1}\mathbf S_j .
\end{aligned}
\end{equation}
We deduce from \eqref{G_componentwise0} that $\Tilde{\mathbf G}_i^{(0)}$ is completely independent of any $R_{j k}$, $\Tilde{\mathbf G}_i^{(1)}$ can only contain elements proportional to DH screening factors $\frac{e^{-\kappa R_{i j}}}{R_{i j}}$ (with $j\ne i$), $\Tilde{\mathbf G}_i^{(2)}$ -- to $\frac{e^{-\kappa R_{i k}}}{R_{i k}} \frac{e^{-\kappa R_{k j}}}{R_{k j}}$ (with $j$ and $k$ such that $k\ne i$, $k\ne j$); similarly for greater $\ell$, $\Tilde{\mathbf G}_i^{(\ell)} \propto \frac{e^{-\kappa R_{i k_{\ell-1}}}}{R_{i k_{\ell-1}}} \frac{e^{-\kappa R_{k_{\ell-1} k_{\ell-2}}}}{R_{k_{\ell-1} k_{\ell-2}}} \cdots \frac{e^{-\kappa R_{k_2 k_1}}}{R_{k_2 k_1}} \frac{e^{-\kappa R_{k_1 j}}}{R_{k_1 j}}$ (with $j,\, k_1,\, \ldots,\, k_{\ell-1}$ such that $k_{\ell-1}\ne i$, $k_{\ell-2}\ne k_{\ell-1}$, \dots, $k_1\ne k_2$, and $k_1\ne j$). Relations \eqref{G_componentwise0} explicitly elucidate the complex nonadditive nature of the involved many-body interactions and reflect iterative ``cascade-type'' character of propagation of polarization effects (an issue which has so far been treated and substantiated mostly numerically, see the very recent work~\cite{Lindgren_jctc2025}, whereas its explicit analytical quantification/evidence seems to date remained elusive).

Plugging \eqref{G_componentwise0_} into \eqref{implement_bc1} we now derive expressions also for the addends of expansion~\eqref{L_componentwise0_}:
\begin{equation}
\label{L_componentwise0}
\begin{aligned}
\mathbf L_i^{(0)} & = \mathsf C_i \Tilde{\mathbf G}_i^{(0)} + \mathbf E_i , \\
\mathbf L_i^{(\ell)} & = \mathsf C_i \Tilde{\mathbf G}_i^{(\ell)} + \sum\nolimits_{j=1,\, j\ne i}^N \mathsf D_{i j} \Tilde{\mathbf G}_j^{(\ell-1)} \\
& = \sum\nolimits_{j=1,\, j\ne i}^N \mathsf P_{i j} \Tilde{\mathbf G}_j^{(\ell-1)}, \qquad \ell\ge1,
\end{aligned}
\end{equation}
where vector $\mathbf E_i \mathrel{:=} \{-\Hat L_{n m,i}/\Tilde a_i^{2 n+1} \}_{n m}$, and matrices $\mathsf C_i \mathrel{:=} \operatorname{diagonal}\left\{k_n(\Tilde a_i) \Upsilon_{n,i}/\Tilde a_i^n \right\}_{n m}$, $\mathsf D_{i j} \mathrel{:=} \left\{i_n(\Tilde a_i)\mathcal H_{n m}^{L M}(\mathbf R_{i j}) \Upsilon_{L,j}/\Tilde a_i^n\right\}_{n m, L M}$. Using subtler properties of the coefficients $\alpha_n$ and $\beta_{n m,L M}$ defined in \eqref{alpha_beta_definitions} we can further represent $\mathbf L_i^{(\ell)}$ at $\ell\ge1$ as $\mathbf L_i^{(\ell)} = \sum_{j=1,\, j\ne i}^N \mathsf P_{i j} \Tilde{\mathbf G}_j^{(\ell-1)}$, where matrix $\mathsf P_{i j} \mathrel{:=} \Bigl\{\frac{\varepsilon_\text{sol} \Upsilon_{L,j} \mathcal H_{n m}^{L M}(\mathbf R_{i j})}{\Tilde a_i^{n+2} \alpha_n(\Tilde a_i,\varepsilon_i)}\Bigr\}_{n m, L M}$ -- see derivations in \eqref{L_l_elementwise} below.

\paragraph*{Element-wise representations.}The above exact general screening-ranged identities, derived in matrix form (see \eqref{G_componentwise0}, \eqref{L_componentwise0}), expand to the following element-wise representations (which we will present just here for ease of use in further calculations) of vectors $\Tilde{\mathbf G}_i^{(\ell)} = \{\Tilde G_{n m,i}^{(\ell)}\}_{n m}$  and  $\mathbf L_i^{(\ell)} = \{L_{n m,i}^{(\ell)}\}_{n m}$ for $\forall\ell\ge0$, $0\le |m|\le n$, $1\le i\le N$:
\begin{subequations}
\label{GL_l_elementwise}
\begin{align}
& \Tilde G^{(\ell)}_{n m,i} = (-1)^\ell\sum_{j=1}^N\sum_{L,M}\bigl((\mathbb K^\ell)_{i j}\bigr)_{n m, L M}\frac{S_{L M,j}}{\alpha_L(\Tilde a_j,\varepsilon_j) \Upsilon_{L,j}} , \label{G_l_elementwise} \\
& L^{(0)}_{n m,i} = \frac{k_n(\Tilde a_i)\Upsilon_{n,i}}{\Tilde a_i^n}\Tilde G_{n m,i}^{(0)}\!+\!E_{n m,i} \!=\! \frac{k_n(\Tilde a_i) S_{n m,i}}{\Tilde a_i^n \alpha_n(\Tilde a_i,\varepsilon_i)}\!+\!E_{n m,i} ,\label{L_0_elementwise} \\
& L^{(\ell)}_{n m,i} = \bigl|\text{use relation $\Tilde{\mathbf G}_i^{(\ell)} = -\sum\nolimits_{j=1,\, j\ne i}^N \mathsf A_i^{-1}\mathsf B_{i j} \Tilde{\mathbf G}_j^{(\ell-1)}$}\bigr| \notag \\
&\quad = \sum_{j=1,\, j\ne i}^N \; \sum_{L,M} \frac{\Tilde G^{(\ell-1)}_{L M,j} \Upsilon_{L,j}}{\Tilde a_i^n} \Bigl( i_n(\Tilde a_i) \mathcal H_{n m}^{L M}(\mathbf R_{i j}) \notag \\
&\qquad - \frac{k_n(\Tilde a_i)}{\alpha_n(\Tilde a_i,\varepsilon_i)}\beta_{n m, L M}(\Tilde a_i,\varepsilon_i,\mathbf R_{i j}) \Bigr) = \bigl|\text{use~\eqref{wronsky_in_kn_corollary_f}}\bigr| \notag \\
&\quad = \sum_{j=1,\, j\ne i}^N \; \sum_{L,M} \frac{\varepsilon_\text{sol} \mathcal H_{n m}^{L M}(\mathbf R_{i j}) \Upsilon_{L,j}}{\Tilde a_i^{n+2} \alpha_n(\Tilde a_i,\varepsilon_i)}\Tilde G^{(\ell-1)}_{L M,j} ,\quad \ell\ge1, \label{L_l_elementwise}
\end{align}
\end{subequations}
where $S_{n m,i}$ and $E_{n m,i}$ are elements of vectors $\mathbf S_i$ and $\mathbf E_i$ (i.e.~$\mathbf S_i = \{S_{n m,i}\}_{n m}$, $\mathbf E_i = \{E_{n m,i}\}_{n m}$). Let us emphasize once again that these general expressions are suitable for arbitrary free charge distributions (described by multipolar expansions~\eqref{varPhi_in_i_multipoles}); below we will instantiate them in some specific cases of particular interest.
\subsubsection{Expansion of total electrostatic energy}
\label{general_expansions_en_}
\noindent 
Similarly to what was just done for the potential coefficients, screened addends \eqref{L_componentwise0_} further allow us to construct the screening-ranged expansion also for the electrostatic energy $\mathcal E = \frac{1}{2}\int_{\mathbb R^3} \rho^\text{fixed}(\mathbf r) \Phi(\mathbf r) d \mathbf r = (\text{see \eqref{Lin_eqs}, \eqref{Lin_eq_tot_pot}}) = \frac{1}{2}\sum_{i=1}^N\int_{\Omega_i} \rho_i^\text{f}(\mathbf r) \Phi_{\text{in},i}(\mathbf r) d \mathbf r$ (here $\rho^\text{fixed}$ is a predetermined fixed charge distribution of any shape):
\begin{equation}
\label{energy_expansion_components_abs_gen}
\mathcal E = \sum_{\ell=0}^{+\infty}\mathcal E^{(\ell)} .
\end{equation}
Indeed, taking into account \eqref{Lin_eq_tot_pot} and \eqref{Phi_in_i_decomposition}, the non-screened ($\mathbf R_{i j}$-independent $\forall i,j$) $\Hat\varPhi_{\text{in},i}$-contributions (if any) are to be ascribed to $\mathcal E^{(0)}$, whereas for $\Tilde\Phi_{\text{in},i}$-contributions we have $\frac{1}{2}\sum_{i=1}^N\int_{\Omega_i} \rho_i^\text{f}(\mathbf r) \Tilde\Phi_{\text{in},i}(\mathbf r) d\mathbf r = \left(\text{see \eqref{Lin_eqs_Phi_in}, \eqref{varPhi_in_i_multipoles_Hat_Lmn}}\right) = \frac{1}{2}\sum_{i=1}^N\frac{\varepsilon_i \varepsilon_0}{\kappa}\sum_{n,m}(2 n+1)\Hat L_{n m,i}^\star L_{n m,i}$; now employing here the expansion $L_{n m,i} = \sum_{\ell=0}^{+\infty}L_{n m,i}^{(\ell)}$ constructed above (see \eqref{L_componentwise0_}, \eqref{L_componentwise0}, \eqref{GL_l_elementwise}), we obtain
\begin{equation}
\label{energy_expansion_components_abs_center}
\mathcal E^{(\ell)} = \frac{1}{2}\sum_{i=1}^N\frac{\varepsilon_i \varepsilon_0}{\kappa}\sum_{n,m}(2 n+1)\Hat L_{n m,i}^\star L_{n m,i}^{(\ell)} .
\end{equation}
For instance, if particles' free distributions are modeled by multipolar moments $\{\Hat L_{n m,i}\}$ co-centered with particles, then, with the exclusion of self-energy terms (in order to prevent the divergence of $\mathcal E$, see e.g.~\cite{RAH1,our_jcp}), non-screened ($\mathbf R_{i j}$-free) addend $\mathcal E^{(0)}$ yields the sum of the one-body solvation energies, whereas 
\begin{equation}
\label{energy_expansion_components_interaction}
\mathcal E^\text{Int} \mathrel{:=} \mathcal E - \mathcal E^{(0)} = \sum\nolimits_{\ell=1}^{+\infty}\mathcal E^{(\ell)}
\end{equation}
represents the full interparticle interaction energy. E.g.~in the case of monopolar charged spheres the starting addend $\mathcal E^{(1)}$ of $\mathcal E^\text{Int}$ just produces the familiar pairwise DLVO expression, while higher-order ($\ell\ge2$)-screened addends have only been described in some particular approximate two-sphere ($N=2$) systems (see Sec.~\ref{application_specific_systems_section} and Sec.~\ref{section_numeric_subsection_2_spheres}).
Their study and accurate quantification are of great importance, since they can reflect mutual polarization-induced effects and, as already mentioned in Sec.~\ref{section-Intro}, can significantly affect the overall interaction landscape -- see \cite{Fish,LLF} and also \cite{our_jcp} for a more detailed description of the physical effects underlying these energy terms. Thus, the analytical formalism proposed here is the first to offer tools for exactly quantifying the corresponding screening-ranged ($\forall\ell\ge0$) contributions and for deriving their expressions in explicit form, without requiring a prior solution of the system of equations governing potential coefficients or surface polarization charges. Interestingly, this formalism works not only for monopolar spheres, but for arbitrary fixed charge distributions.
\begin{remark}
\label{energy_expansion_fast_convergence}
Energy expansion \eqref{energy_expansion_components_abs_gen} converges absolutely \textcolor{red}{\cite{supplem_pre_math}} and quickly (especially thanks to the rapid growth of quantities \eqref{alpha_definition}, which are ubiquitously present in denominators of the constructed potential coefficients' expansions -- see Appendix~\ref{appendix_properties_asympt_alpha_beta} for a nuanced discussion of their asymptotics and properties) as $\ell$ grows, so that the first few easily-calculable addends are usually enough for accurately approximating the correct energy $\mathcal E$ --- in Sec.~\ref{section_numeric} we show this over a range of numerical examples and system parameters and demonstrate that calculations up to the double-screened addend $\mathcal E^{(2)}$ already yield very accurate results capable of capturing key features of the full-fledged energy profile (such as like-charge attraction or opposite-charge repulsion arising for certain combinations of system parameters).
\end{remark}

\subsection{Fixed charge at the boundaries between dielectric media}
\label{nonzero_sigma_free_surfacic}
\noindent
The expansion theory constructed in Sec.~\ref{general_expansions_pot_} can be easily extended to the case where fixed charges are present on the boundaries. Indeed, in the case of a nonzero inhomogeneous surface density $\sigma_i^\text{f}$ in \eqref{Lin_eq_standard_bc_2nd}, expanding it in Fourier series in spherical harmonics, that is $\sigma_i^\text{f}(\Hat{\mathbf r}_i) = \sum_{n,m} \sigma_{n m,i}^\text{f} Y_n^m(\Hat{\mathbf r}_i)$ with expansion coefficients 
\begin{equation}
\label{sigma_nm_i_def}
\sigma_{n m,i}^\text{f} = a_i^{-2} \oint_{\partial\Omega_i} \sigma_i^\text{f}(\Hat{\mathbf s}_i) Y_n^m(\Hat{\mathbf s}_i)^\star d s_i ,
\end{equation}
we then obtain that just an additional term $\sigma_{n m,i}^\text{f}a_i/\varepsilon_0$ must be appended to the right-hand side of \eqref{main_eq_G}; accordingly, $\mathbf S_i$ defined in \eqref{various_matrix_definitions_0} then also modifies~to 
\begin{equation}
\label{F_for_surface_charge_vec_}
\mathbf S_i = \biggl\{\frac{(2 n+1)\varepsilon_i\Hat L_{n m,i}}{\Tilde a_i^{n+2}} + \frac{\sigma_{n m,i}^\text{f}}{\varepsilon_0\kappa}\biggr\}_{n m} .
\end{equation}
\begin{remark}
\label{remark_point_uniform_charges_the_same}
In particular, it also easily follows from \eqref{F_for_surface_charge_vec_} the useful observation that either having spheres with centrally-located free point charges $q_i$ solely (that is
\begin{equation}
\label{uniform_point_charge_distr_}
\Hat L_{n m,i} = \kappa q_i \delta_{n,0} /(\sqrt{4\pi}\varepsilon_0\varepsilon_i) 
\end{equation}
with $\delta_{\cdot,\cdot}$ being a Kronecker delta, and $\sigma_{n m,i}^\text{f} =0$) or having the same free charge amounts but smeared out uniformly over spherical surfaces (that is $\sigma_i^\text{f}(\Hat{\mathbf r}_i)=q_i/(4\pi a_i^2)$,~hence 
\begin{equation}
\label{uniform_surf_charge_distr_}
\sigma_{n m,i}^\text{f} = q_i \delta_{n,0}/(\sqrt{4\pi}a_i^2) ,
\end{equation} 
and $\Hat L_{n m,i}=0$) produce the same system of equations (see \eqref{eqs_G_intermediate_}, \eqref{global_lin_sys1}) governing the coefficients of DH potentials \eqref{Lin_eqs_Phi_out} -- indeed, in both these cases one arrives at the same right-hand side vector $\mathbf S_i = \{q_i \delta_{n,0}/(\sqrt{4\pi}\varepsilon_0\kappa a_i^2)\}_{n m}$.
\end{remark}
\subsubsection{An important particular case: $N$ spheres with only fixed surface charge distribution}
\label{Appendix_screened_potential_coefficients_surface}
\noindent
Since the case of surface free charges is of particular practical interest and is often treated in the literature, let us customize the above expansions to it. As coming to (arbitrary heterogeneous) free charge distributions given at the boundaries of particles and described by general expansion coefficients \eqref{sigma_nm_i_def}, for instance, plugging relation \eqref{F_for_surface_charge_vec_} (while putting $\Hat L_{n m,i}=0$ there, since we are now considering a situation where free charges are present only at the dielectric boundaries) into the general formulas \eqref{GL_l_elementwise} immediately yields the following exact equalities:
\begin{subequations}
\label{L_nmi_single_screened_components}
\begin{align}
\!\!\! \Tilde G^{(0)}_{n m,i} & = \frac{\sigma_{n m,i}^\text{f}}{\varepsilon_0 \kappa \alpha_n(\Tilde a_i,\varepsilon_i) \Upsilon_{n,i}} , \label{L_nmi_single_screened_components_G0}\\
\!\!\! L^{(0)}_{n m,i} & = \frac{k_n(\Tilde a_i) \sigma_{n m,i}^\text{f}}{\varepsilon_0 \kappa \Tilde a_i^n \alpha_n(\Tilde a_i,\varepsilon_i)} ,\label{L_nmi_single_screened_components_L0}\\  
\!\!\! \Tilde G^{(1)}_{n m,i} &  = \frac{- \Upsilon_{n,i}^{-1}}{\alpha_n(\Tilde a_i,\varepsilon_i)}\! \sum_{j=1, j\ne i}^N \: \sum_{L,M}\!\frac{\beta_{n m, L M}(\Tilde a_i,\varepsilon_i,\mathbf R_{i j}) \sigma_{L M,j}^\text{f}}{\alpha_L(\Tilde a_j,\varepsilon_j) \varepsilon_0 \kappa},\label{L_nmi_single_screened_components_G1}\\
\!\!\! L^{(1)}_{n m,i} & = \frac{\varepsilon_\text{sol}}{\alpha_n(\Tilde a_i,\varepsilon_i) \Tilde a_i^{n+2}}\sum_{j=1,\, j\ne i}^N \; \sum_{L,M}\frac{\mathcal H_{n m}^{L M}(\mathbf{R}_{i j}) \sigma_{L M,j}^\text{f}}{\alpha_L(\Tilde a_j,\varepsilon_j)\varepsilon_0 \kappa} , \label{L_nmi_single_screened_components_L1} \\
\!\!\! \Tilde G^{(2)}_{n m,i} &  = \frac{1}{\varepsilon_0\kappa \Upsilon_{n,i}} \sum_{j=1,\, j\ne i}^N \: \sum_{L, M} \frac{\beta_{n m, L M}(\Tilde a_i,\varepsilon_i,\mathbf R_{i j})}{\alpha_n(\Tilde a_i,\varepsilon_i)} \notag \\
&\qquad \times \sum_{p=1,\, p\ne j}^N \: \sum_{L', M'} \frac{\beta_{L M, L' M'}(\Tilde a_j,\varepsilon_j,\mathbf R_{j p}) \sigma_{L' M',p}^\text{f}}{\alpha_L(\Tilde a_j,\varepsilon_j) \alpha_{L'}(\Tilde a_p,\varepsilon_p)}, \notag\\
\!\!\! L^{(2)}_{n m,i} &  = \frac{-\varepsilon_\text{sol}}{\varepsilon_0\kappa \Tilde a_i^{n+2}} \sum_{j=1,\, j\ne i}^N \: \sum_{L, M} \frac{\mathcal H_{n m}^{L M}(\mathbf{R}_{i j})}{\alpha_n(\Tilde a_i,\varepsilon_i)} \notag \\
&\qquad \times \sum_{p=1,\, p\ne j}^N \: \sum_{L', M'} \frac{\beta_{L M, L' M'}(\Tilde a_j,\varepsilon_j,\mathbf R_{j p}) \sigma_{L' M',p}^\text{f}}{\alpha_L(\Tilde a_j,\varepsilon_j) \alpha_{L'}(\Tilde a_p,\varepsilon_p)} , \notag\\
\!\!\! \ldots \  . & \notag
\end{align}
\end{subequations}
Expressions \eqref{L_nmi_single_screened_components} will be further employed in Sec.~\ref{subsection_janus_particles} when considering Janus/patchy-charge particles (and they will be instantiated there by the Janus-specific expansion coefficients~\eqref{Janus_sigma_nm_i_rotated}).

As for the energy expansion addends $\mathcal E^{(\ell)}$ in the expansion of $\mathcal E$ (see \eqref{energy_expansion_components_abs_gen}), we arrive in this case at the following expressions:
\begin{equation}
\label{energy_expansion_components_abs_surface}
\mathcal E^{(\ell)} = \frac{1}{2}\sum_{i=1}^N a_i^2 \sum_{n,m} \sigma_{n m,i}^{\text{f} \ \star} \Tilde a_i^n L_{n m,i}^{(\ell)} .
\end{equation}

\section{Application to special cases of many-body systems of particular interest}
\label{application_specific_systems_section}
\noindent
Here we apply the general theory of screening-ranged expansions developed in Sec.~\ref{statements_basics_section} to some relevant cases of many-body (many-sphere) systems and discuss how it improves on results existing in the literature. Particularly detailed attention will also be given to quantifying the double-screened energy addend $\mathcal E^{(2)}$ owing to its importance as being the leading corrector to the single-screened (pairwise DLVO---in the case of monopolar spheres) interaction term~$\mathcal E^{(1)}$ (see Remark~\ref{energy_expansion_fast_convergence}).

\subsection{Special case: $N$ spheres with centrally-located point charges}
\label{Appendix_screened_potential_coefficients_point}
\subsubsection{Screened energy terms at non-zero ionic strength}
\label{Appendix_screened_potential_coefficients_point_kappa_nonzero}
\noindent
For $\rho_i^\text{f}$ defined by point charges $q_i$ at the centers of particles, we will have $\Hat L_{n m,i} = \kappa q_i \delta_{n,0} /(\sqrt{4\pi}\varepsilon_0\varepsilon_i)$ (see \eqref{uniform_point_charge_distr_}); plugging it into general relations \eqref{GL_l_elementwise} we immediately arrive at the following exact equalities:
\begin{align*}
\Tilde G^{(0)}_{n m,i} & = \frac{q_i \delta_{n,0}}{\sqrt{4\pi}\varepsilon_0\varepsilon_\text{sol}(1+\Tilde a_i)} , \\  
L_{n m,i}^{(0)}  & = \frac{q_i \delta_{n,0}}{\sqrt{4\pi}\varepsilon_0 a_i}\!\biggl(\frac{1}{(1+\Tilde a_i)\varepsilon_\text{sol}}-\frac{1}{\varepsilon_i}\biggr) , \\ 
\Tilde G^{(1)}_{n m,i} & = \frac{-(2 n+1)k_n(\Tilde a_i)\Tilde a_i}{\sqrt{4\pi}\varepsilon_0\varepsilon_\text{sol} \alpha_n(\Tilde a_i,\varepsilon_i)}\!\sum_{j=1,\, j\ne i}^N \frac{q_j e^{\Tilde a_j}}{1+\Tilde a_j} \beta_{n m, 0 0}(\Tilde a_i,\varepsilon_i,\mathbf R_{i j}) , \\ 
L^{(1)}_{n m,i}  & = \frac{\kappa}{\varepsilon_0\Tilde a_i^{n+2}\alpha_n(\Tilde a_i,\varepsilon_i)}\sum_{j=1,\, j\ne i}^N \frac{q_j e^{\Tilde a_j}}{1+\Tilde a_j}k_n(\Tilde R_{i j}) Y_n^m(\Hat{\mathbf R}_{i j})^\star , \\ 
\Tilde G^{(2)}_{n m,i} & = \frac{(2 n+1) k_n(\Tilde a_i)\Tilde a_i}{\sqrt{4\pi}\varepsilon_0\varepsilon_\text{sol} \alpha_n(\Tilde a_i,\varepsilon_i)}\sum_{j=1,\, j\ne i}^N \; \sum_{L, M} \frac{\beta_{n m, L M}(\Tilde a_i,\varepsilon_i,\mathbf R_{i j})}{\alpha_L(\Tilde a_j,\varepsilon_j)}\\ 
&\qquad\times\sum_{p=1,\, p\ne j}^N \frac{q_p e^{\Tilde a_p}}{1+\Tilde a_p} \beta_{L M, 0 0}(\Tilde a_j,\varepsilon_j,\mathbf R_{j p}) ,  \\
L^{(2)}_{n m,i} & = \frac{-\kappa}{\sqrt{4\pi}\varepsilon_0\Tilde a_i^{n+2}\alpha_n(\Tilde a_i,\varepsilon_i)} \sum_{j=1,\, j\ne i}^N \; \sum_{L, M} \frac{\mathcal H_{n m}^{L M}(\mathbf{R}_{i j})}{\alpha_L(\Tilde a_j,\varepsilon_j)} \\ 
&\qquad\times\sum_{p=1,\, p\ne j}^N \frac{q_p e^{\Tilde a_p}}{1+\Tilde a_p} \beta_{L M, 0 0}(\Tilde a_j,\varepsilon_j,\mathbf R_{j p}) ,\\
\ldots \ . &
\end{align*}
Additionally employing identities of Sec.~\ref{list_of_some_H_nmLM_values_} and \eqref{Ynm_addition_theorem}, expression $L^{(2)}_{0 0,i}$ (we will need it for calculation the double-screened energy $\mathcal E^{(2)}$) can also be recast in the following form:
\begin{align}
L^{(2)}_{0 0,i} & = \frac{-\kappa e^{\Tilde a_i}}{\sqrt{4\pi}\varepsilon_0\varepsilon_\text{sol}(1+\Tilde a_i)}\sum_{j=1,\, j\ne i}^N \, \sum_{p=1,\, p\ne j}^N \frac{q_p e^{\Tilde a_p}}{1+\Tilde a_p}\sum_{l=0}^{+\infty}(-1)^l \notag\\
&\quad\times \biggl( \frac{(\varepsilon_j-\varepsilon_\text{sol}) l}{\Tilde a_j} i_l(\Tilde a_j) -\varepsilon_\text{sol}i_{l+1}(\Tilde a_j) \! \biggr) \frac{2 l+1}{\alpha_l(\Tilde a_j,\varepsilon_j)} \notag\\
&\quad\times k_l(\Tilde R_{i j}) k_l(\Tilde R_{j p}) P_l(\cos\gamma_{i j,j p}), \label{L00_point_charges_N_spheres_s}
\end{align}
where $\alpha_l(\cdot)$ is determined by \eqref{alpha_definition} and $\gamma_{i j,j p}$ is the angle between $\Hat{\mathbf R}_{i j}$ and $\Hat{\mathbf R}_{j p}$ ($\cos\gamma_{i j,j p} = \cos\theta_{i j} \cos\theta_{j p} + \sin\theta_{i j} \sin\theta_{j p} \cos(\varphi_{i j}-\varphi_{j p})$; here $\theta_{i j}$ and $\varphi_{i j}$ are spherical angles of $\Hat{\mathbf R}_{i j}$); note that in the very particular case of two spheres ($N=2$) this angle is equal to $\pi$, so the Legendre polynomial $P_l(\cos\gamma_{i j,j p})$ then simply yields another factor~$(-1)^l$. 

Now employing these in 
\begin{equation}
\label{energy_N_monopoles_ell_screened_}
\mathcal E^{(\ell)} = \frac{1}{2\sqrt{4\pi}}\sum_{i=1}^N q_i L_{0 0, i}^{(\ell)}
\end{equation}
(which immediately follows from \eqref{energy_expansion_components_abs_center} with the above-specified coefficients $\Hat L_{n m,i}$ of \eqref{uniform_point_charge_distr_}; as follows from the results of \textcolor{red}{\cite{supplem_pre_math}}, the energy expansion \eqref{energy_expansion_components_abs_gen} with addends \eqref{energy_N_monopoles_ell_screened_} converges absolutely) we readily obtain the familiar non-screened $\mathcal E^{(\ell=0)}$ and single-screened $\mathcal E^{(\ell=1)}$ energetic contributions, namely 
\begin{equation}
\label{Born_energy_N_spheres_central_point_charges}
\mathcal E^{(0)} = \frac{1}{8\pi\varepsilon_0} \sum_{i=1}^N \frac{q_i^2}{a_i}\Bigl(\frac{1}{(1+\kappa a_i)\varepsilon_\text{sol}}-\frac{1}{\varepsilon_i}\Bigr)
\end{equation}
(the sum of the individual/one-body solvation/Born-Kirkwood energy expressions \cite{our_jcp, Fish, Qin2019, CheDzu2008}) and 
\begin{equation}
\label{DLVO_energy_N_spheres_central_point_charges}
\mathcal E^{(1)} = \frac{1}{8\pi\varepsilon_0\varepsilon_\text{sol}}\sum_{i=1}^N \sum_{j=1,\,j\ne i}^N\frac{q_i q_j e^{\Tilde a_i+\Tilde a_j-\Tilde R_{i j}}\kappa}{(1+\Tilde a_i)(1+\Tilde a_j)\Tilde R_{i j}} 
\end{equation}
(the sum of the direct two-body/pairwise DLVO interaction energies). However, already putting $\ell=2$ in \eqref{energy_N_monopoles_ell_screened_} and plugging \eqref{L00_point_charges_N_spheres_s} into it leads to the following \emph{exact} explicit many-body relation:
\begin{align}
&\!\!\!\! \mathcal E^{(2)} = \frac{-\kappa}{8\pi\varepsilon_0\varepsilon_\text{sol}}\sum_{i=1}^N \frac{q_i e^{\Tilde a_i}}{1+\Tilde a_i}\sum_{j=1,\, j\ne i}^N \, \sum_{p=1,\, p\ne j}^N \frac{q_p e^{\Tilde a_p}}{1+\Tilde a_p} \notag\\
&\qquad \times\sum_{l=0}^{+\infty} \biggl( \frac{(\varepsilon_j-\varepsilon_\text{sol}) l}{\Tilde a_j} i_l(\Tilde a_j) -\varepsilon_\text{sol}i_{l+1}(\Tilde a_j) \! \biggr) \frac{2 l+1}{\alpha_l(\Tilde a_j,\varepsilon_j)} \notag\\
&\qquad\quad \times k_l(\Tilde R_{i j}) k_l(\Tilde R_{j p}) P_l(\cos\gamma_{j i,j p}) \label{energy_N_monopoles_2_screened_legendre_full}
\intertext{(or, simplifying the first two addends $l=0, 1$)}
&\!\!\!\! = \frac{\kappa}{16\pi\varepsilon_0\varepsilon_\text{sol}}\sum_{i=1}^N \frac{q_i e^{\Tilde a_i}}{1+\Tilde a_i} \sum_{j=1,\, j\ne i}^N \; \sum_{p=1,\, p\ne j}^N \frac{q_p e^{\Tilde a_p}}{1+\Tilde a_p} \frac{e^{-\Tilde R_{i j}}}{\Tilde R_{i j}} \frac{e^{-\Tilde R_{j p}}}{\Tilde R_{j p}} \notag \\ 
&\times\!\biggl(\!d_{0,j} \!-\! 3 d_{2,j} \Bigl(\!1\!+\!\frac{1}{\Tilde R_{i j}}\!\Bigr)\!\Bigl(\!1\!+\!\frac{1}{\Tilde R_{j p}}\!\Bigr)\!\cos\gamma_{j i,j p}\!\biggr)\! + \text{H.O.T.} 
\label{energy_N_monopoles_2_screened_legendre}
\end{align} 
with $d_{0,i} \mathrel{:=} e^{2\Tilde a_i} \frac{\Tilde a_i-1}{\Tilde a_i+1}+1$, $d_{2,i} \mathrel{:=} e^{2\Tilde a_i}\frac{(\varepsilon_i+2\varepsilon_{\mathsf{sol}})(\Tilde a_i-1)-\varepsilon_{\mathsf{sol}}\Tilde a_i^2}{(\varepsilon_i+2\varepsilon_{\mathsf{sol}})(1+\Tilde a_i)+\varepsilon_{\mathsf{sol}}\Tilde a_i^2}+1$. The terms displayed in the last equality of \eqref{energy_N_monopoles_2_screened_legendre} originate from $l\le 1$ addends (calculated by employing exact relations for Bessel functions, i.e.~with no approximation/accuracy losses) of the sum $\sum_{l=0}^{+\infty}$ in \eqref{energy_N_monopoles_2_screened_legendre_full}; they are of order $O(\Tilde a_j^3)$ if $a_j$ is small, while Higher-Order Terms (H.O.T.) corresponding to $l\ge2$ are then at least of order $O(\Tilde a_j^5)$ -- indeed, as follows from asymptotics \eqref{asymptotics_Kij_small_spheres_radii_1} applied to the expression $\frac{(\varepsilon_j-\varepsilon_\text{sol}) l \Tilde a_j^{-1} i_l(\Tilde a_j) -\varepsilon_\text{sol}i_{l+1}(\Tilde a_j)}{\alpha_l(\Tilde a_j,\varepsilon_j)}$ (in turn coming from factor $\frac{\beta_{L M, 0 0}(\Tilde a_j,\varepsilon_j,\mathbf R_{j p})}{\alpha_L(\Tilde a_j,\varepsilon_j)}$ -- see above for the general expression of coefficient $L_{n m,i}^{(2)}$) present in that sum, its $l$-th addend is of order $O(\Tilde a_j^{\max(3,2 l+1)})$. 

Many-body relations \eqref{energy_N_monopoles_2_screened_legendre_full}-\eqref{energy_N_monopoles_2_screened_legendre} were not previously known in the literature except for the particular two-sphere ($N=2$) case recently considered in \cite{our_jcp} and for some earlier very rough approximations of \eqref{energy_N_monopoles_2_screened_legendre} (still for $N=2$) -- see Sec.~\ref{section_numeric_subsection_2_spheres} where such approximations are examined numerically in detail.

In the same way as before, by successively applying the above-stated particular coefficients $\Hat L_{n m,i}$ in \eqref{GL_l_elementwise} and using \eqref{energy_N_monopoles_ell_screened_}, one may explicitly derive any higher-order energy addends $\mathcal E^{(\ell)}$ with $\ell>2$. E.g.~to obtain an exact relation for the triple-screened energy $\mathcal E^{(3)}$, by employing the above $\Tilde G_{n m,i}^{(2)}$ values in \eqref{L_l_elementwise} we find the coefficients~$L_{0 0,i}^{(3)}$: 
\begin{align*}
& L_{0 0,i}^{(3)} = \sum_{j=1,\,j\ne i}^N \, \sum_{L,M}\frac{(-1)^L \kappa e^{\Tilde a_i} k_L(\Tilde R_{i j}) Y_L^M(\Hat{\mathbf R}_{i j})}{(1+\Tilde a_i)\alpha_L(\Tilde a_j,\varepsilon_j)\varepsilon_0\varepsilon_\text{sol}} \sum_{p=1,\,p\ne j}^N \, \sum_{L',M'} \\
& \frac{\beta_{L M, L' M'}(\Tilde a_j,\varepsilon_j,\mathbf R_{j p})}{\alpha_{L'}(\Tilde a_p,\varepsilon_p)} \sum_{s=1,\,s\ne p}^N \frac{q_s e^{\Tilde a_s}}{1+\Tilde a_s}\beta_{L' M', 0 0}(\Tilde a_p,\varepsilon_p,\mathbf R_{p s})
\end{align*}
(here we also used identities of Sec.~\ref{list_of_some_H_nmLM_values_} to express $\mathcal H_{0 0}^{L M}(\mathbf{R}_{i j})$), plugging which into \eqref{energy_N_monopoles_ell_screened_} we readily arrive~at
\begin{align}
&\!\!\!\!\! \mathcal E^{(3)} = \frac{1}{2\sqrt{4\pi}}\sum_{i=1}^N q_i L_{0 0, i}^{(3)} = \left|\text{use identities of Appendix~\ref{list_of_some_H_nmLM_values_}}\right| \notag \\ 
&\!\!\!\!\! = \frac{\kappa}{32\pi\varepsilon_0\varepsilon_\text{sol}}\sum_{i=1}^N \, \sum_{j=1,\,j\ne i}^N \, \sum_{p=1,\,p\ne j}^N \, \sum_{s=1,\,s\ne p}^N \frac{q_i e^{\Tilde a_i}}{1+\Tilde a_i}\frac{q_s e^{\Tilde a_s}}{1+\Tilde a_s} \notag \\ 
&\!\!\!\!\! \times \frac{e^{-\Tilde R_{i j}}}{\Tilde R_{i j}} \frac{e^{-\Tilde R_{j p}}}{\Tilde R_{j p}} \frac{e^{-\Tilde R_{p s}}}{\Tilde R_{p s}} \biggl[d_{0,j} d_{0,p} + 3 d_{0,j} d_{2,p} \Bigl(\!1 \!+\! \frac{1}{\Tilde R_{j p}}\Bigr)\!\Bigl(\!1 \!+\! \frac{1}{\Tilde R_{p s}}\Bigr) \notag\\ 
&\!\!\!\!\! \times (\Hat{\mathbf R}_{j p}\cdot\Hat{\mathbf R}_{p s}) + 3 d_{2,j} d_{0,p} \Bigl(\!1 \!+\! \frac{1}{\Tilde R_{i j}}\Bigr)\!\Bigl(\!1 \!+\! \frac{1}{\Tilde R_{j p}}\Bigr)(\Hat{\mathbf R}_{i j}\cdot\Hat{\mathbf R}_{j p}) \notag \\
&\!\!\!\!\! + 3 d_{2,j} d_{2,p} \Bigl(\!1 \!+\! \frac{1}{\Tilde R_{i j}}\Bigr)\!\Bigl(\!1 \!+\! \frac{1}{\Tilde R_{p s}}\Bigr) \! \biggl\{\!(\Hat{\mathbf R}_{i j}\cdot\Hat{\mathbf R}_{p s}) \! - \!\Bigl(\!1\!+\!\frac{3}{\Tilde R_{j p}}\!+\!\frac{3}{\Tilde R_{j p}^2}\Bigr) \notag \\ 
&\!\!\!\!\! \times\bigl((\Hat{\mathbf R}_{i j}\cdot\Hat{\mathbf R}_{p s})-3(\Hat{\mathbf R}_{i j}\cdot\Hat{\mathbf R}_{j p})(\Hat{\mathbf R}_{p s}\cdot\Hat{\mathbf R}_{j p})\bigr)
\biggr\} \biggr] + \text{H.O.T.}, \label{E_triply_screened_0}
\end{align}
The terms displayed in the last equality of \eqref{E_triply_screened_0} originate from $L, L'\le1$ addends of the corresponding sums in the above $L_{0 0,i}^{(3)}$ (accordingly, H.O.T.~represent addends beyond this range); let us note that the addends of that $L_{0 0,i}^{(3)}$ are of orders $O(\Tilde a_j^{\max(3,2 L+1)} \Tilde a_p^{\max(3,2 L'+1)})$ for small particle radii as immediately follows from asymptotics \eqref{asymptotics_Kij_small_spheres_radii_1} applied to factor $\frac{\beta_{L M, L' M'}(\Tilde a_j,\varepsilon_j,\mathbf R_{j p})}{\alpha_L(\Tilde a_j,\varepsilon_j)}\frac{\beta_{L' M', 0 0}(\Tilde a_p,\varepsilon_p,\mathbf R_{p s})}{\alpha_{L'}(\Tilde a_p,\varepsilon_p)}$ present in the above $L_{0 0,i}^{(3)}$ expression. Many-body relation \eqref{E_triply_screened_0} was not previously known in the literature, excepting for the particular two-sphere ($N=2$) case recently derived in \cite[Eqs.~(28) \& (32)]{our_jcp} (however, the leading term of $\mathcal E^{(3)}$ in the two-sphere case, i.e.~$\frac{q_1 q_2 \kappa e^{\Tilde a_1 + \Tilde a_2} d_{0,1} d_{0,2} e^{-3\Tilde R_{1 2}}}{16\pi\varepsilon_0 \varepsilon_\text{sol}(1+\Tilde a_1) (1+\Tilde a_2) \Tilde R_{1 2}^3}$, was also obtained earlier in recent \cite[Eq.~(E14)]{Yu3} within the novel approach developed there based on the energy reciprocity calculations). The $N=2$ case will also be outlined in more detail below in the context of numerical benchmarks (see Sec.~\ref{section_numeric_subsection_2_spheres}).
\begin{remark}
\label{remark_two-sphere-kappa-nonzero}
Although the two-body formalism of our earlier work~\cite{our_jcp} (technically built on matrix relation \eqref{global_lin_sys1_two_particles_explicit_G} -- see Appendix~\ref{appendix_2_spheres_case} for details), in principle, allows one to analytically treat two-sphere ($N=2$) systems and derive the corresponding exact expressions of $\Tilde G_{n m,i}^{(\ell)}\bigr|_{N=2}$ (and thereby $L_{n m,i}^{(\ell)}\bigr|_{N=2}$ and $\mathcal E^{(\ell)}\bigr|_{N=2}$) at arbitrary $\ell\ge0$, in practice, however, the complicated form of relation \eqref{global_lin_sys1_two_particles_explicit_G} makes the necessary calculations extremely laborious and cumbersome (especially for $\ell>3$). The novel many-body analytical formalism proposed in the current study, even in the case of $N=2$, leads to the desired results in a more straightforward and plain way -- see Appendix~\ref{two_spheres_kappa_nonzero_appendix} where we apply our novel formalism to derive the general patterns of $\mathcal E^{(\ell)}\bigr|_{N=2}$ expressions at even and odd $\ell$ values and explicitly express them in terms of basic system parameters.
\end{remark}
\begin{remark}
\label{remark_asymm_diel_screening}
It is easy to see that in the two-sphere case ($N=2$), many-body relation~\eqref{energy_N_monopoles_2_screened_legendre_full} reduces to expression $\mathcal E^{(2)} = \mathcal E^{(2)}\bigr|_{N=2}(R) = \frac{-\kappa}{8\pi\varepsilon_0\varepsilon_\text{sol}}\sum\limits_{i=1}^2 \frac{q_i^2 e^{2\Tilde a_i}}{(1+\Tilde a_i)^2}\sum\limits_{j=1,\, j\ne i}^2 \: \sum\limits_{l=0}^{+\infty} \Bigl(\frac{(\varepsilon_j-\varepsilon_\text{sol}) l}{\Tilde a_j} i_l(\Tilde a_j) -\varepsilon_\text{sol}i_{l+1}(\Tilde a_j)\Bigr) \frac{2 l+1}{\alpha_l(\Tilde a_j,\varepsilon_j)} k_l(\Tilde R)^2$ (where $\Tilde R = \kappa R$, $R=R_{1 2} = R_{2 1}$). By taking into account \eqref{alpha_is_positive_ineq} and \eqref{in_kn_are_positive_}, we can easily see that this expression is of firmly \emph{repulsive} character for particles less polarisable than the medium (more precisely, as $\varepsilon_1 \le \varepsilon_\text{sol}$ and $\varepsilon_2 \le \varepsilon_\text{sol}$) with arbitrary-sign fixed charges $q_1$ and $q_2$ irrespective of the behavior of the leading interaction energy addend~\eqref{DLVO_energy_N_spheres_central_point_charges} (DLVO term) $\mathcal E^{(1)}(R) = \frac{q_1 q_2 e^{\Tilde a_1+\Tilde a_2}}{4\pi\varepsilon_0\varepsilon_\text{sol}(1+\Tilde a_1)(1+\Tilde a_2)}\frac{e^{-\Tilde R}}{R}$ in this situation. The above explicit expression for the double-screened energy thus exhibits that $\mathcal E^{(2)}\bigr|_{N=2}(R)$ tends to strengthen repulsive / weaken attractive interactions resulted from $\mathcal E^{(1)}(R)$. This phenomenon corresponds to the effect termed \emph{asymmetric dielectric screening} in earlier works~\cite{DY_2006,DY_amjp_2014,Yu2021}, where it was primarily identified in numerical modeling. Furthermore, it can be derived (see Remark~\ref{appendix_lca_isnt_possible} below for technical details) that, for two-sphere systems in this situation, the screened energy addends behave in pairs: terms $\mathcal E^{(\ell)}\bigr|_{N=2}(R)$ with odd $\ell=\overline{3,5,\ldots}$ qualitatively follow the behavior of $\mathcal E^{(1)}\bigr|_{N=2}(R)$, whereas terms with even $\ell=\overline{4,6,\ldots}$ resemble $\mathcal E^{(2)}\bigr|_{N=2}(R)$. As a direct consequence, and to the best of our knowledge for the first time in the literature, it follows rigorously that \emph{no like-charge attraction can occur under such conditions} (i.e.~as $\varepsilon_1 \le \varepsilon_\text{sol}$ and $\varepsilon_2 \le \varepsilon_\text{sol}$). In contrast, when $q_1 q_2<0$, a delicate balance between the attractive odd-$\ell$ and repulsive even-$\ell$ screened energy terms $\mathcal E^{(\ell)}\bigr|_{N=2}(R)$ may lead to the occurrence of a short-range opposite-charge repulsion under certain combinations of system parameters. Finally, when $\varepsilon_1>\varepsilon_\text{sol}$, $\varepsilon_2>\varepsilon_\text{sol}$, the behavior of energy terms of $\mathcal E^{(\ell)}\bigr|_{N=2}(R)$ becomes more involved (see Remark~\ref{appendix_lca_isnt_possible} for technical details) which further expands the variety of possible types of interactions in this situation and allows for the possibility of both like-charge attraction and opposite-charge repulsion. 
Numerical illustrations are given in Sec.~\ref{section_numeric_subsection_2_spheres} below. Our present formalism thus provides, for the first time, a rigorous analytical explanation of the appearance and origin of these effects by explicitly decomposing the interaction energy into its screened component contributions with analyzable behavior. See also Remark~\ref{attr_rep_remark0} below for a treatment of the counterpart effects in the context of zero ionic strength limit case, and Table~\ref{tab:Ex1_2_spheres} for a summary of possible particle behavior scenarios for both the $\kappa>0$ and $\kappa=0$ cases.
\end{remark}
\subsubsection{Behavior of energy terms in the $\kappa\to0$ limit} 
\label{Appendix_screened_potential_coefficients_point_kappa0}
\noindent
Although the case of zero ionic strength lies beyond the primary scope of this study, it is worth emphasizing in view of the renewed interest in systems of dielectric spheres under the pure Poisson ($\kappa=0$) case (see e.g.~recent papers \cite{Chan3, DuanGan2025, LiLiGH2024, GAMR2025} discussing counterintuitive like-charge attraction / opposite-charge repulsion, and also a very recent study \cite{DGC_softmat} computationally investigating the polarization-driven mechanisms of such phenomena in the zero ionic strength case and references therein). The proposed screening-ranged expansions for energy (and also for forces~\textcolor{red}{\cite{supplem_pre_force}}) remain well-posed and meaningful in the $\kappa \to 0$ limit. This can be seen by elaborating expressions~\eqref{GL_l_elementwise} together with the asymptotics of Appendix~\ref{appendix_properties_asympt_alpha_beta} and relations~\eqref{small_Bessel_i_k}. In this way, the expansions correctly and uniformly reproduce their pertinent ``Coulombic" counterparts and, moreover, provide rigorous tools to analyze physical phenomena whose analytical understanding has so far remained incomplete (see e.g.~Remark~\ref{attr_rep_remark0} below).

For instance, in the $\kappa\to0$ limit by employing \eqref{small_Bessel_i_k} and the asymptotics of $\alpha_l(\cdot)$ (see \eqref{alpha_asymptotic_small_arg_0} in Appendix~\ref{appendix_properties_asympt_alpha_beta}) we obtain that the right-hand side of \eqref{energy_N_monopoles_2_screened_legendre_full} yields new relation
\begin{equation}
\label{E2_point_charges_N_spheres_s_kappa_zero}
\begin{aligned}
& \mathcal E^{(2)}\bigr|_{\kappa\to0} = \frac{-1}{8\pi\varepsilon_0\varepsilon_\text{sol}}\sum_{i=1}^N q_i \sum_{j=1,\, j\ne i}^N \, \sum_{p=1,\, p\ne j}^N q_p \\ 
&\qquad \times\sum_{l=1}^{+\infty}\frac{l (\varepsilon_j-\varepsilon_\text{sol}) a_j^{2 l+1}}{(l\varepsilon_j+(l+1)\varepsilon_\text{sol}) R_{i j}^{l+1} R_{j p}^{l+1}} P_l(\cos\gamma_{j i,j p}) .
\end{aligned}
\end{equation}

In the particular two-sphere case ($N=2$) many-body expression \eqref{E2_point_charges_N_spheres_s_kappa_zero} reduces~to 
\begin{align}
& \mathcal E^{(2)}\bigr|_{\kappa\to0, \; N=2} = \frac{-1}{8\pi\varepsilon_0\varepsilon_\text{sol}}\sum\limits_{l=1}^{+\infty} \frac{l}{R^{2 l+2}} \biggl(\frac{(\varepsilon_2-\varepsilon_\text{sol})q_1^2 a_2^{2 l+1}}{l\varepsilon_2+(l+1)\varepsilon_\text{sol}} \notag \\ 
&\quad + \frac{(\varepsilon_1-\varepsilon_\text{sol})q_2^2 a_1^{2 l+1}}{l\varepsilon_1+(l+1)\varepsilon_\text{sol}} \biggr) = \frac{-1}{8\pi\varepsilon_0\varepsilon_\text{sol} R^4}\biggl(\frac{(\varepsilon_2-\varepsilon_\text{sol})q_1^2 a_2^3}{\varepsilon_2+2\varepsilon_\text{sol}} \notag \\ 
&\quad + \frac{(\varepsilon_1-\varepsilon_\text{sol})q_2^2 a_1^3}{\varepsilon_1+2\varepsilon_\text{sol}} \biggr) + \text{H.O.T.} \label{E2_point_charges_N_spheres_s_kappa_zero_N_2}
\end{align}
(where $R_{1 2} = R_{2 1} = R$ and Higher-Order Terms (H.O.T.) here are of order $O(R^{-6})$ as $R\to+\infty$), so that the expression in the last equality of \eqref{E2_point_charges_N_spheres_s_kappa_zero_N_2} reproduces the proper $O(R^{-4})$-form of the leading correction to the conventional $O(R^{-1})$-order Coulombic energy $\frac{q_1 q_2}{4\pi\varepsilon_0\varepsilon_\text{sol}R} = \mathcal E^{(1)}\bigr|_{\kappa\to0, \; N=2}$ already described in \cite{Fish,LLF}. If $a_2\to0$, i.e.~one has a situation of the presence of a free/fixed point charge $q_2$ outside sphere~$1$ of radius $a_1$, then denoting $g_1\mathrel{:=}\frac{\varepsilon_\text{sol}}{\varepsilon_1+\varepsilon_\text{sol}}$ relation \eqref{E2_point_charges_N_spheres_s_kappa_zero_N_2} gives $\mathcal E^{(2)}\bigr|_{\kappa\to0, \; N=2} = \frac{-q_2^2 (\varepsilon_1-\varepsilon_\text{sol})}{8\pi\varepsilon_0\varepsilon_\text{sol}}\sum\limits_{l=1}^{+\infty}\frac{l a_1^{2 l+1} R^{-2 l-2}}{l\varepsilon_1+(l+1)\varepsilon_\text{sol}} = \Bigl|\text{use geometric progressions $\frac{1/2}{R+a_1}-\frac{1/2}{R-a_1}= \frac{-a_1}{R^2}\sum\limits_{l=0}^{+\infty}\bigl(\frac{a_1}{R}\bigr)^{2 l}$}\Bigr| = \frac{q_2^2}{8\pi\varepsilon_0\varepsilon_\text{sol}}\frac{\varepsilon_1-\varepsilon_\text{sol}}{\varepsilon_1+\varepsilon_\text{sol}}\Bigl(\frac{1/2}{R+a_1}-\frac{1/2}{R-a_1} + \frac{a_1}{R^2}\sum\limits_{l=0}^{+\infty}\frac{g_1}{l+g_1}\frac{a_1^{2 l}}{R^{2 l}}\Bigr)$, which exactly reproduces one of the key results of very recent work \cite{DuanGan2025} -- a novel three-point image formula for the polarization energy (see \cite[Eq.~(10)]{DuanGan2025}) derived there and used to quantitatively establish the conditions under which the like-charge attraction effect in polarizable spheres occurs. The series appearing in the last equality converges faster than that of the original expression $\mathcal E^{(2)}\bigr|_{\kappa\to0, \; N=2}$ above (where $l$ was also present in the numerator) and can be expressed in terms of the incomplete beta function. However, the derivation of this explicit formula is carried out in \cite{DuanGan2025} in a completely different way (based on a subtle application of the image-charge principle and of the properties of the incomplete beta function) which is specific to the particular situation of a point charge outside a sphere and it does not seem to be readily generalizable to more complex situations. Finally, let us note that in the considered case of two spheres, if either $a_1\to0$ or $a_2\to0$, then the ($\ell\ge3$)-addends in \eqref{energy_expansion_components_interaction} vanish (see expressions for $\mathcal E^{(\ell)}\bigr|_{\kappa\to0, \; N=2}$ in Appendix~\ref{two_spheres_kappa0_appendix}), hence $\mathcal E^\text{Int}$ is fully determined by the Coulombic term and the above $\mathcal E^{(2)}\bigr|_{\kappa\to0, \; N=2}$ in this case.

Let us note that similarly to what has just been discussed, convergence in the general many-body case can also be accelerated --- namely, by representing $\frac{l(\varepsilon_\text{sol}-\varepsilon_j)}{l\varepsilon_j+(l+1)\varepsilon_\text{sol}} = \frac{\varepsilon_\text{sol}-\varepsilon_j}{\varepsilon_\text{sol}+\varepsilon_j}\bigl(1-\frac{g_j}{l+g_j}\bigr)$ with $g_j\mathrel{:=} \frac{\varepsilon_\text{sol}}{\varepsilon_j+\varepsilon_\text{sol}}$ and using the generating function for Legendre polynomials~\cite{QPF} (see \eqref{leg_pol_gen_func}), the infinite series in many-body relation \eqref{E2_point_charges_N_spheres_s_kappa_zero} can also be transformed into an equivalent but more rapidly converging form: $\sum\limits_{l=1}^{+\infty}\frac{(\varepsilon_\text{sol}-\varepsilon_j) l a_j^{2 l+1} P_l(\cos\gamma_{j i,j p})}{(l\varepsilon_j+(l+1)\varepsilon_\text{sol}) R_{j i}^{l+1} R_{j p}^{l+1}} = \frac{\varepsilon_\text{sol}-\varepsilon_j}{\varepsilon_\text{sol}+\varepsilon_j} \frac{a_j}{R_{j i} R_{j p}}\Bigl(\bigl(1-2 t_{i j p} \cos\gamma_{j i,j p} + t^2_{i j p}\bigr)^{-1/2} - g_j\sum\limits_{l=0}^{+\infty}\frac{t^l_{i j p}}{l+g_j}P_l(\cos\gamma_{j i,j p})\Bigr)$, where $t_{i j p}\mathrel{:=}\frac{a_j^2}{R_{j i} R_{j p}}<1$. Indeed, the last series does not contain $l$ in the numerator and hence converges faster than the original~one.

A further interesting case is that of a conducting sphere, sphere~$1$ having $\varepsilon_1\to+\infty$, while all other spheres have null radii ($a_j\to0$ for $\forall j>1$). This corresponds to a cloud of point charges scattered around a charged conducting sphere. Then, in this case the many-body relation \eqref{E2_point_charges_N_spheres_s_kappa_zero} reduces to relation $\mathcal E^{(2)}\bigr|_{\kappa\to0} = \frac{-1}{8\pi\varepsilon_0\varepsilon_\text{sol}}\sum\limits_{i=2}^N q_i \sum\limits_{p=2}^N q_p \sum\limits_{l=1}^{+\infty} \frac{a_1^{2 l+1} P_l(\cos\gamma_{1 i,1 p})}{R_{1 i}^{l+1} R_{1 p}^{l+1}} = \frac{-1}{8\pi\varepsilon_0\varepsilon_\text{sol}}\sum\limits_{i=2}^N \frac{q_i^2}{a_1}\frac{(a_1/R_{1 i})^4}{1-(a_1/R_{1 i})^2} + \frac{-1}{8\pi\varepsilon_0\varepsilon_\text{sol}}\sum\limits_{i=2}^N q_i \sum\limits_{p=2,\,p\ne i}^N q_p \Bigl(\frac{-a_1}{R_{1 i} R_{1 p}}+\sum\limits_{l=0}^{+\infty} \frac{a_1^{2 l+1} P_l(\cos\gamma_{1 i,1 p})}{R_{1 i}^{l+1} R_{1 p}^{l+1}} \Bigr)$, which recovers the expressions very recently derived in \cite{RPSA_2024} and properly takes into account the correlations arising through the polarization/induction at the surface. According to the extensive literature overview found in \cite[Sec.~1]{RPSA_2024}, they can be comparable to the direct Coulombic interaction magnitude but have not yet been deeply considered, especially in many-body simulations and modelling ionization of nanospheres~\cite{SLT_2022}.
\begin{remark}
\label{attr_rep_remark0}
Let us examine the two-sphere ($N=2$) system more closely. In Appendix~\ref{two_spheres_kappa0_appendix} we derive the general expression of $\mathcal E^{(\ell)}\bigr|_{\kappa\to0, \, N=2}$. While the lowest-order Coulombic energy addend $\mathcal E^{(1)}\bigr|_{\kappa\to0, \; N=2} = \frac{q_1 q_2}{4\pi\varepsilon_0\varepsilon_\text{sol}R}$ patently cannot exhibit any counterintuitive interaction effect (such as like-charge attraction / opposite-charge repulsion or interaction between a charged particle and a neutral one), higher-order addends $\mathcal E^{(\ell)}$ of $\mathcal E^\text{Int}$ can. For instance, the next addend $\mathcal E^{(2)}\bigr|_{\kappa\to0, \; N=2}$ (see \eqref{E2_point_charges_N_spheres_s_kappa_zero_N_2}) can drive such effects depending on the values of the ratios $k_i\mathrel{:=}\varepsilon_i/\varepsilon_\text{sol}$, $i=\overline{1, 2}$: in particular, opposite-charge repulsion in the case of low polarizable spheres ($k_1<1$, $k_2<1$) and like-charge attraction in the case of highly polarizable spheres ($k_1>1$, $k_2>1$). It can also be directly observed in relation~\eqref{E2_point_charges_N_spheres_s_kappa_zero_N_2} that it strives to strengthen Coulombic like-charge repulsion in the former case and opposite-charge attraction in the latter case. Furthermore, we derive (see Remark~\ref{appendix_lca_isnt_possible_kappa_zero} for details) that energy addends $\mathcal E^{(\ell)}\bigr|_{\kappa\to0, \; N=2}(R)$ with odd $\ell=\overline{3, 5, \ldots}$ behave qualitatively like $\mathcal E^{(1)}\bigr|_{\kappa\to0, \; N=2}(R)$ while those with even $\ell=\overline{4, 6, \ldots}$ behave qualitatively like $\mathcal E^{(2)}\bigr|_{\kappa\to0, \; N=2}(R)$, provided that $\varepsilon_1$ and $\varepsilon_2$ lie on the same side of $\varepsilon_\text{sol}$. These findings rigorously entail \emph{the impossibility of opposite-charge repulsion in a less polarizable medium ($k_1>1$, $k_2>1$) and of like-charge attraction in a more polarizable medium ($k_1<1$, $k_2<1$)}. This behavior was observed very recently in the work \cite{DGC_softmat}, when computationally analyzing inter-sphere interactions for different fixed charges $q_i$ and ratios $k_i$. However, an exhaustive and analytical evidence/proof has been lacking so far. The impossibility of opposite-charge repulsion for polarisable spheres in vacuum was already derived in another recent work, \cite{Chan}, based on empirical polarizability-volume relationships.
\end{remark}
\begin{remark}
\label{energy_components_irregular_kappa_0}
\noindent
It is known (see \cite{Fish,LLF}) that in a two-sphere system ($N=2$) in the $\kappa=0$ case the corrections to the conventional Coulombic $O(R^{-1})$-term (i.e.~$\frac{q_1 q_2}{4\pi\varepsilon_0\varepsilon_\text{sol}R}$) will asymptotically behave like 
\begin{equation}
\label{Energy_correction_to_Coulombic_kappa_zero}
\mathfrak C_4/R^4 + \mathfrak C_6/R^6 + \mathfrak C_7/R^7 + \mathfrak C_8/R^8 + \cdots
\end{equation}
as $R\to+\infty$ (with some coefficients $\mathfrak C_i$, the explicit determination of which is quite involved, see \cite{Fish} and references therein). This significantly differs from the behavior of the interaction energy expansion \eqref{energy_expansion_components_interaction} at non-zero $\kappa>0$, which appears to be more regular in the sense that its addends are strictly ranked by screening order $\ell$ with no gaps in the inverse powers of $R$ -- indeed, in this case one has the corrections $\sum_{\ell=2}^{+\infty} \mathcal E^{(\ell)}$ to the leading interaction addend (DLVO term) $\mathcal E^{(1)} = \frac{q_1 q_2 e^{\Tilde a_1+\Tilde a_2}}{4\pi\varepsilon_0\varepsilon_\text{sol}(1+\Tilde a_1)(1+\Tilde a_2)}\frac{e^{-\Tilde R}}{R}$ (see \eqref{DLVO_energy_N_spheres_central_point_charges}) with all non-trivial addends $\mathcal E^{(\ell)}\propto (e^{-\Tilde R}/R)^\ell$ (where as usual $\Tilde R \mathrel{:=} \kappa R$). In contrast, the $\kappa=0$ case does not contain either $O(R^{-2})$ or $O(R^{-3})$ corrective terms, so that relation \eqref{E2_point_charges_N_spheres_s_kappa_zero_N_2} immediately elucidates the leading (namely, of $O(R^{-4})$-type, as it was envisaged by \eqref{Energy_correction_to_Coulombic_kappa_zero}) correction to the Coulomb law and provides its explicit expression. All other terms from \eqref{Energy_correction_to_Coulombic_kappa_zero} can also be reproduced  from the energy expansion addends $\mathcal E^{(\ell)}\bigr|_{\kappa\to0, \; N=2}$ obtained using our formalism (see Appendix~\ref{two_spheres_kappa0_appendix}). The $\mathcal E^{(\ell)}\bigr|_{\kappa\to0, \, N=2}$ expressions derived in Appendix~\ref{two_spheres_kappa0_appendix} provide more detailed insight into the origin of asymptotic correction \eqref{Energy_correction_to_Coulombic_kappa_zero} and explain where the $\{\mathfrak C_i\}$ coefficients come from. Besides, this formulation provides a new practical way for calculating any coefficient $\mathfrak C_i$ (it should be noted here that although the low-order coefficients $\mathfrak C_i$ are known at least up to $\mathfrak C_{11}$ as evidenced by \cite{Fish}, a general explicit construction of the $\mathfrak C_i$ is, to the best of our knowledge, still missing).
\end{remark}
\begin{remark}
\label{remark_MSF}
Finally, let us note that concerning the total electrostatic energy at zero ionic strength ($\kappa=0$), the recently developed image-based MSF of \cite{Qin2019,GXFQ,QPF,Freed2014,QLLJPF,LQF_2018} allows for the construction of perturbative energy expansions. The originally developed in \cite{Freed2014} for spheres with centrally-located point charges MSF (it was generalized later to handle particles also with embedded dipoles and multipoles and with the application of external fields \cite{GXFQ,LQF_2018,Qin2019}) iterates the integral equation for self-consistent potential and in essence reduces an original many-sphere problem to local one-body problems and self-consistently (by iterating the procedure) determines the field induced on a sphere by other spheres. To formalize the solution process, MSF introduces the notion of ``string'' like $\{i;j_n,\ldots,j_1;k\}$, where the centrally-located point charge of source sphere $k$ polarizes sphere $j_1$, that in turn polarizes sphere $j_2$, etc., until the propagated polarization field is measured by interaction of that of sphere $j_n$ with the centrally-located point charge of destination sphere $i$; such individual strings then serve as building blocks of the MSF's perturbative energy expansion which eventually consists of ``one-body'' contributions, in turn followed by ``two-body'', ``three-body'', ``four-body'', etc., corrections. The MSF approach yields a formal solution to the underlying Poisson equation and numerically reduces computational cost, but may suffer from convergence issues \cite{BesleyACR} (especially at short inter-sphere separations). The original approach of \cite{Freed2014} was later improved and extended in \cite{QPF}, where it was recognized that the strings $\{j; k; i\}$, $\{j; k, k; i\}$, $\{j; k, k, k; i\}$, etc., corresponding to the surface polarization continually re-polarizing itself, can be summed analytically (and in this way, an expression resembling the $\sum_l$-part of expression \eqref{E2_point_charges_N_spheres_s_kappa_zero} may be derived). However there are currently no known extensions of MSF for handling the LPBE case, and the terms of MSF expansions are in general expressed through (quite nontrivial) iterative integrals / special functions which also significantly complicates the analysis of the corresponding expansion terms (especially those caused by high-order polarization contributions) and obtaining explicit/closed-form formulas for them. The approach proposed in this study does not reduce many-body problem to one-body ones, but starts directly from the global many-body boundary-value problem and yields exact operator-series-based expansions, rigorously quantified at each Debye screening order and well-posed also in the $\kappa\to0$ limit.
\end{remark}

\subsubsection{Conditions for like-charge attraction / opposite-charge repulsion in two-sphere systems}
\label{regimes_lca_ocr_two_spheres_subs}
\begin{table*}
\begin{center}
\caption{The appearance of like-charge attraction / opposite-charge repulsion effects in two-sphere ($N=2$) systems.}
\label{tab:Ex1_2_spheres}
\begin{tabular}{|p{0.495\textwidth}|p{0.495\textwidth}|}
\hline
\multicolumn{2}{|c|}{\textbf{$\kappa=0$}}\\
\hline
\multicolumn{1}{|c|}{$\varepsilon_1$ and $\varepsilon_2 <\varepsilon_\text{sol}$ (typical of aqueous solutions)} & \multicolumn{1}{|c|}{$\varepsilon_1$ and $\varepsilon_2 >\varepsilon_\text{sol}$ (e.g.~conducting particles in solution, static fields)} \\
\hline
\multicolumn{2}{|c|}{Addends $\mathcal E^{(\ell)}|_{\kappa\to0, N=2}(R)$ with \textbf{odd} $\ell=\overline{3,5,\ldots}$ qualitatively behave like the Coulombic energy $\mathcal E^{(1)}|_{\kappa\to0, N=2}(R) = \frac{q_1 q_2}{4\pi\varepsilon_0\varepsilon_\text{sol} R}$.}\\
Addends $\mathcal E^{(\ell)}|_{\kappa\to0, N=2}(R)$ with \textbf{even} $\ell=\overline{2,4,\ldots}$ are always \textbf{repulsive} ($\forall q_1 q_2$), therefore: \emph{no like-charge attraction, but opposite-charge repulsion may occur.}
&
Addends $\mathcal E^{(\ell)}|_{\kappa\to0, N=2}(R)$ with \textbf{even} $\ell=\overline{2,4,\ldots}$ are always \textbf{attractive} ($\forall q_1 q_2$), therefore: \emph{no opposite-charge repulsion, but like-charge attraction may occur.} \\
\hline
\multicolumn{2}{|c|}{\textbf{$\kappa>0$}}\\
\hline
\multicolumn{1}{|c|}{$\varepsilon_1$ and $\varepsilon_2 \le\varepsilon_\text{sol}$ (typical of aqueous electrolytic solutions)} & \multicolumn{1}{|c|}{$\varepsilon_1$ and $\varepsilon_2 >\varepsilon_\text{sol}$ (e.g.~conducting particles in solution, static fields)} \\
\hline
Addends $\mathcal E^{(\ell)}|_{N=2}(R)$ with \textbf{odd} $\ell=\overline{3,5,\ldots}$ qualitatively behave like the DLVO energy $\mathcal E^{(1)}|_{N=2}(R) = \frac{q_1 q_2 e^{\kappa a_1} e^{\kappa a_2} e^{-\kappa R}}{4\pi\varepsilon_0\varepsilon_\text{sol} (1+\kappa a_1)(1+\kappa a_2) R}$ and addends $\mathcal E^{(\ell)}|_{N=2}(R)$ with \textbf{even} $\ell=\overline{2,4,\ldots}$ are always \textbf{repulsive} ($\forall q_1 q_2$), therefore: \emph{no like-charge attraction, but opposite-charge repulsion may occur.}
&
Addends $\mathcal E^{(\ell)}|_{N=2}(R)$ with \textbf{odd} $\ell=\overline{3,5,\ldots}$ not necessary qualitatively behave like the DLVO energy term $\mathcal E^{(1)}|_{N=2}(R)$ and addends $\mathcal E^{(\ell)}|_{N=2}(R)$ with \textbf{even} $\ell=\overline{2,4,\ldots}$ also do not possess a definite behavior (can be \textbf{attractive} as well as \textbf{repulsive} in general), therefore: \emph{like-charge attraction as well as opposite-charge repulsion may occur.} \\
\hline
\end{tabular}
\end{center}
\end{table*}
\noindent
Based on the analytical results provided by our formalism, namely the exact expressions of energy addends derived in Appendices \ref{two_spheres_kappa_nonzero_appendix} ($\kappa>0$) and \ref{two_spheres_kappa0_appendix} ($\kappa=0$), and taking into account the discussions of Remarks \ref{remark_asymm_diel_screening} and \ref{attr_rep_remark0} above, we summarize the possible behavior scenarios in two-sphere systems in Table~\ref{tab:Ex1_2_spheres}. It can be seen thence that, in addition to the polarizability of the particles and the surrounding medium conventionally considered as the drivers of such unusual/counterintuitive effects (see \cite{DGC_softmat} and references therein), the ionic strength factor (as $\kappa>0$) matters and can expand the variety of possible interaction types.

\subsection{$N$ spheres with centrally-located point charges plus dipoles}
\label{Appendix_screened_potential_coefficients_point_dipoles}
\noindent
The situation of the presence of free centrally located point charges (considered in Sec.~\ref{Appendix_screened_potential_coefficients_point} in details) can be easily extended by adding higher-order multipole (dipolar, quadrupolar, octupolar, etc.)~moments of the free charge distribution. Assuming point free (fixed) charges $q_i$ and dipoles $\mathbf p_i$ at the centers of particles (so that $\Hat\varPhi_{\text{in},i} = \frac{q_i\kappa}{4\pi\varepsilon_0\varepsilon_i}\frac{1}{\Tilde r_i} + \frac{\kappa^2}{4\pi\varepsilon_0\varepsilon_i}\frac{\mathbf p_i \cdot \Hat{\mathbf r}_i}{\Tilde r_i^2}$), the corresponding multipolar coefficients \eqref{varPhi_in_i_multipoles} will then be $\Hat L_{0 0,i} = \frac{q_i \kappa}{\sqrt{4\pi}\varepsilon_0\varepsilon_i}$, $\Hat L_{1 0,i} = \frac{1}{\sqrt{3}}\frac{\kappa^2 p_i \cos\Bar\theta_i}{\sqrt{4\pi}\varepsilon_0\varepsilon_i}$, $\Hat L_{1, -1,i} = \frac{1}{\sqrt{6}}\frac{\kappa^2 p_i e^{\imath\Bar\varphi_i} \sin\Bar\theta_i}{\sqrt{4\pi}\varepsilon_0\varepsilon_i}$, $\Hat L_{1 1,i} = \frac{-1}{\sqrt{6}}\frac{\kappa^2 p_i e^{-\imath\Bar\varphi_i} \sin\Bar\theta_i}{\sqrt{4\pi}\varepsilon_0\varepsilon_i}$, and all other $\Hat L_{n m,i}$ are zero; here the triple $(p_i, \Bar\theta_i, \Bar\varphi_i)$ describes the dipole $\mathbf p_i$ in spherical coordinates and $\imath$ is the complex unit. Now, using the above-specified coefficients $\{\Hat L_{n m,i}\}$ in the general relations~\eqref{GL_l_elementwise}, we arrive at the screening-ranged potential coefficients $\Tilde G_{n m,i}^{(\ell)}$ and $L_{n m,i}^{(\ell)}$ for this situation and we can further calculate the corresponding screening-ranged energy addends $\mathcal E^{(\ell)}$. We do not dwell here on the detailed calculation of the non-screened ($\ell=0$) and single-screened ($\ell=1$) energy contributions, as they are a quite straightforward generalization of those of the two-sphere ($N=2$) case (see \cite[Sec.~II~C]{our_jcp}); 

E.g.~to calculate $\mathcal E^{(2)}$ we need $L_{n m,i}^{(2)}$ (see \eqref{energy_expansion_components_abs_center}); to this end we obtain in this case:
\begin{align}
& L_{n m,i}^{(2)} =  \frac{-\varepsilon_\text{sol}}{\sqrt{4\pi}\varepsilon_0\Tilde a_i^{n+2}\alpha_n(\Tilde a_i,\varepsilon_i)} \sum_{j=1,j\ne i}^N \, \sum_{L,M} \frac{\mathcal H_{n m}^{L M}(\mathbf R_{i j})}{\alpha_L(\Tilde a_j,\varepsilon_j)} \notag \\
& \times \sum_{s=1, s\ne j}^N \biggl( \frac{q_s \kappa \beta_{L M, 0 0}(\Tilde a_j,\varepsilon_j,\mathbf R_{j s})}{\alpha_0(\Tilde a_s,\varepsilon_s) \Tilde a_s^2} + \frac{3\kappa^2 p_s}{\sqrt{6} \alpha_1(\Tilde a_s,\varepsilon_s) \Tilde a_s^3} \Bigl( \sqrt{2} \notag \\ 
& \times \beta_{L M, 1 0}(\Tilde a_j,\varepsilon_j,\mathbf R_{j s})\cos\Bar\theta_s + \beta_{L M, 1, -1}(\Tilde a_j,\varepsilon_j,\mathbf R_{j s})e^{\imath \Bar\varphi_s} \sin\Bar\theta_s \notag \\
& - \beta_{L M, 1 1}(\Tilde a_j,\varepsilon_j,\mathbf R_{j s}) e^{-\imath \Bar\varphi_s} \sin\Bar\theta_s \Bigr)\!\biggr) . \label{mon_dip_N_spheres_L2}
\end{align}
Now employing \eqref{mon_dip_N_spheres_L2} in equality 
\begin{equation}
\label{energy_N_mon_dip_2_screened_gen}
\begin{aligned}
\mathcal E^{(2)} = & \frac{1}{2\sqrt{4\pi}}\sum_{i=1}^N \bigl( q_i L_{0 0,i}^{(2)} + \frac{\sqrt{3}}{\sqrt{2}}\kappa p_i (\sqrt{2} L_{1 0,i}^{(2)} \cos\Bar\theta_i \\ 
& + L_{1, -1,i}^{(2)} e^{-\imath\Bar\varphi_i} \sin\Bar\theta_i - L_{1 1,i}^{(2)} e^{\imath\Bar\varphi_i} \sin\Bar\theta_i )\!\bigr) 
\end{aligned}
\end{equation}
(energy expression \eqref{energy_N_mon_dip_2_screened_gen} follows immediately from the general relation \eqref{energy_expansion_components_abs_center} when substituting there our custom coefficients $\{\Hat L_{n m,i}\}$ indicated above), we thus obtain the full exact quantification of the double-screened energy~$\mathcal E^{(2)}$. It is instructive to write down several starting terms of this $\mathcal E^{(2)}$ in expanded form -- namely, we obtain
\begin{equation}
\label{energy_double_screened_mon_dip_general}
\mathcal E^{(2)} = \mathcal E^{(2)}_\text{mon,mon} + \mathcal E^{(2)}_\text{mon,dip} + \mathcal E^{(2)}_\text{dip,dip} + \text{H.O.T.} ,
\end{equation}
where the leading monopolar-monopolar $\mathcal E^{(2)}_\text{mon,mon}$, monopolar-dipolar $\mathcal E^{(2)}_\text{mon,dip}$, and dipolar-dipolar $\mathcal E^{(2)}_\text{dip,dip}$ interactions terms are determined~by
\begin{widetext}
\begin{align*}
& \mathcal E^{(2)}_\text{mon,mon} \mathrel{:=} \sum_{i=1}^N \frac{q_i e^{\Tilde a_i}}{8\pi\varepsilon_0(1+\Tilde a_i)} \sum_{j=1,j\ne i}^N \, \sum_{p=1,p\ne j}^N \frac{q_p\kappa e^{\Tilde a_p}}{(1+\Tilde a_p)\varepsilon_\text{sol}} \frac{e^{-\Tilde R_{i j}}}{\Tilde R_{i j}} \frac{e^{-\Tilde R_{j p}}}{\Tilde R_{j p}}\biggl( \frac{d_{0,j}}{2} + \Bigl(1+\frac{1}{\Tilde R_{i j}}\Bigr)\Bigl(1+\frac{1}{\Tilde R_{j p}}\Bigr)\frac{3 d_{2,j}}{2} (\Hat{\mathbf R}_{i j}\cdot\Hat{\mathbf R}_{j p}) \biggr), \\
& \mathcal E^{(2)}_\text{mon,dip} \mathrel{:=} \sum_{i=1}^N \frac{-q_i e^{\Tilde a_i} \kappa^2}{8\pi\varepsilon_0(1+\Tilde a_i)} \sum_{j=1,j\ne i}^N \, \sum_{p=1,p\ne j}^N 3 e^{\Tilde a_p} d_{1,p}^\prime \frac{e^{-\Tilde R_{i j}}}{\Tilde R_{i j}} \frac{e^{-\Tilde R_{j p}}}{\Tilde R_{j p}}\biggl[\Bigl(1+\frac{1}{\Tilde R_{j p}}\Bigr)\frac{d_{0,j}}{2}(\mathbf p_p\cdot\Hat{\mathbf R}_{j p}) - \Bigl(1+\frac{1}{\Tilde R_{i j}}\Bigr)\frac{d_{2,j}}{2} \biggl(\!\Bigl(1+\frac{3}{\Tilde R_{j p}}+\frac{3}{\Tilde R_{j p}^2}\Bigr) \\ 
&\quad \times \bigl((\mathbf p_p\cdot\Hat{\mathbf R}_{i j})-3(\mathbf p_p\cdot\Hat{\mathbf R}_{j p})(\Hat{\mathbf R}_{i j}\cdot\Hat{\mathbf R}_{j p})\bigr) - (\mathbf p_p\cdot\Hat{\mathbf R}_{i j}) \biggr) \biggr] + \sum_{i=1}^N \frac{3\kappa^2 e^{\Tilde a_i} d_{1,i}^\prime}{8\pi\varepsilon_0} \sum_{j=1,j\ne i}^N \, \sum_{p=1,p\ne j}^N \frac{q_p e^{\Tilde a_p}}{1+\Tilde a_p} \frac{e^{-\Tilde R_{i j}}}{\Tilde R_{i j}} \frac{e^{-\Tilde R_{j p}}}{\Tilde R_{j p}}\biggl[\!\Bigl(1+\frac{1}{\Tilde R_{i j}}\Bigr)\frac{d_{0,j}}{2} \\ 
&\quad \times (\mathbf p_i\cdot\Hat{\mathbf R}_{i j}) - \Bigl(1+\frac{1}{\Tilde R_{j p}}\Bigr)\frac{d_{2,j}}{2}\biggl(\!\!\Bigl(1 +\frac{3}{\Tilde R_{i j}} +\frac{3}{\Tilde R_{i j}^2}\Bigr) \bigl((\mathbf p_i\cdot\Hat{\mathbf R}_{j p})-3(\mathbf p_i\cdot\Hat{\mathbf R}_{i j})(\Hat{\mathbf R}_{i j}\cdot\Hat{\mathbf R}_{j p})\bigr) - (\mathbf p_i\cdot\Hat{\mathbf R}_{j p})\!\biggr)\!\biggr], \\
& \mathcal E^{(2)}_\text{dip,dip} \mathrel{:=} \sum_{i=1}^N\frac{-3\varepsilon_\text{sol} \kappa^3 e^{\Tilde a_i} d_{1,i}^\prime}{8\pi\varepsilon_0} \sum_{j=1,j\ne i}^N \, \sum_{p=1,p\ne j}^N e^{\Tilde a_p} d_{1,p}^\prime \frac{e^{-\Tilde R_{i j}}}{\Tilde R_{i j}} \frac{e^{-\Tilde R_{j p}}}{\Tilde R_{j p}} \biggl[ \frac{3 d_{0,j}}{2}\Bigl(1+\frac{1}{\Tilde R_{i j}}\Bigr)\Bigl(1+\frac{1}{\Tilde R_{j p}}\Bigr) (\mathbf p_i\cdot\Hat{\mathbf R}_{i j})(\mathbf p_p\cdot\Hat{\mathbf R}_{j p}) + \frac{d_{2,j}}{2}\biggl\{ (\mathbf p_i\cdot\mathbf p_p) \\
&\quad - \Bigl(1 +\frac{3}{\Tilde R_{i j}} +\frac{3}{\Tilde R_{i j}^2}\Bigr) \bigl((\mathbf p_i\cdot\mathbf p_p)-3(\mathbf p_i\cdot\Hat{\mathbf R}_{i j}) (\mathbf p_p\cdot\Hat{\mathbf R}_{i j})\bigr)
- \Bigl(1 +\frac{3}{\Tilde R_{j p}} +\frac{3}{\Tilde R_{j p}^2}\Bigr) \bigl((\mathbf p_i\cdot\mathbf p_p)-3(\mathbf p_i\cdot\Hat{\mathbf R}_{j p}) (\mathbf p_p\cdot\Hat{\mathbf R}_{j p})\bigr) + \Bigl(1 +\frac{3}{\Tilde R_{i j}} +\frac{3}{\Tilde R_{i j}^2}\Bigr) \\ 
&\quad \times \Bigl(1 +\frac{3}{\Tilde R_{j p}} +\frac{3}{\Tilde R_{j p}^2}\Bigr) \bigl( (\mathbf p_i\cdot\mathbf p_p)-3(\mathbf p_i\cdot\Hat{\mathbf R}_{i j})(\mathbf p_p\cdot\Hat{\mathbf R}_{i j}) - 3(\mathbf p_i\cdot\Hat{\mathbf R}_{j p})(\mathbf p_p\cdot\Hat{\mathbf R}_{j p}) + 9 (\mathbf p_i\cdot\Hat{\mathbf R}_{i j})(\mathbf p_p\cdot\Hat{\mathbf R}_{j p})(\Hat{\mathbf R}_{i j}\cdot\Hat{\mathbf R}_{j p}) \bigr) \biggr\} \biggr] ,
\end{align*}
\end{widetext}
where as before it is denoted $d_{0,i} \mathrel{:=} e^{2\Tilde a_i} \frac{\Tilde a_i-1}{\Tilde a_i+1}+1$, $d_{2,i} \mathrel{:=} e^{2\Tilde a_i}\frac{(\varepsilon_i+2\varepsilon_{\mathsf{sol}})(\Tilde a_i-1)-\varepsilon_{\mathsf{sol}}\Tilde a_i^2}{(\varepsilon_i+2\varepsilon_{\mathsf{sol}})(1+\Tilde a_i)+\varepsilon_{\mathsf{sol}}\Tilde a_i^2}+1 $, and $d_{1,i}^\prime \mathrel{:=} \left((\varepsilon_i+2\varepsilon_\text{sol})(1+\Tilde a_i)+\varepsilon_\text{sol}\Tilde a_i^2\right)^{-1}$. The terms presented in $\mathcal E^{(2)}_\text{mon,mon}$, $\mathcal E^{(2)}_\text{mon,dip}$ and $\mathcal E^{(2)}_\text{dip,dip}$ originate from ($L\le1$)-addends of the series in \eqref{mon_dip_N_spheres_L2} calculated using the exact (with no truncations/approximations) relations for all the quantities appearing in~\eqref{mon_dip_N_spheres_L2} (see Appendix~\ref{appendix_properties_asympt_alpha_beta_H_nmLM} for the explicit values of $\alpha$, $\beta$, $\mathcal H$ used in \eqref{mon_dip_N_spheres_L2}); accordingly, Higher-Order Terms (H.O.T.) in \eqref{energy_double_screened_mon_dip_general}, caused by the ($L\ge2$)-addends of \eqref{mon_dip_N_spheres_L2}, rapidly decay with increasing $L$ thanks to the presence of factor $\alpha_L(\Tilde a_j,\varepsilon_j)^{-1}$ in \eqref{mon_dip_N_spheres_L2} (see Appendix~\ref{appendix_properties_asympt_alpha_beta}) and are formally at least of order $O(\Tilde a_j^5)$ as $a_j$ is small (which immediately follows from asymptotics \eqref{asymptotics_Kij_small_spheres_radii_1} applied to $\beta_{L M,\ldots}(\Tilde a_j,\varepsilon_j,\mathbf R_{j s})/\alpha_L(\Tilde a_j,\varepsilon_j)$ in~\eqref{mon_dip_N_spheres_L2}). 

Many-body formulas \eqref{mon_dip_N_spheres_L2}-\eqref{energy_double_screened_mon_dip_general} were not previously known, to the best of our knowledge; in the particular case of two bodies ($N=2$) relation \eqref{energy_double_screened_mon_dip_general} recovers the recent \cite[Eq.~(27)]{our_jcp}.

\subsection{A distinguished class of surface charge distributions: Janus particles}
\label{subsection_janus_particles}
\begin{figure}
\subfloat[\label{Janus_sigma_figure_canonical}Example of $\sigma_i^\text{f}$: AS about the canonical (global) axis~$\mathbf z$.]
{
\includegraphics[trim=2.65cm 2.9cm 1.65cm 1.5cm,clip=true,width=0.99\linewidth]{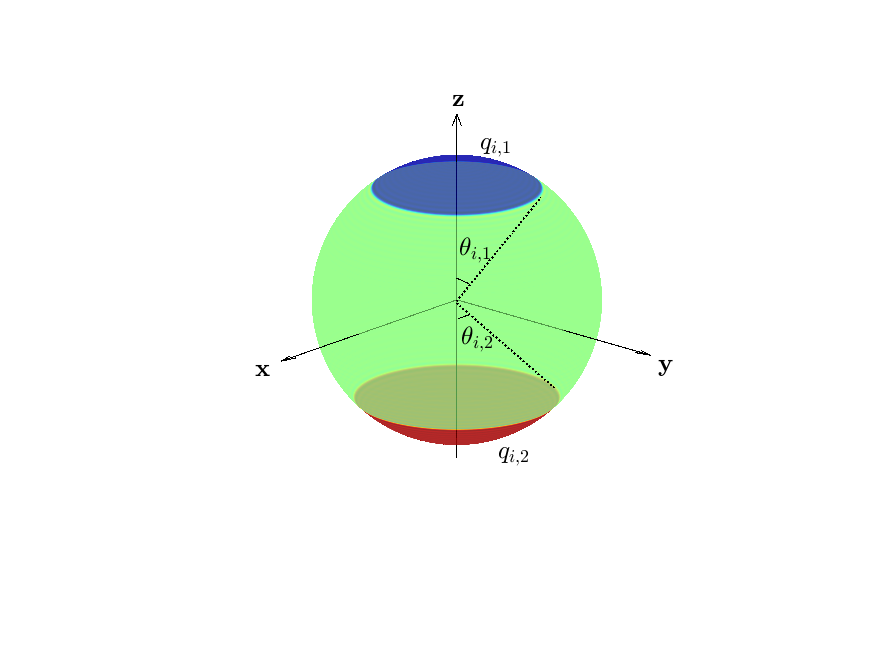}
}\hfill
\subfloat[\label{Janus_sigma_figure_rotated}Example of $\sigma_i^\text{f}$: a rotated coordinate system with the new axis $\mathbf z_i^\prime$ of AS having arbitrary orientation (spherical angles $\theta_i=\theta_i^r$ and $\varphi_i = \varphi_i^r$ give the orientation of the unit vector $\mathbf z_i^\prime$ with respect to the global coordinate system).]
{
\includegraphics[trim=2.7cm 3.0cm 1.7cm 1.5cm,clip=true,width=0.99\linewidth]{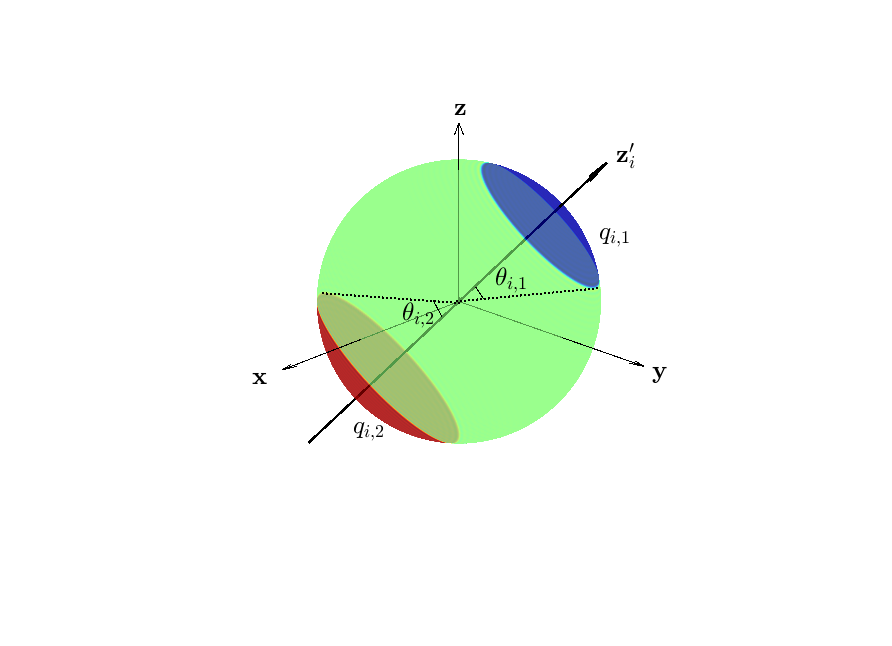}
}
\caption{Examples of $\sigma_i^\text{f}$ for (generalized) Janus particles: free charges $q_{i,1}$ and $q_{i,2}$ are uniformly distributed over spherical caps of polar angles $[0,\, \theta_{i,1}]$ and $[\pi-\theta_{i,2},\,  \pi]$, respectively, while the intermediate surface (i.e.~$(\theta_{i,1},\, \pi-\theta_{i,2})$) has no fixed charge (in particular, putting $q_{i,1}=-q_{i,2}$ and $\theta_{i,1}=\theta_{i,2}=\pi/2$ one arrives at the classical bipatchy Janus particles consisting of oppositely charged hemispheres).}
\label{Janus_sigma_figures}
\end{figure}

\noindent
An important class of heterogeneous ``patchy" surface distributions is the one characterizing Janus particles \cite{PopovHernan2023, NasGol2020, BoonJanus2010, GraafBoon2012, HRH2016, Mehr2020, Sciortino2009, Yu2021, ChungSW, SMM_omega_2024} (see Fig.~\ref{Janus_sigma_figures}). It turned out that electrostatic interactions of patchy/inhomogeneously-charged particles play a crucial role in, for instance, controlled aggregation, self-assembly and chain formation, facilitating the building of target superstructures of increased complexity (see e.g.~recent papers~\cite{Mehr2020,ChungSW,ZapClem_jpcb,WHO_sm_24} and also \cite{AM_nat2019,BianchiPCCP11,YKBD_lang2017,ASM_softm_2017,BYAD_pre_2013} and references therein). However, the corresponding DH-rigorous many-body analytical theory for their explicit quantification has to date remained elusive (see e.g.~recent studies \cite{PopovHernan2023,GLCBB_2025} aimed at constructing the interaction between two spherical particles, references therein and also earlier works \cite{Yu2021,GraafBoon2012,BoonJanus2010,HRH2016,BP1,BOLK2017,DemCruz2017,YHD_jcp_2015,SKB2015}).
\begin{figure}
\includegraphics[trim=2.7cm 3.5cm 1.7cm 1.65cm,clip=true,width=0.99\linewidth]{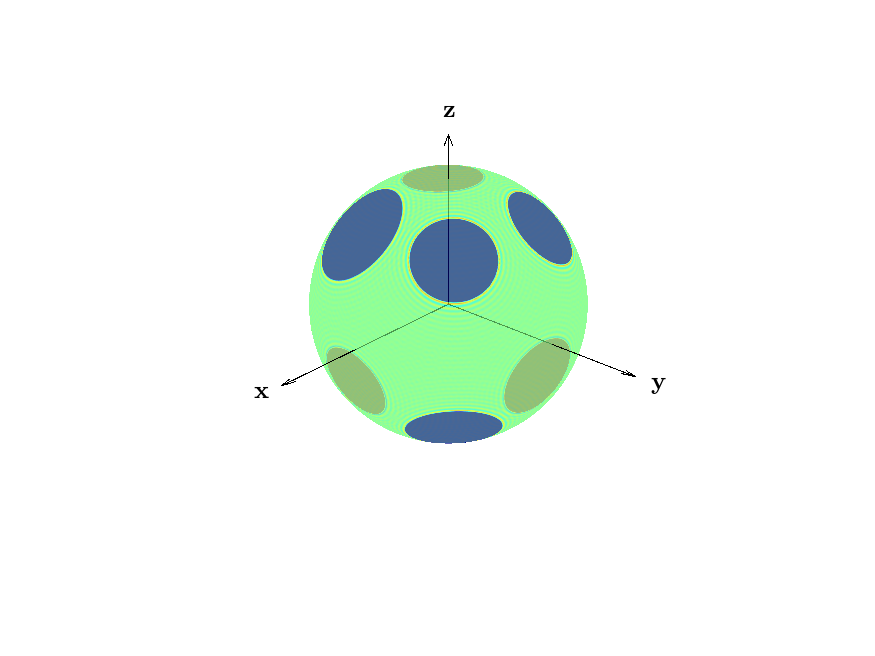}
\caption{Example of $\sigma_i^\text{f}$: a patchy charge distribution.}
\label{Janus_sigma_figure_arbitrary}
\end{figure}

Janus particles treatment leads to the adoption of specifying Fourier coefficients \eqref{sigma_nm_i_def} and can be easily treated within the analytical framework built here. Indeed, in the canonical case of a charge distribution with AS (azimuthal symmetry) with respect to a global axis $\mathbf z$ (Fig.~\ref{Janus_sigma_figure_canonical}), coefficients \eqref{sigma_nm_i_def} are equal to (see Appendix~\ref{appendix_Janus_proofs_canonical})
\begin{equation}
\label{Janus_sigma_nm_i_canonical}
\!\!\begin{aligned}
\sigma_{0 0,i}^\text{f} & = \frac{q_{i,1}+q_{i,2}}{\sqrt{4\pi} a_i^2} ;\quad\text{and  for  $n>0$:}\\
\sigma_{n m,i}^\text{f} & = \sqrt{\frac{\pi}{2 n+1}} \Bigl(\!\bigl(P_{n-1}(\cos\theta_{i,1})-P_{n+1}(\cos\theta_{i,1})\bigr)\varsigma_{i,1} \\
 &+ (-1)^n \bigl(P_{n-1}(\cos\theta_{i,2})-P_{n+1}(\cos\theta_{i,2})\bigr)\varsigma_{i,2}\!\Bigr)\delta_{m,0} ,
\end{aligned}
\end{equation}
where constants $\varsigma_{i,1} \mathrel{:=} \frac{q_{i,1}}{2\pi a_i^2 (1-\cos\theta_{i,1})}$ and $\varsigma_{i,2} \mathrel{:=} \frac{q_{i,2}}{2\pi a_i^2 (1-\cos\theta_{i,2})}$ (forming the overall distribution $\sigma_i^\text{f}$ in this case) are surface densities of free (fixed) charges $q_{i,1}$ and $q_{i,2}$ smeared out uniformly over the corresponding spherical caps (see Fig.~\ref{Janus_sigma_figure_canonical}); $P_n(\cdot) \; ( = P_n^0(\cdot) )$ are the usual Legendre polynomials (see \eqref{Ynm_definition}). Interestingly, definition \eqref{Janus_sigma_nm_i_canonical} remains finite in the limit $\theta_{i,1},\theta_{i,2}\to0$ (i.e.~when spherical caps collapse to point charges) yielding $\sigma_{n m,i}^\text{f} = \frac{\sqrt{2 n+1}}{2\sqrt{\pi} a_i^2}\bigl(q_{i,1}+(-1)^n q_{i,2}\bigr)$ (see \eqref{small_theta_asymptotic_Leg}). In the more general case of spherical caps centered around an arbitrary axis of AS ($\mathbf z_i^\prime$ -- see Fig.~\ref{Janus_sigma_figure_rotated}) coefficients \eqref{sigma_nm_i_def} can be calculated~as
\begin{equation}
\label{Janus_sigma_nm_i_rotated}
\sigma_{n m,i}^\text{f} = \sqrt{\frac{4\pi}{2 n+1}}(-1)^m Y_n^{-m}(\theta_i^r, \varphi_i^r)\sigma_{n 0,i}^\text{f, can} ,
\end{equation}
where the ``canonical'' coefficients $\sigma_{n 0,i}^\text{f, can}$ are defined by the previous identities \eqref{Janus_sigma_nm_i_canonical}; let us note that, while 3D rotations of spherical harmonics are in general described by Wigner D-matrices, the AS about $\mathbf z_i^\prime$ allows us to get rid of them (see Appendix~\ref{appendix_Janus_proofs_general}) and thus to express coefficients $\sigma_{n m,i}^\text{f}$ of the rotated charge distribution solely in terms of spherical harmonics (so that $Y_n^{-m}(\mathbf z_i^\prime) = Y_n^{-m}(\theta_i^r, \varphi_i^r)$ in \eqref{Janus_sigma_nm_i_rotated} take care of the proper transformation of canonical coefficients under rotations). Note that, due to the linearity of Fourier coefficients $\sigma_{n m,i}^\text{f}$ with respect to the surface free charge given, particles with arbitrary patchy charge distributions (consisting of any number of spherical surface caps of arbitrary sizes and charges) can be modeled by superpositions/sums of coefficients \eqref{Janus_sigma_nm_i_canonical}-\eqref{Janus_sigma_nm_i_rotated} corresponding to individual patches/caps -- see Fig.~\ref{Janus_sigma_figure_arbitrary} as an illustrative example of such a superposition. Note also that, for each particle $\Omega_i$, $\sigma_i^\text{f}$ is set individually (this fact is actually emphasized by subscript $i$ in relations \eqref{sigma_nm_i_def}, \eqref{Janus_sigma_nm_i_canonical}, \eqref{Janus_sigma_nm_i_rotated}, as well as in various charge parameters included therein).

Now the use of Janus-specific expressions \eqref{Janus_sigma_nm_i_rotated} in element-wise relations \eqref{L_nmi_single_screened_components} (which, recall, readily follow from the general screening-ranged operator relations for potential coefficients constructed in Sec.~\ref{statements_basics_section}) in turn opens the way to an exact quantification of the screening-ranged energy components~\eqref{energy_expansion_components_abs_surface} --- namely:
\begin{subequations}
\label{Janus_Born_all_}
\begin{align}
& \mathcal E^{(0)} = \left|\text{see comments and details in Appendix~\ref{appendix_Janus_proofs_Born}}\right| \notag \\ 
& = \frac{1}{2\varepsilon_0\kappa}\sum_{i=1}^N a_i^2 \sum_{n=0}^{+\infty} \frac{k_n(\Tilde a_i) (\sigma_{n 0,i}^\text{f, can})^2}{\alpha_n(\Tilde a_i,\varepsilon_i)} \label{Janus_Born_general_expr} \\ 
& = \underbrace{\frac{1}{8\pi\varepsilon_0\varepsilon_\text{sol}} \! \sum_{i=1}^N \frac{q_i^2}{(1\!+\!\Tilde a_i) a_i}}_{n=0} \! + \! \underbrace{\frac{3}{8\pi\varepsilon_0}\!\sum_{i=1}^N\frac{(1\!+\!\Tilde a_i)d_{1,i}^\prime}{a_i^3} \mathbf p_i \!\cdot\! \mathbf p_i}_{n=1} \label{Janus_Born_n=0_1_expr} \\ 
& + \  \cdots  \notag 
\end{align}
\end{subequations}
for $\ell=0$, 
\begin{subequations}
\label{Janus_DLVO_expr}
\begin{align}
& \mathcal E^{(1)} = \bigl|\text{employ expressions listed in Sec.~\ref{list_of_some_H_nmLM_values_}}\bigr| \notag \\
& = \frac{1}{8\pi\varepsilon_0\varepsilon_\text{sol}}\sum_{i=1}^N \sum_{j=1, j\ne i}^N \frac{q_i q_j e^{\Tilde a_i + \Tilde a_j} e^{-\Tilde R_{i j}}}{(1+\Tilde a_i) (1+\Tilde a_j) R_{i j}} \label{Janus_DLVO_MM_expr} \\
& - \frac{3}{8\pi\varepsilon_0} \sum_{i=1}^N \sum_{j=1, j\ne i}^N \frac{q_i e^{\Tilde a_i + \Tilde a_j} d_{1,j}^\prime e^{-\Tilde R_{i j}}}{(1+\Tilde a_i) R_{i j}}\Bigl(\kappa+\frac{1}{R_{i j}}\Bigr) \mathbf p_j\cdot\Hat{\mathbf R}_{i j} \label{Janus_DLVO_MD_expr_1} \\
& + \frac{3}{8\pi\varepsilon_0} \sum_{i=1}^N \sum_{j=1, j\ne i}^N \frac{q_j e^{\Tilde a_i + \Tilde a_j} d_{1,i}^\prime e^{-\Tilde R_{i j}}}{(1+\Tilde a_j) R_{i j}}\Bigl(\kappa+\frac{1}{R_{i j}}\Bigr) \mathbf p_i\cdot\Hat{\mathbf R}_{i j} \label{Janus_DLVO_MD_expr_2} \\
& + \frac{9\varepsilon_\text{sol}}{8\pi\varepsilon_0} \sum_{i=1}^N \sum_{j=1, j\ne i}^N e^{\Tilde a_i + \Tilde a_j} d_{1,i}^\prime d_{1,j}^\prime \biggl(\!\Bigl(\frac{\kappa}{R_{i j}}+\frac{1}{R_{i j}^2}\Bigr) \mathbf p_i \cdot \mathbf p_j \label{Janus_DLVO_DD_expr_1} \\
& \quad - \Bigl(\kappa^2+\frac{3\kappa}{R_{i j}}+\frac{3}{R^2_{i j}}\Bigr)(\mathbf p_i\cdot\Hat{\mathbf R}_{i j})(\mathbf p_j\cdot\Hat{\mathbf R}_{i j}) \biggr) \frac{e^{-\Tilde R_{i j}}}{R_{i j}} \label{Janus_DLVO_DD_expr_2} \\
& + \  \cdots \notag
\end{align}
\end{subequations}
for $\ell=1$, and so on. There, we denoted $d_{1,i}^\prime \mathrel{:=} \bigl((\varepsilon_i+2\varepsilon_\text{sol})(1+\Tilde a_i)+\varepsilon_\text{sol}\Tilde a_i^2 \bigr)^{-1}$, surface net free charge $q_i \mathrel{:=} q_{i,1}+q_{i,2}$ and vector $\mathbf p_i \mathrel{:=} \frac{1}{2}\bigl(q_{i,1}(1+\cos\theta_{i,1})-q_{i,2}(1+\cos\theta_{i,2})\bigr) a_i \mathbf z_i^\prime$ (note that one would get $\mathbf p_i = (2 a_i)q_{i,1}\mathbf z_i^\prime$ if $q_{i,2}=-q_{i,1}$ and $\theta_{i,1},\theta_{i,2}\to0$, therefore it is reasonable to call $\mathbf p_i$ here an effective physical dipole moment of the $i$-th Janus particle). Expressions \eqref{Janus_Born_n=0_1_expr} and \eqref{Janus_DLVO_MM_expr}-\eqref{Janus_DLVO_DD_expr_2} display the energy addends corresponding to $n, L\le 1$ (see \eqref{L_nmi_single_screened_components}, \eqref{energy_expansion_components_abs_surface}) -- they precisely reproduce, with no approximation, the corresponding $R_{i j}$-independent and single-screened terms of the true/analytical energy series for $n, L\le 1$, respectively. By analogy with pairwise interactions of particles with centrally located point multipoles (see recent paper \cite{our_jcp} for an exhaustive analytical description of such interactions) the terms \eqref{Janus_DLVO_MM_expr}, \eqref{Janus_DLVO_MD_expr_1}-\eqref{Janus_DLVO_MD_expr_2} and \eqref{Janus_DLVO_DD_expr_1}-\eqref{Janus_DLVO_DD_expr_2} may be identified with monopole-monopole, monopole-dipole and dipole-dipole interactions, respectively. Let us note that an extension of DLVO theory to describe the electrostatic pair-interaction of dielectric particles with a substantial dipolar contribution to the surface charge, such as Janus particles, was recently derived in \cite{GraafBoon2012} -- namely, \cite[Sec.~III~C]{GraafBoon2012} considered equal-sized Janus particles consisting of hemispheres with possibly different charge. It is easy to show that in this particular case our relations \eqref{Janus_DLVO_MM_expr}-\eqref{Janus_DLVO_DD_expr_2} subsume/reproduce the corresponding result of \cite[Eqs.~(19), (20), (A16), (A18)]{GraafBoon2012} as well (with a minor difference that \cite[Eq.~(19b)]{GraafBoon2012} approximates the exact factor $\frac{e^{-\Tilde R_{i j}}}{R_{i j}}\bigl(\kappa+\frac{1}{R_{i j}}\bigr)$ by $\frac{e^{-\Tilde R_{i j}}}{R_{i j}^2}$ in monopole-dipole interactions, see~\eqref{Janus_DLVO_MD_expr_1}-\eqref{Janus_DLVO_MD_expr_2}). Considering the further terms (i.e.~those with $n, L > 1$) in \eqref{L_nmi_single_screened_components} and \eqref{energy_expansion_components_abs_surface} one gets the exact analytical quantification of higher-order interactions (which may be identified with effective quadrupolar, octupolar, and so on, interactions, similarly to what was done in relations \eqref{Janus_Born_all_}-\eqref{Janus_DLVO_expr} above). In a fully similar manner, employing Janus-specific coefficients \eqref{Janus_sigma_nm_i_rotated} in the general screening-ranged expansions derived above (see Sec.~\ref{Appendix_screened_potential_coefficients_surface}), the corresponding energy components $\mathcal E^{(\ell)}$ for higher screening orders $\ell\ge2$ are also readily obtained --- e.g.~it is easy to show (see Appendix~\ref{appendix_ell_2_screened_proof}) that for $\ell=2$ the double-screened energy $\mathcal E^{(2)}$, up to monopole-monopole contributions, takes exactly the form of equation \eqref{energy_N_monopoles_2_screened_legendre_full} in which the charges $q_i$ are now expressed in terms of the Janus surface net charges $q_i = q_{i,1}+q_{i,2}$. Expressions for higher-order multipolar contributions such as monopole-dipole/dipole-dipole (similar to \eqref{energy_double_screened_mon_dip_general}), etc., can also be easily obtained within the same framework. Let us note that while some analytical explicit results towards pairwise interactions ($\ell=1$) of Janus particles are known \cite{GraafBoon2012}, no such results were to date known, to the best of our knowledge, for higher-order screened ($\ell\ge2$) DH-rigorous contributions.

\subsection{Some other special cases of fixed charge distributions}
\label{Appendix_screened_potential_coefficients_other}
\subsubsection{$N$ spheres with clouds of point charges}
\label{Appendix_screened_potential_coefficients_cloud}
\noindent
A straightforward generalization of the case considered in Sec.~\ref{Appendix_screened_potential_coefficients_point} is to assume that each sphere $i$ ($i=\overline{1,\ldots,N}$) contains $N_{i,q}$ free point charges $q_{i,p}$ ($p=\overline{1,\ldots,N_{i,q}}$) located at points $\mathbf r_{i,p}$ inside this sphere, i.e.~$\rho_i^\text{f}(\mathbf r) = \sum_{p=1}^{N_{i,q}} q_{i,p} \delta(\mathbf r - \mathbf r_{i,p})$ ($\delta(\cdot)$ is a Dirac delta). Indeed, in this case we will just have multipolar moments $\{\Hat L_{n m,i}\}$ determined by $$\Hat L_{n m,i} = \frac{\kappa}{(2 n+1)\varepsilon_i\varepsilon_0}\sum_{p=1}^{N_{i,q}} q_{i,p} \Tilde r_{i,p}^n Y_n^m(\Hat{\mathbf r}_{i,p})^\star$$ in the corresponding local coordinate frames and the general construction of screening-ranged expansions developed in Sec.~\ref{statements_basics_section} steps in with no changes; see also numerical illustrations in Sec.~\ref{section_numeric_subsection_N_spheres} below.
\subsubsection{$N$ spheres with uniform surface charge distributions}
\label{Appendix_screened_potential_coefficients_surf_uni}
\noindent
Taking into account Remark~\ref{remark_point_uniform_charges_the_same} and employing the constant densities $\sigma_i^\text{f} = q_i/(4\pi a_i^2)$ (see \eqref{uniform_surf_charge_distr_}) in \eqref{energy_expansion_components_abs_surface} and in the screening-ranged expressions of Sec.~\ref{Appendix_screened_potential_coefficients_surface}, it is then easy to deduce that the values of $\Tilde G^{(\ell)}_{n m,i}$ (as $\ell\ge0$), $L^{(\ell)}_{n m,i}$ (as $\ell\ge1$), and $\mathcal E^{(\ell)}$ (as $\ell\ge1$) coincide with those of the situation of centrally-located point charges considered in Sec.~\ref{Appendix_screened_potential_coefficients_point}; as for $L^{(0)}_{n m,i}$ we obtain $L_{n m,i}^{(0)}  = \frac{q_i \delta_{n,0}}{\sqrt{4\pi}\varepsilon_0\varepsilon_\text{sol} a_i (1+\Tilde a_i)}$, from which then immediately follows $\mathcal E^{(0)}$ representing the sum of the known individual/one-body solvation/Born-Kirkwood energy expressions,
\begin{equation}
\label{Born_energy_N_spheres_central_surf_unif}
\mathcal E^{(0)} = \frac{1}{8\pi\varepsilon_0\varepsilon_\text{sol}} \sum_{i=1}^N \frac{q_i^2}{(1+\kappa a_i) a_i} ,
\end{equation}
of spheres with uniform surfacic free distribution (see e.g.~\cite{Sokoloff2021} for the corresponding expression at zero ionic strength).

\section{Numerical validation of the proposed formalism} 
\label{section_numeric}
\begin{figure*}
\subfloat[\label{full_dlvo_equal_mon}$a_1=a_2$.]
{
\includegraphics[trim=0.0cm 0.0cm 0.98cm 0.5cm,clip=true,width=0.45\linewidth]{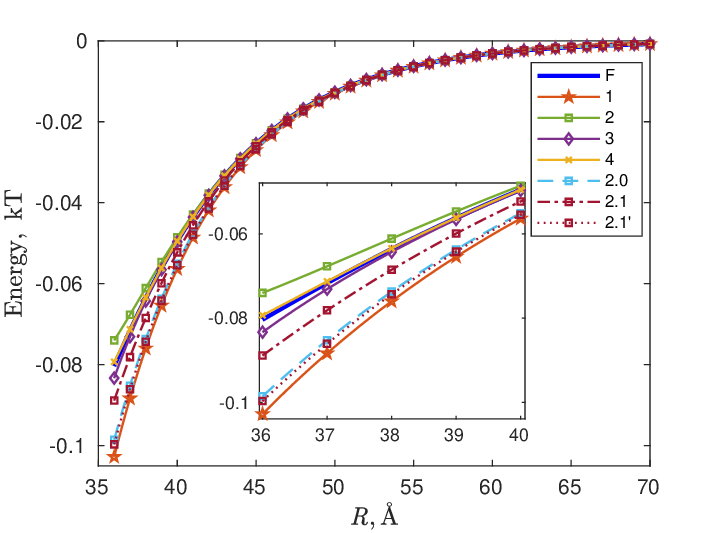}
}
\subfloat[\label{full_dlvo_unequal_2to1_mon}$a_1=2 a_2$.]
{
\includegraphics[trim=0.0cm 0.0cm 0.98cm 0.5cm,clip=true,width=0.45\linewidth]{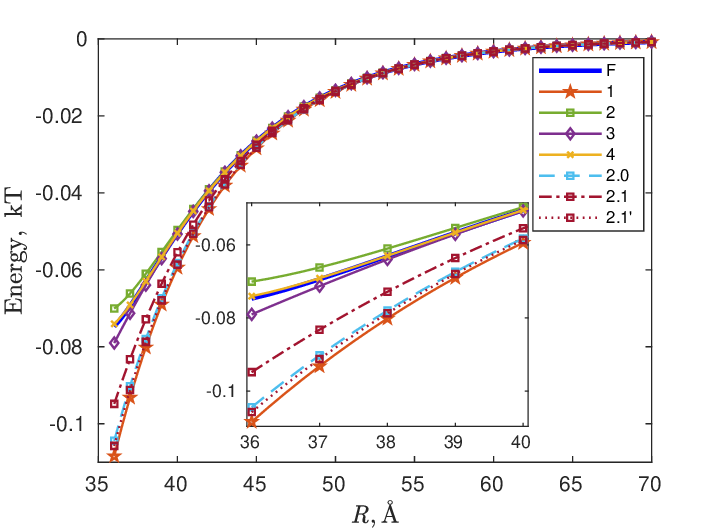}
}\hfill
\subfloat[\label{full_dlvo_unequal_10to1_mon}$a_1=10 a_2$.]
{
\includegraphics[trim=0.16cm 0.0cm 0.98cm 0.64cm,clip=true,width=0.45\linewidth]{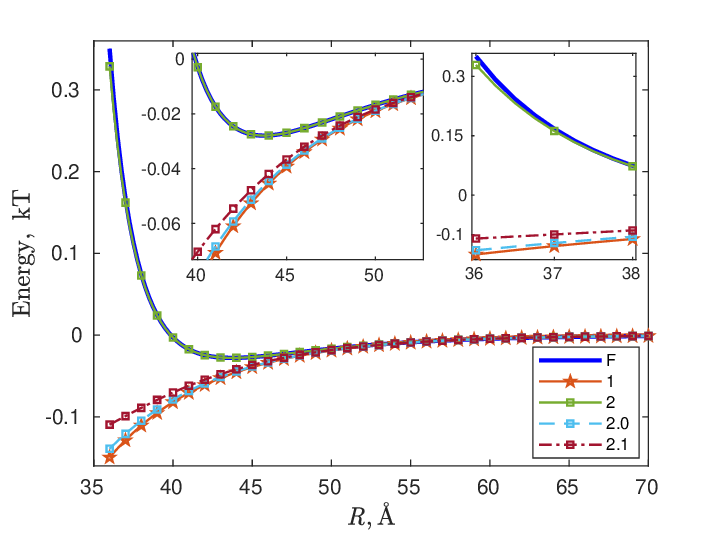}
}
\subfloat[\label{full_dlvo_unequal_100to1_mon}$a_1=100 a_2$.]
{
\includegraphics[trim=0.16cm 0.0cm 0.98cm 0.64cm,clip=true,width=0.45\linewidth]{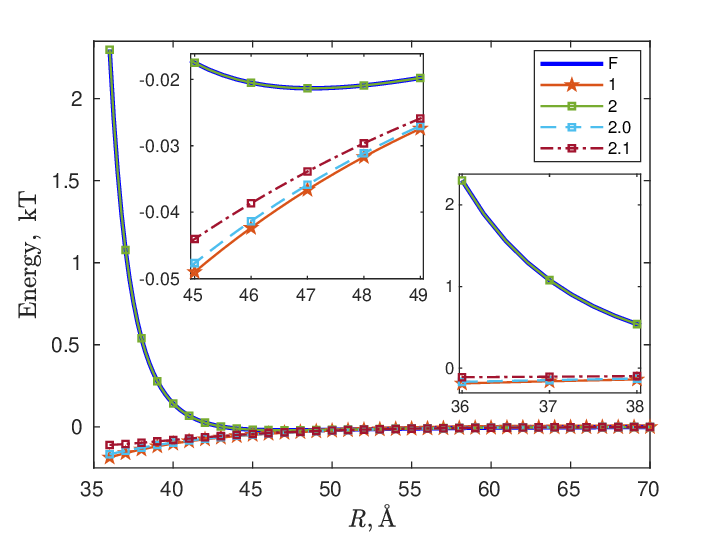}
}
\caption{Interaction energy, $a_1+a_2=35$~\AA. Line~F depicts the full (true) $\mathcal E^\text{Int}$. Lines 1, 2, 3, and 4 depict $\mathcal E^{(1)}$ (i.e.~DLVO in this example), $\mathcal E^{(1)}+\mathcal E^{(2)}$, $\mathcal E^{(1)}+\mathcal E^{(2)}+\mathcal E^{(3)}$, and $\mathcal E^{(1)}+\mathcal E^{(2)}+\mathcal E^{(3)}+\mathcal E^{(4)}$, respectively (in Figures \ref{full_dlvo_unequal_10to1_mon} and \ref{full_dlvo_unequal_100to1_mon}, Lines 3 and 4 are difficult to distinguish visually from Lines 2 and F, and are therefore not drawn). Finally, Lines 2.0 (MZ approximation), 2.1, and 2.1' (FLL approximation) depict $\mathcal E^{(1)}+\mathcal E^{(2)}$ with $\mathcal E^{(2)}$ being here approximated by employing the first addend of \eqref{L00_monopolar_two_spheres_}, the first two addends of \eqref{L00_monopolar_two_spheres_}, and the simplified (see text for details) first two addends of \eqref{L00_monopolar_two_spheres_}, respectively. All energy curves shown are measured in $\mathrm{k T}$ ($\mathrm k$ is the Boltzmann constant, $\mathrm T$ is the temperature). (Embedded insets show close-up views.)}
\label{full_dlvo_mon}
\end{figure*}
\begin{figure}
\includegraphics[trim=0.2cm 0.0cm 1.05cm 0.56cm,clip=true,width=0.93\linewidth]{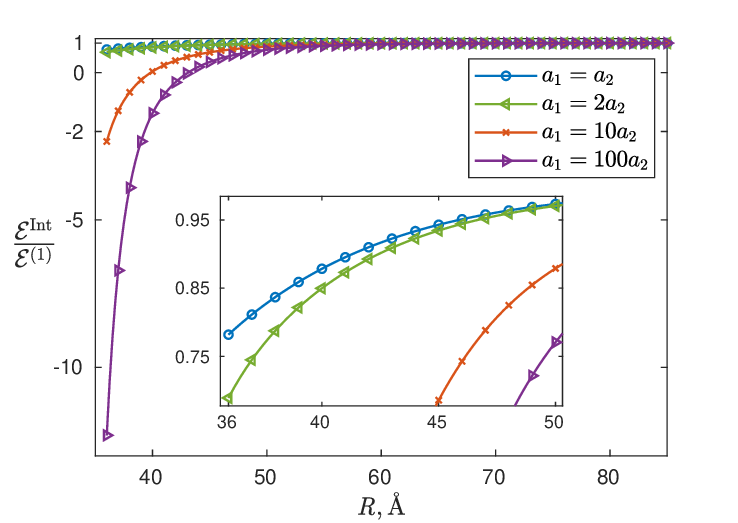}
\caption{The ratio of the total (full) interaction energy $\sum_{\ell=1}^{+\infty}\mathcal E^{(\ell)} = \mathcal E^\text{Int}$ to its leading DLVO addend~$\mathcal E^{(1)}$. (Embedded inset shows a close-up view.)}
\label{ratio_full_dlvo_mon}
\end{figure}
\begin{figure*}
\subfloat[\label{full_dlvo_unequal_10to1_mon_attr1}$q_2=-2e$.]
{
\includegraphics[trim=0.25cm 0.0cm 0.96cm 0.54cm,clip=true,width=0.45\linewidth]{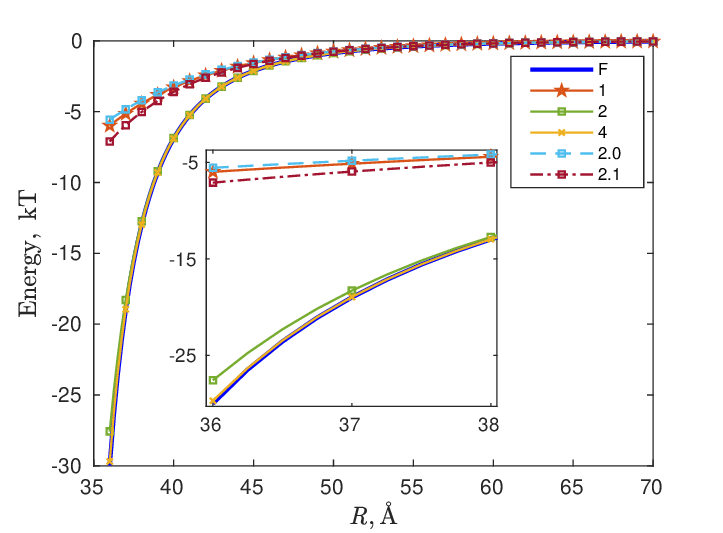}
}
\subfloat[\label{full_dlvo_unequal_10to1_mon_attr2_tr}$q_2=2e$.]
{
\includegraphics[trim=0.25cm 0.0cm 0.96cm 0.64cm,clip=true,width=0.45\linewidth]{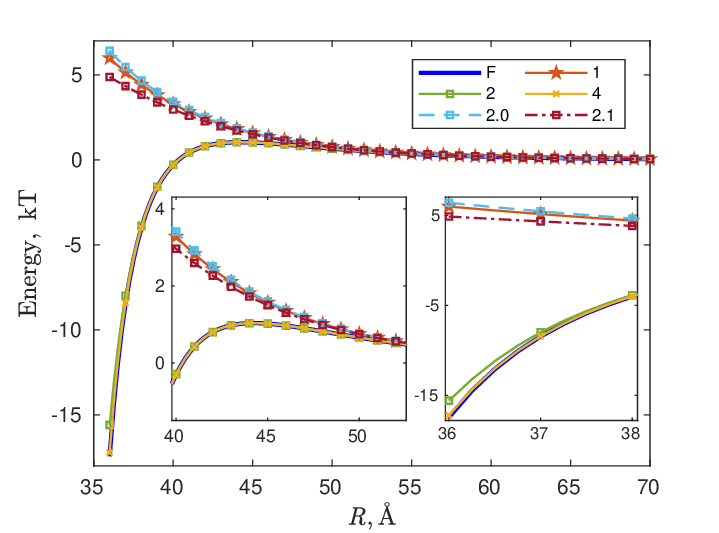}
}
\caption{The case of dielectric constants $\varepsilon_1=\varepsilon_2=80$, $\varepsilon_\text{sol}=2$; $q_2=\pm2e$. All other parameters and line legends are as in Fig.~\ref{full_dlvo_unequal_10to1_mon}.}
\label{full_dlvo_unequal_10to1_mon_attr}
\end{figure*}
\begin{figure}
\includegraphics[trim=0.14cm 0.0cm 1.05cm 0.62cm,clip=true,width=0.93\linewidth]{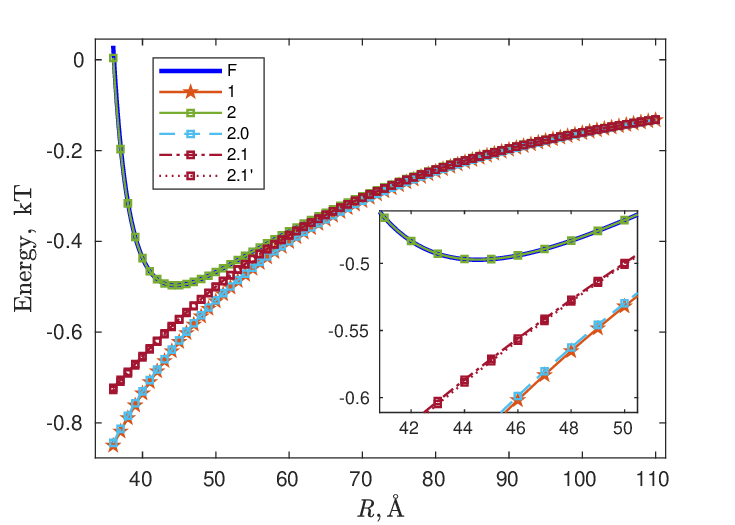}
\caption{$\kappa^{-1}=10^2$~\AA~(all other parameters and line legend are as in Fig.~\ref{full_dlvo_unequal_10to1_mon}).}
\label{full_dlvo_mon_kappa_1e-2}
\end{figure}
\begin{figure*}
\subfloat[\label{full_dlvo_dip_10}Interaction energy $\mathcal E^\text{Int}$ and its approximations.]
{
\includegraphics[trim=0.15cm 0.0cm 1.0cm 0.6cm,clip=true,width=0.47\linewidth]{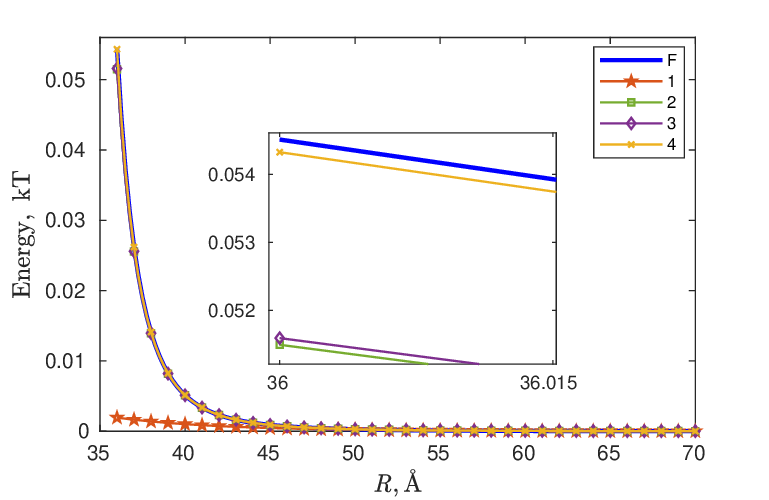}
}\hfill
\subfloat[\label{ratio_full_dlvo_dip}The ratio of $\sum_{\ell=1}^{+\infty}\mathcal E^{(\ell)} = \mathcal E^\text{Int}$ to its leading addend~$\mathcal E^{(1)}$.]
{
\includegraphics[trim=0.15cm 0.0cm 1.0cm 0.5cm,clip=true,width=0.43\linewidth]{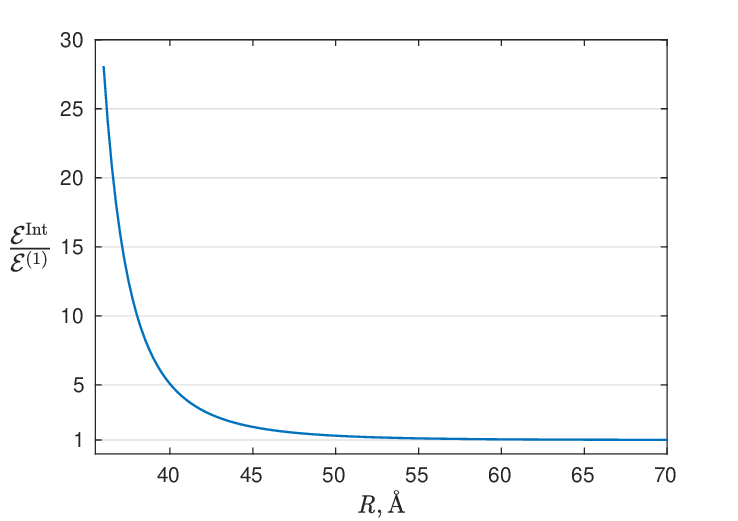}
}
\caption{Dipole-dipole interaction: dipole moment magnitudes $p_1=3$~$e$\AA, $p_2=2$~$e$\AA, spherical angles $\Bar\theta_1=\pi/6$, $\Bar\varphi_1=\pi/3$, $\Bar\theta_2=3\pi/4$, $\Bar\varphi_2=\pi/4$, centers $\mathbf x_1$ and $\mathbf x_2$ are $z$-aligned; all other parameters and line legend are as in Fig.~\ref{full_dlvo_unequal_10to1_mon}. (Embedded insets show close-up views.)}
\label{full_dlvo_unequal_10_dip}
\end{figure*}
\noindent
In numerical calculations, due to the practical impossibility of operating with infinite size matrices, it is necessary to put an upper bound $n_\text{max}$ to the index $n$ in \eqref{Lin_eqs_Phi_in_out}. This reduces the size of the assembled linear systems \eqref{eqs_G_intermediate_} and \eqref{global_lin_sys1} governing the potential coefficients in \eqref{Lin_eqs_Phi_in_out}. In practical calculations, the choice of a suitable value can be performed by e.g.~gradually increasing $n_\text{max}$ and monitoring the convergence of the potentials and/or energy values~\cite{our_jcp,our_jpcb,DiFlorio2025NextGenPB}. Various computational aspects of convergence of numerical solutions in two-particle ($N=2$) systems with respect to increasing the user-defined cutoff parameter $n_\text{max}$ have already been systematically benchmarked and described in detail in \cite{our_jcp} (spherical particles) and also in~\cite{our_jpcb}. Therefore, here we solely focus on how higher-order screened energy addends $\mathcal E^{(\ell)}$ impact on the total interparticle interaction energy (while using in our numerical assessment an $n_\text{max}$ already at convergence)---an issue that has never been studied in detail before. Indeed, since the formalism built in this study provides, for the first time in the many body framework, analytical expressions for the exact quantification of the $\mathcal E^{(\ell)}$ addends of arbitrary screening order $\ell$ in the full interaction energy expansion $\mathcal E^\text{Int}=\sum_{\ell=1}^{+\infty}\mathcal E^{(\ell)}$ (see Sec.~\ref{general_expansions_en_} and formula~\eqref{energy_expansion_components_interaction}), thus we will take advantage to observe how various screened energy terms contribute to $\mathcal E^\text{Int}$. It will emerge that the double-screened energy $\mathcal E^{(2)}$ indeed plays a decisive role in the fundamental corrections to the lowest-order single-screened $\mathcal E^{(1)}$ addend (see Remark~\ref{energy_expansion_fast_convergence}).

\subsection{The case of two ($N=2$) monopolar spheres: interaction energy beyond the leading single-screened DLVO term~$\mathcal E^{(1)}$}
\label{section_numeric_subsection_2_spheres}
\noindent
For two monopolar spheres $i$ and $j$ bearing charges $q_i$ and $q_j$ (with no loss of generality we can consider here centrally-located point charges, since the same charges uniformly distributed over surfaces lead to an equivalent formulation and essentially analogous expressions for the interaction potentials -- see Remark~\ref{remark_point_uniform_charges_the_same}), the single-screened addend $\mathcal E^{(1)}$ is actually the familiar DLVO pairwise interaction energy term (see Sec.~\ref{Appendix_screened_potential_coefficients_point_kappa_nonzero}): $\mathcal E^{(1)} = \mathcal E^\text{DLVO} = \frac{1}{4\pi\varepsilon_0\varepsilon_\text{sol}}\frac{q_i q_j e^{\Tilde a_i+\Tilde a_j-\Tilde R_{i j}}\kappa}{(1+\Tilde a_i)(1+\Tilde a_j)\Tilde R_{i j}}$. 

Higher-order screened energy addends $\mathcal E^{(\ell)}$ (with $\ell\ge2$) are determined by $\mathcal E^{(\ell)} = \frac{1}{2\sqrt{4\pi}}\sum_{i=1}^2 q_i L_{0 0, i}^{(\ell)}$ (see \eqref{energy_N_monopoles_ell_screened_}), where $L_{0 0, i}^{(\ell)}$ can be obtained putting $N=2$ in the general (many-body) expressions given in Sec.~\ref{Appendix_screened_potential_coefficients_point_kappa_nonzero} -- e.g.~for calculating the leading correction $\mathcal E^{(2)}$ to DLVO we obtain (see~\eqref{L00_point_charges_N_spheres_s}) 
\begin{equation}
\label{L00_monopolar_two_spheres_}
\begin{aligned}
L_{0 0, i}^{(2)} & = \frac{- q_i e^{2 \Tilde a_i } \kappa }{\sqrt{4\pi}\varepsilon_0\varepsilon_\text{sol}(1+\Tilde a_i)^2}\sum_{l=0}^{+\infty} \left(k_l(\Tilde R)\right)^2 \biggl( \frac{(\varepsilon_j-\varepsilon_\text{sol}) l}{\Tilde a_j} \\
& \times i_l(\Tilde a_j) -\varepsilon_\text{sol}i_{l+1}(\Tilde a_j) \! \biggr) \frac{2 l+1}{\alpha_l(\Tilde a_j,\varepsilon_j)} , 
\end{aligned}
\end{equation}
$i=\overline{1,2}$, $j=3-i$, $\Tilde R \mathrel{:=} \Tilde R_{i j}$. For instance, writing out the first two addends of series \eqref{L00_monopolar_two_spheres_} immediately details the leading terms of $\mathcal E^{(2)}=\frac{q_i L_{0 0, i}^{(2)} + q_j L_{0 0, j}^{(2)}}{2\sqrt{4\pi}}$ as follows:
\begin{equation}
\label{double_screened_energy_mon_2sphere}
\mathcal E^{(2)} = \frac{\kappa e^{-2\Tilde R}}{8\pi\varepsilon_0\varepsilon_\text{sol}\Tilde R^2}\bigl(\mathfrak E_i + \mathfrak E_j + \text{H.O.T.}\bigr),
\end{equation}
where $\mathfrak E_i \mathrel{:=} \frac{q_i^2 e^{2\Tilde a_i}}{2(1+\Tilde a_i)^2}\Bigl(d_{0,j}-3\bigl(1+\frac{1}{\Tilde R}\bigr)^2 d_{2,j}\Bigr)$ exactly represents the first two addends ($l=0, 1$) of \eqref{L00_monopolar_two_spheres_}, $d_{0,i} = e^{2\Tilde a_i} \frac{\Tilde a_i-1}{\Tilde a_i+1}+1$, $d_{2,i} = e^{2\Tilde a_i}\frac{(\varepsilon_i+2\varepsilon_\text{sol})(\Tilde a_i-1)-\varepsilon_\text{sol}\Tilde a_i^2}{(\varepsilon_i+2\varepsilon_\text{sol})(1+\Tilde a_i)+\varepsilon_\text{sol}\Tilde a_i^2}+1$, while Higher-Order Terms (H.O.T.) represent addends $l\ge2$ of \eqref{L00_monopolar_two_spheres_} (note that, as immediately follows from \eqref{asymptotics_Kij_small_spheres_radii}, the general $l$-th addend of \eqref{L00_monopolar_two_spheres_} is of order $O(\Tilde a_j^{2 l+1})$ for small $a_j$). Interestingly, if only the ($l=0$)-addend of \eqref{L00_monopolar_two_spheres_} were taken into account, we would arrive at $\mathcal E^{(2)}\approx \frac{\kappa e^{-2\Tilde R}}{8\pi\varepsilon_0\varepsilon_\text{sol}\Tilde R^2}\Bigl(\frac{q_i^2 e^{2\Tilde a_i} d_{0,j}}{2(1+\Tilde a_i)^2}+\frac{q_j^2 e^{2\Tilde a_j} d_{0,i}}{2(1+\Tilde a_j)^2}\Bigr)$, which recovers the expression derived by McClurg \& Zukoski \cite{McClurg} (MZ approximation, see \cite[Eq.~(37)]{McClurg}). Moreover, by approximating $d_{0,i}$ and $d_{2,i}$ in \eqref{double_screened_energy_mon_2sphere} by simplified factors\footnote{Apart from the general asymptotics \eqref{asymptotics_Kij_small_spheres_radii_1} readily capable to tell us the asymptotic infinitesimal order (i.e.~$O(\Tilde a_i^3)$) for $d_{0,i}$ and $d_{2,i}$, let us examine their behavior more closely. Employing Laurent series around the zero point we see that small parameter expansions with respect to small $\Tilde a_i$ hold: $d_{0,i} = \frac{2}{3} \Tilde a_i^3 + \frac{2}{5} \Tilde a_i^5 + O(\Tilde a_i^6)$ (formally valid as $\left|\Tilde a_i\right| < 1$) and $d_{2,i} = \frac{2(\varepsilon_i-\varepsilon_\text{sol})}{3(\varepsilon_i+2\varepsilon_\text{sol})}\Tilde a_i^3 - \frac{2(\varepsilon_i\varepsilon_\text{sol} - \varepsilon_i^2 +\varepsilon^2_\text{sol})}{5(\varepsilon_i+2\varepsilon_\text{sol})^2}\Tilde a_i^5 + O(\Tilde a_i^6)$ (formally valid as $\left|\Tilde a_i\right| < \Tilde a_i^* \mathrel{:=} \sqrt{2+\varepsilon_i/\varepsilon_\text{sol}}$ if $\varepsilon_i<2\varepsilon_\text{sol}$ and as $\left|\Tilde a_i\right| < \Tilde a_i^* \mathrel{:=} 2\big/\bigl( 1+\sqrt{(\varepsilon_i-2\varepsilon_\text{sol})/(\varepsilon_i+2\varepsilon_\text{sol})} \bigr)$ if $\varepsilon_i\ge2\varepsilon_\text{sol}$; note also that always $\Tilde a_i^* > 1$ for finite dielectric constants). To be noted also the useful fact that (as it can be concluded from \eqref{alpha_beta_definitions}, \eqref{in_kn_are_positive_}, \eqref{alpha_is_positive_ineq}) $\forall\Tilde a_i>0$ it holds $d_{0,i}>0$, and $d_{2,i}<0$ (in the ``biomolecular'' case $\varepsilon_i<\varepsilon_\text{sol}$).\label{small_parameter_expansions_d0_d2}} $d_{0,i} \approx \frac{2}{3}\Tilde a_i^3$ and $d_{2,i} \approx \frac{2(\varepsilon_i-\varepsilon_\text{sol})}{3(\varepsilon_i+2\varepsilon_\text{sol})}\Tilde a_i^3$, it leads to the simplified expression for $\mathcal E^{(2)}$ previously derived by Fisher \& Levin \& Li \cite{Fish} (FLL approximation for small/weakly-screened particles, see \cite[Eq.~(4.11)]{Fish}):
\begin{align*}
&\mathcal{E}^{(2)} \approx \frac{\kappa e^{-2\Tilde R}}{8\pi\varepsilon_0\varepsilon_\text{sol} \Tilde R^2}  \biggl[\frac{q_i^2 e^{2\Tilde a_i} \Tilde a_j^3}{(1+\Tilde a_i)^2} \biggl( \frac{1}{3} + \Bigl(1+\frac{1}{\Tilde R}\Bigr)^2 
 \frac{\varepsilon_\text{sol}-\varepsilon_j}{2\varepsilon_\text{sol}+\varepsilon_j}
 \biggr) \notag \\
 &\qquad\quad +  \frac{q_j^2 e^{2\Tilde a_j} \Tilde a_i^3}{(1+\Tilde a_j)^2} \biggl( \frac{1}{3} + \Bigl(1+\frac{1}{\Tilde R}\Bigr)^2 
 \frac{\varepsilon_\text{sol}-\varepsilon_i}{2\varepsilon_\text{sol}+\varepsilon_i}
 \biggr)\biggr] . 
\end{align*}
Even for the two-sphere case, to our knowledge, a complete exact expression for $\mathcal E^{(2)}$ in explicit form (as the one yielded by series~\eqref{L00_monopolar_two_spheres_}) was not previously known until our very recent paper~\cite{our_jcp}.

Besides the leading correction $\mathcal E^{(2)}$ to DLVO, higher-order energy corrections $\mathcal E^{(\ell)}$ with $\ell\ge3$ can also be easily derived (using the corresponding exact relations for $L_{0 0, i}^{(\ell)}$, see Sec.~\ref{Appendix_screened_potential_coefficients_point_kappa_nonzero})  -- e.g. for the triple-screened energy~$\mathcal E^{(3)}$ we have from~\eqref{E_triply_screened_0} 
\begin{align}
\mathcal{E}^{(3)} & = \frac{q_i q_j \kappa e^{\Tilde a_i + \Tilde a_j}}{16\pi\varepsilon_0 \varepsilon_\text{sol}(1+\Tilde a_i) (1+\Tilde a_j)}\biggl[ d_{0,i} d_{0,j} - \Bigl(1+\frac{1}{\Tilde R}\Bigr)^2 \notag\\
&\times 3\bigl(d_{0,i} d_{2,j} + d_{2,i} d_{0,j}\bigr) + \Bigl(1+\frac{1}{\Tilde R}\Bigr)^2 \Bigl(1+\frac{2}{\Tilde R}+\frac{2}{\Tilde R^2}\Bigr) \notag\\
&\times 9 d_{2,i} d_{2,j} + \text{H.O.T.}\biggr]\frac{e^{-3\Tilde R}}{\Tilde R^3} \label{energy_2_spheres_monopoles_3screened}
\end{align}
(H.O.T.~in \eqref{energy_2_spheres_monopoles_3screened} are of order $O(\Tilde a_i^3 \Tilde a_j^3)$ for small particle radii and are not listed here for simplicity, but they can also be easily obtained -- see Sec.~\ref{Appendix_screened_potential_coefficients_point_kappa_nonzero}).

Consider now the example of two spheres with $\varepsilon_1=2$, $\varepsilon_2=3$, centrally-located point charges $q_1=3 e$, $q_2=-2 e$, the sum of radii $a_1+a_2=35$~\AA~(we will test a diverse range of radius ratios -- namely, $a_1/a_2 = 1, 2, 10, 100$), and the solvent parameters (quite typical for systems of biophysical interest) $\varepsilon_\text{sol} = 80$ (aqueous solution), $\kappa^{-1}= 8.071$~\AA~($0.145$~M physiological NaCl concentration at the room temperature $25^{\circ}$ Celsius). Then Fig.~\ref{full_dlvo_mon} shows the full interaction energy $\mathcal E^\text{Int}$ (calculated by solving the linear system \eqref{global_lin_sys1} numerically using the standard GMRES method with a tolerance at least $10^{-12}$) and its approximations by partial sums of addends $\mathcal E^{(\ell)}$ calculated explicitly using the constructed operator series / exact analytical expressions yielded by the developed theory (see Sec.~\ref{general_expansions_pot_en} and Sec.~\ref{Appendix_screened_potential_coefficients_point_kappa_nonzero}), depending on intercenter distance $R=R_{i j}$ for $R\ge36$~\AA. Regarding the series cutoff threshold $n_\text{max}$ let us note that while relatively small $n_\text{max}$ are usually sufficient to produce well-converged solutions~\cite{our_jcp,our_jpcb,DiFlorio2025NextGenPB} (e.g.~tested values in the range of $15$-$20$ demonstrated very accurate results in the considered examples), we set a more conservative value of $n_\text{max}=25$; higher $n_\text{max}$ values resulted in negligible quantitative (and no qualitative) changes in the graph behavior -- e.g.~increasing $n_\text{max}$ to $26$ leads to perturbations that are two (in the extreme case of $a_1=100 a_2$) to ten (at $a_1=a_2$) orders of magnitude smaller than $\mathcal E^\text{Int}$. In addition, Fig.~\ref{ratio_full_dlvo_mon} also depicts the corresponding ratios $\mathcal E^\text{Int}/\mathcal E^\text{(1)}$, showing that at short interparticle separations the higher-order screened DLVO-beyond tail $\sum_{\ell=2}^{+\infty}\mathcal E^{(\ell)}$ can significantly exceed in magnitude the DLVO energy addend $\mathcal E^{(1)}$, and this imbalance grows as the asymmetry/dissimilarity between the particles increases. Notwithstanding these limitations of the approximation capabilities of $\mathcal E^{(1)}$, Fig.~\ref{full_dlvo_mon} clearly demonstrates the fast convergence of the screening-ranged energy expansion $\sum_{\ell=1}^{+\infty}\mathcal E^{(\ell)}$. Remarkably, while the DLVO energy $\mathcal E^{(1)}$ fails at short range, one additional addend beyond DLVO (i.e.~$\mathcal E^{(2)}$) already yields an accurate approximation of the full-fledged $\mathcal E^\text{Int}$ even in cases of very dissimilar particles and captures the features of $\mathcal E^\text{Int}$ such as short-range opposite-charge repulsion effects (observed for certain ratios of particle radii -- see Fig.~\ref{full_dlvo_unequal_10to1_mon},~\ref{full_dlvo_unequal_100to1_mon}). However, as one can observe in Fig.~\ref{full_dlvo_mon}, the same gain in accuracy is not achieved when using low-order approximations to $\mathcal E^{(2)}$ such as FLL or MZ and they lead only to minor improvements to the DLVO result.

\begin{figure*}
\subfloat[\label{ocr_kappa_non_zero_eps81sc_25_kappa}$\kappa\approx\frac{1}{8.071\text{~\AA}}$.]
{
\includegraphics[trim=0.1cm 0.0cm 1.0cm 0.45cm,clip=true,width=0.47\linewidth]{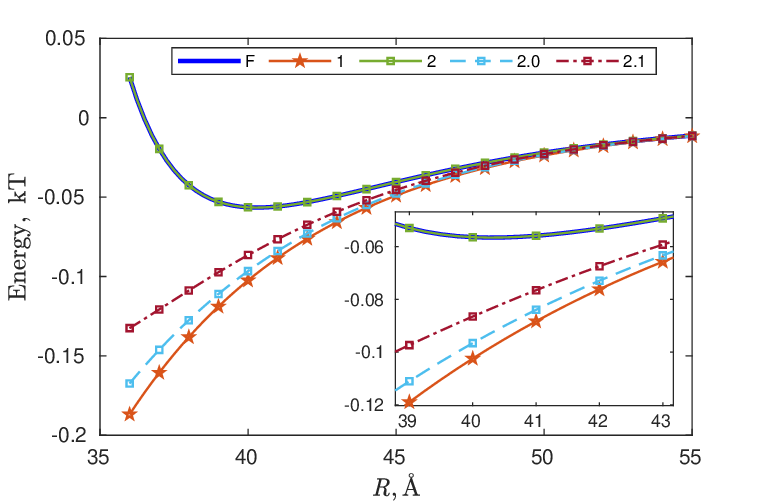}
}
\subfloat[\label{ocr_kappa_non_zero_eps81sc_25_half_kappa}$\kappa\approx\frac{1}{2}\frac{1}{8.071\text{~\AA}}$.]
{
\includegraphics[trim=0.1cm 0.0cm 1.0cm 0.5cm,clip=true,width=0.467\linewidth]{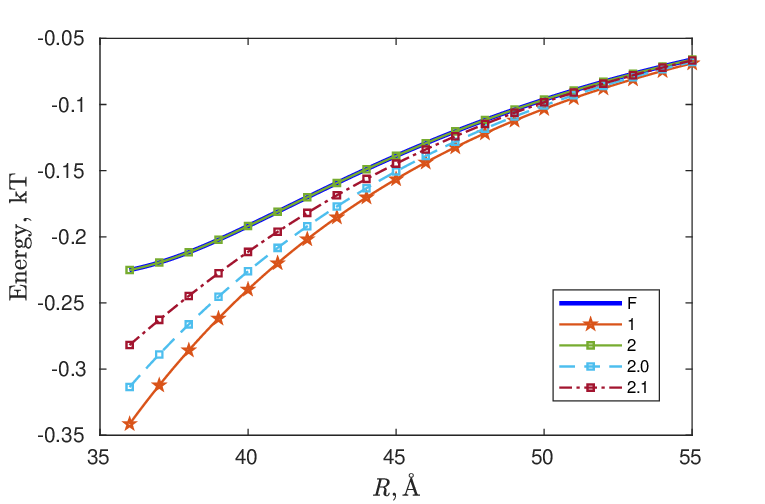}
}
\caption{The case of dielectric constants $\varepsilon_1=\varepsilon_2=81$, $\varepsilon_\text{sol}=80$. All other parameters and line legends are as in Fig.~\ref{full_dlvo_unequal_100to1_mon}.}
\label{ocr_kappa_non_zero_eps81sc_25}
\end{figure*}
With $\varepsilon_i > \varepsilon_\text{sol}$, like-charge attraction effects may occur -- the corresponding situation is provided in Fig.~\ref{full_dlvo_unequal_10to1_mon_attr}, from which one observes that the features of $\mathcal E^\text{Int}$ (especially, the short-range like-charge attraction as $q_2=2e$ and $q_1=3e$ in Fig.~\ref{full_dlvo_unequal_10to1_mon_attr2_tr}) are well-captured by $\mathcal E^{(1)}+\mathcal E^{(2)}$. Furthermore, since opposite-charge repulsion / like-charge attraction effects cannot manifest themselves in the leading single-screened DLVO energy term $\mathcal E^{(1)}$ (just like in its Coulombic counterpart $\frac{q_i q_j}{4\pi\varepsilon_0\varepsilon_\text{sol}R} = \mathcal E^{(1)}\bigr|_{\kappa\to0}$ at zero ionic strength) of $\mathcal E^\text{Int}$, but require higher-order screened terms for their appearance, it can thus be expected that a decrease in $\kappa$ should strengthen the role of these DLVO-beyond terms over $\mathcal E^{(1)}$ and, hence, deepen these effects; the corresponding illustration is provided in Fig.~\ref{full_dlvo_mon_kappa_1e-2}. Lastly, at non-zero ionic strength, opposite-charge repulsion may also arise due to competing energy expansion terms behavior (see Sec.~\ref{regimes_lca_ocr_two_spheres_subs}) -- the corresponding situation is illustrated in Fig.~\ref{ocr_kappa_non_zero_eps81sc_25_kappa} in which particle dielectric constants $\varepsilon_1$ and $\varepsilon_2$ slightly exceed $\varepsilon_\text{sol}$ but the particles (carrying charges $q_1=3 e$, $q_2=-2 e$) exhibit a short-range opposite-charge repulsion effect, a situation that is impossible in the limit of zero ionic strength (see Table~\ref{tab:Ex1_2_spheres}); this effect quickly diminishes and disappears with increasing particle dielectric constants (e.g.~it is not observed if putting $\varepsilon_1=\varepsilon_2=90$) or decreasing $\kappa$ (see Fig.~\ref{ocr_kappa_non_zero_eps81sc_25_half_kappa} with the halved $\kappa$, where the effect is absent).

Finally, Fig.~\ref{full_dlvo_unequal_10_dip} shows an example where point charges are replaced by point dipoles (see Sec.~\ref{Appendix_screened_potential_coefficients_point_dipoles}) located in the centers of spheres. Although Fig.~\ref{ratio_full_dlvo_dip} indicates the significant dominance of the higher-order screened terms of $\mathcal E^\text{Int}$ over $\mathcal E^{(1)}$ for short interparticle separations, Fig.~\ref{full_dlvo_dip_10} still reaffirms the rapid convergence of the $\mathcal E^\text{Int}=\sum_{\ell=1}^{+\infty}\mathcal E^{(\ell)}$ series, as well as the efficacy and robustness of the ($\mathcal E^{(1)}+\mathcal E^{(2)}$)-approximation for~$\mathcal E^\text{Int}$.

Let us emphasize once again that, employing the easily computable (according to the explicit analytical relations derived in this study) sum $\mathcal E^{(1)}+\mathcal E^{(2)}$ (or, if higher accuracy is critical, also adding the further addends $\mathcal E^{(3)}$ and $\mathcal E^{(4)}$), the need to numerically solve linear systems (such as \eqref{global_lin_sys1}) that determine the potentials' expansion coefficients is avoided.
\subsection{The case of $N$ monopolar spheres}
\label{section_numeric_subsection_N_spheres}
\begin{figure}
\subfloat[\label{cubic_125_20_scheme_}Arrangement of particles at $R=1$~\AA~(the color of each particle corresponds to the value of its free charge (in $e$) according to the colormap on the right).]
{
\includegraphics[trim=2.8cm 0.05cm 1.8cm 0.8cm,clip=true,width=0.94\linewidth]{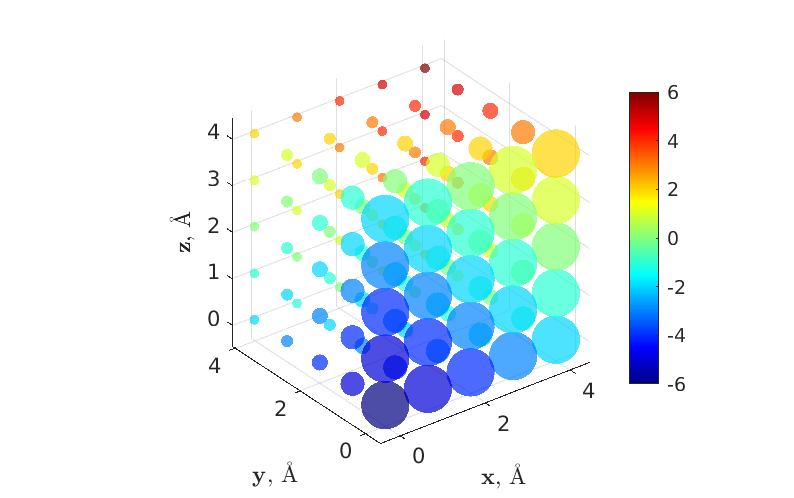}
}\hfill
\subfloat[\label{cubic_125_20_energy_}Dependence of the interaction energy on $R$ varying from 1~\AA~(the densest packing -- the separation between the surfaces of the front spheres $j'=0$ in Fig.~\ref{cubic_125_20_scheme_} is 0.1~\AA~along $i'$, $k'$) to 3~\AA~with step 0.125~\AA; $n_\text{max}=20$. (Line legend is as in Fig.~\ref{full_dlvo_mon}; continuous lines are used to guide the eye; embedded inset shows a close-up view.)]
{
\includegraphics[trim=0.3cm 0cm 0.97cm 0.17cm,clip=true,width=0.92\linewidth]{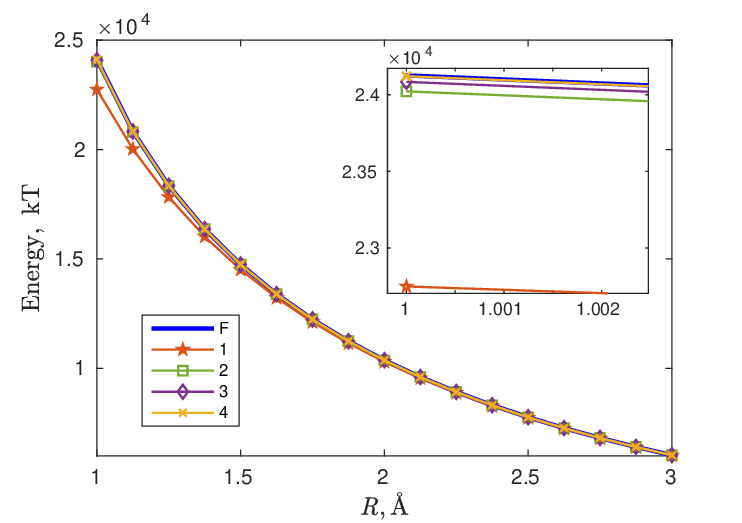}
}
\caption{Cubic lattice of 125 non-overlapping spheres of variable radii, charges, and dielectric constants.}
\label{cubic_125_20_fig}
\end{figure}
\begin{figure}
\subfloat[\label{4_spheres_ArgGlu_scheme}Arrangement of spheres at $R=5.3$~\AA, $R=10$~\AA. Spheres~1-3 (blue) contain glutamate charge distribution, while sphere~4 (red) contains arginine charge distribution; the corresponding (partial) charges assigned are represented as points.]
{
\includegraphics[trim=0.9cm 1.0cm 1.4cm 0.66cm,clip=true,width=0.99\linewidth]{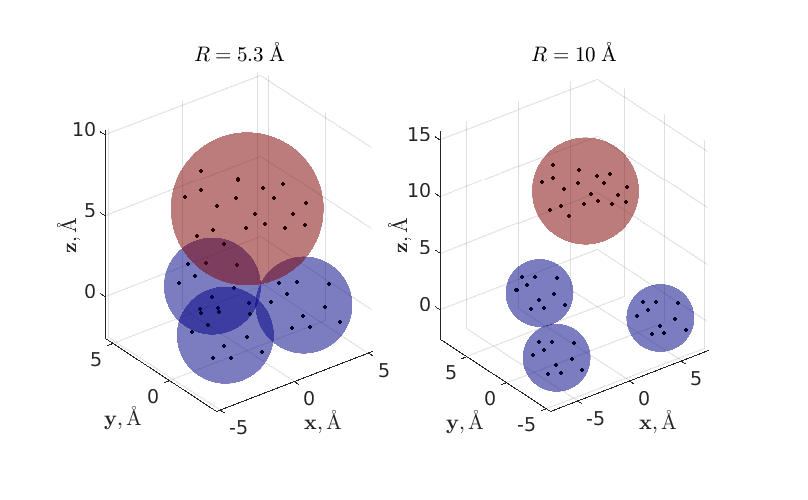}
}\hfill
\subfloat[\label{4_spheres_ArgGlu_energy}Dependence of the interaction energy on $R$ varying from 5.3~\AA~to 10~\AA~with step 0.1~\AA; $n_\text{max}=25$. (Line legend is as in Fig.~\ref{full_dlvo_mon}; continuous lines are used to guide the eye; embedded inset shows a close-up view.)]
{
\includegraphics[trim=0.3cm 0.0cm 0.97cm 0.58cm,clip=true,width=0.92\linewidth]{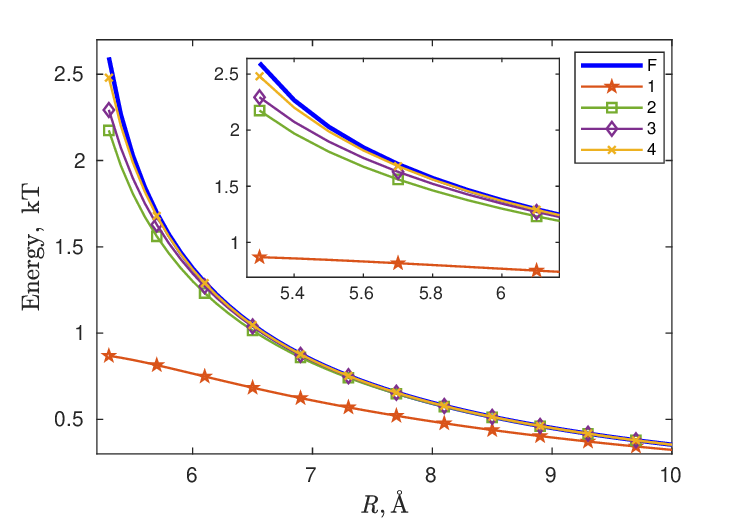}
}
\caption{Tetrahedron of spheres with arginine and glutamate charge distributions.}
\label{4_spheres_ArgGlu_}
\end{figure}
\noindent
To assess whether the leading correction $\mathcal E^{(2)}$ to the DLVO energy $\mathcal E^{(1)}$ is still a mostly sufficient for a robust approximation of $\mathcal E^\text{Int}$ also in the situation of many particles ($N>2$), let us consider a cubic lattice of 125 non-overlapping spheres (see Fig.~\ref{cubic_125_20_scheme_}) centered at points $(R i', R j', R k')$ with variable radii $\frac{0.45~\text{\AA}}{j'+1}$, dielectric constants $i'+1$ and centrally-located point charges $(i'+j'+k'-6)e$, where $i',j',k'=\overline{0,\ldots,4}$, and $R$ is the distance between two adjacent particle centers in every coordinate direction. The solvent parameters are as in Fig.~\ref{full_dlvo_mon}. Then Fig.~\ref{cubic_125_20_energy_} shows how interaction energy depends on $R$ -- one clearly observes that the sum $\mathcal E^{(1)}+\mathcal E^{(2)}$ performs much better than $\mathcal E^{(1)}$ alone and yields a quite accurate approximation of the true interaction energy $\mathcal E^\text{Int}$ (even at small inter-particle separations).

In our next many-body example, we consider four spheres (see Fig.~\ref{4_spheres_ArgGlu_scheme}) with centers $\mathbf x_1=\bigl(-\frac{R}{2},-\frac{R}{2\sqrt{3}},0\bigr)$, $\mathbf x_2=\bigl(\frac{R}{2},-\frac{R}{2\sqrt{3}},0\bigr)$, $\mathbf x_3=\bigl(0,\frac{R}{\sqrt{3}},0\bigr)$, $\mathbf x_4=\bigl(0,0,R \frac{\sqrt{407}}{\sqrt{300}}\bigr)$, radii $a_1=a_2=a_3=2.6$~\AA, $a_4=4.1$~\AA, and dielectric constants $\varepsilon_1=2$, $\varepsilon_2=3$, $\varepsilon_3=4$, $\varepsilon_4=5$; the solvent parameters are as in Fig.~\ref{full_dlvo_mon}. One sees that the distances between $\mathbf x_1$, $\mathbf x_2$, $\mathbf x_3$ are equal to $R$, while the distance from these centers to $\mathbf x_4$ is equal to $1.3 \cdot R$. The corresponding solutes' charge distributions are provided by clouds of point charges (hence, the theory of Sec.~\ref{Appendix_screened_potential_coefficients_cloud} steps in) representing those of the Arginine-Glutamate pair modeled in \cite{MS2013,our_jpcb} (CHARMM22 force field); glutamate charge distribution is embedded into spheres~1-3, while arginine charge distribution is embedded into sphere~4. Let us note that these charge distributions are quite tightly inscribed into the corresponding spheres (the distance from the extreme points of the distributions to the proper spherical surfaces is less than $0.1$~\AA) and are geometrically co-centered with the corresponding spheres' centers (see Fig.~\ref{4_spheres_ArgGlu_scheme}). Then Fig.~\ref{4_spheres_ArgGlu_energy} shows the dependence of the interparticle interaction energy $\mathcal E^\text{Int}$ on $R$ varying from $5.3\text{ \AA}$ to $10\text{ \AA}$ for this system -- as previously, one clearly sees that $\mathcal E^{(1)}$ alone is not sufficient to robustly approximate $\mathcal E^\text{Int}$, but adding the double-screened contribution $\mathcal E^{(2)}$ considerably improves the situation. Let us finally note that at $R=5.3\text{ \AA}$ the spheres are tightly packed (see Fig.~\ref{4_spheres_ArgGlu_scheme}) -- the separation between spherical surfaces is less than $0.2\text{ \AA}$ (=~$0.1\text{ \AA}$ among spheres~1-3).

\section{Summary and outlook}
\label{Discuss_Conclusions_section}
\noindent
We have introduced a novel analytical formalism that provides a rigorous and overarching description of many-body electrostatic interactions at the LPBE level. This framework allows one to construct exact screening-ranged expansions of electrostatic quantities\footnote{In the current paper we mainly focus on building the screening-ranged expansions for the electrostatic \emph{potential} and \emph{energy}, see Sec.~\ref{general_expansions_en_}, whereas the corresponding theory for the interparticle electrostatic \emph{forces} is developed in the joint paper~\textcolor{red}{\cite{supplem_pre_force}}.} and to quantify their terms in explicit form. While the search for exact analytical solutions to electrostatic problems using multipolar expansions has a long history, even the classical case of two interacting dielectric spheres poses considerable mathematical challenges: mutual interaction, polarization, and ionic effects become strongly coupled~\cite{Fish,our_jcp}, giving rise to algebraic systems with involved, infinite-dimensional matrices that typically require truncation and numerical treatment. As aptly noted in~\cite{BSL2014}, ``An implicit series expansion is known for the system of two dielectric spheres. For more than two dielectric spheres, numerical treatment is required." To overcome this limitation and to construct an \emph{explicit} analytical solution for many interacting spheres---while avoiding the need for laborious numerical solutions of multipolar systems and numerical inversion of the corresponding matrices---we start from a global formulation coupling solely external DH potentials, and then obtain its solution explicitly through a specific operator Neumann-type series (whose convergence and spectral properties are rigorously analyzed\footnote{Formulations to the PE-LPBE coupled problem that eliminate the expansion coefficients of internal (or external) potentials—like the one employed in system \eqref{global_lin_sys1}—have not been previously spectrally studied, to the best of our knowledge. In \textcolor{red}{\cite{supplem_pre_math}}, we establish a kind of isomorphism of the ``matrix of infinite sub-matrices" $\mathbb K$ to specifically-built composite Neumann--Poincar\'e-type boundary operators and study them as elements of Banach algebras of bounded operators acting in suitable infinite-dimensional spaces. Let us finally note that row-/column-norms of elements of $\mathbb K$ can acquire values larger than $1$, which puts substantiating \eqref{G_neumann_series0} by, e.g., conventional methods of infinite matrix analysis or Gershgorin-type spectrum localization theorems out of reach.} in~\textcolor{red}{\cite{supplem_pre_math}}).

On this basis, we derive exact screening-ranged expansions of electrostatic quantities (Sec.~\ref{general_expansions_pot_en}) that explicitly reveal the nonadditive nature of many-body interactions and the complex propagation of polarization with increasing expansion order. Asymptotic analysis (see Remark~\ref{energy_expansion_fast_convergence} and Appendix~\ref{appendix_properties_asympt_alpha_beta}) indicates rapid convergence: already the first few addends $\mathcal E^{(\ell)}$ yield highly accurate energy approximations. Numerical modelling (Sec.~\ref{section_numeric}) confirms that calculations up to the double-screened addend ($\mathcal E^{(2)}$) are sufficient to reproduce key features of the full energy profile, including short-range like-charge attraction and opposite-charge repulsion occurring under certain parameter combinations. The developed formalism is general: it handles arbitrary inhomogeneous charge distributions $\rho_i^\text{f}$ and $\sigma_i^\text{f}$ by multipolar expansions (see \eqref{varPhi_in_i_multipoles_general} and \eqref{sigma_nm_i_def}). Applications to different charge distributions are presented in Sec.~\ref{application_specific_systems_section}. For instance, for centrally located point monopoles (and dipoles, if present), we generalize in explicit form the previously known two-body energy results (FLL~\cite{Fish,LLF}, MZ~\cite{McClurg}), as well as more recent triple-screened expressions~\cite{our_jcp,Yu3}. For Janus-like/patchy-charged particles (Sec.~\ref{subsection_janus_particles}), the single-screened term $\mathcal E^{(1)}$ reduces naturally to the DLVO-like theory of~\cite{GraafBoon2012}, while higher-order terms $\mathcal E^{(\ell\geq2)}$ provide systematic corrections capturing polarization-induced and ionic effects (not addressed in the theory of~\cite{GraafBoon2012}).

The framework also streamlines the analysis of the long-studied two-sphere case (see Remark~\ref{remark_two-sphere-kappa-nonzero} and Appendix~\ref{appendix_2_spheres_case}), offering a much clearer way (compared to \cite{our_jcp}) to obtain screened energy components $\mathcal E^{(\ell)}$ of arbitrary order (Appendix~\ref{two_spheres_kappa_nonzero_appendix}), as well as their $\kappa \to 0$ limits (Appendix~\ref{two_spheres_kappa0_appendix}). In the latter limit, we establish a new method for quantifying irregular corrections to the Coulomb energy (see Remark~\ref{energy_components_irregular_kappa_0}). Numerical experiments (Sec.~\ref{section_numeric_subsection_2_spheres}) show that the accurate $\mathcal E^{(\ell)}$ components derived here, especially $\mathcal E^{(2)}$, outperform the known FLL and MZ approximations. Analytically, the formalism opens the way to the first rigorous analysis of phenomena previously treated mostly numerically or within simplified models, such as asymmetric dielectric screening (Remark~\ref{remark_asymm_diel_screening}) and like-charge attraction and opposite-charge repulsion effects (see Table~\ref{tab:Ex1_2_spheres}). In particular, we rigorously show (see Table~\ref{tab:Ex1_2_spheres} and Remark~\ref{attr_rep_remark0} for detailed comments) that, at zero ionic strength, opposite-charge repulsion cannot occur in a less polarizable medium, nor like-charge attraction in a more polarizable one. 

Beyond theory, the analytical expressions developed here can be directly integrated into modelling approaches such as MC or MD\footnote{See e.g.~\cite{NLV_jctc}, which highlights the importance of many-body interactions in molecular simulations, especially three-body contributions.}, thereby accelerating simulations and improving polarizable force field calculations. The approach also provides an efficient algorithm for electrostatics (and is actually used in Sec.~\ref{section_numeric}), where polarization/off-sphere contributions are treated in a fully analytical manner, unlike recent many-sphere boundary-integral methods \cite{LinQS,Lindgren_jcis,LSB2018,Lindgren_jctc2025,Hassan_Stamm_jctc} relying on numerical integration by spherical Lebedev quadratures or hybrid methods~\cite{GJLX,GWJXL} employing Gauss-Jacobi quadratures. Moreover, this grid-free approach is not subject to numerical artifacts owing to discretization of the underlying differential equations or dielectric particle surfaces. Although a detailed numerical study is beyond the scope of this article, the theory offers accurate analytical benchmarks for testing PB solvers~\cite{DiFlorio2025NextGenPB}. An extension to non-spherical particles---an active research area~\cite{TDB_2017,EvertsSciAdv,OMHOOC,WGT,GFBK,EvertsPRR,GRN2024,JR2025,GRLN2022,VSSAN,Derb5}---also remains a promising direction. Preliminary work in~\cite{our_jpcb} suggests that grid-free analytical treatments can substantially improve both accuracy and efficiency over grid-based solvers such as DelPhi~\cite{RAH1,RSN1}.

Altogether, this work establishes a mathematically rigorous and practically versatile framework for many-body electrostatics within the LPBE, opening the way for improved theoretical analyses, efficient simulation methods, and extensions to more complex particle geometries.

\section{Acknowledgments}
\label{sec:Acknowledgements}

We acknowledge the financial support from the European 
Union - NextGenerationEU and the Ministry of University and Research (MUR), 
National Recovery and Resilience Plan (NRRP): 
Research program CN00000013 “National Centre for HPC, 
Big Data and Quantum Computing”, CUP: J33C22001180001, funded by the D.D. n.1031, 17.06.2022 and Mission 4, Component 2, Investment 1.4 - Avviso “Centri Nazionali” - D.D. n. 3138, 16 December 2021.

\section*{Data availability}
\noindent
Codes/data can be found in~\textcolor{red}{\cite{our_github_rep}}.

\appendix

\counterwithin{remark}{section}

\section{A particular case of two dielectric particles ($N=2$)}
\label{appendix_2_spheres_case_general}
\subsection{On the connections with the analytical theory of paper~\cite{our_jcp}}
\label{appendix_2_spheres_case}
\noindent
In the particular case of $N=2$ system \eqref{global_lin_sys1} boils down to 
\begin{equation}
\label{global_lin_sys1_two_particles}
\begin{pmatrix}
\mathsf I & \mathsf A_1^{-1} \mathsf B_{1 2} \\
\mathsf A_2^{-1} \mathsf B_{2 1} & \mathsf I
\end{pmatrix}\!
\begin{pmatrix}
\Tilde{\mathbf G}_1 \\
\Tilde{\mathbf G}_2
\end{pmatrix}
=
\begin{pmatrix}
\mathsf A_1^{-1} \mathbf S_1 \\
\mathsf A_2^{-1} \mathbf S_2 \\
\end{pmatrix}\!,
\end{equation}
where $\mathsf I$ denotes an identity matrix (matrix block). Utilizing now the Frobenius $2\times2$ block matrix inversion formula~\cite{HornJohnson}
\begin{align}
\begin{pmatrix}
\mathsf A & \mathsf B \\
\mathsf C & \mathsf D
\end{pmatrix}^{\!-1}
&=
\begin{pmatrix}
(\mathsf A-\mathsf B \mathsf D^{-1} \mathsf C)^{-1} & \mathsf 0 \\
\mathsf 0 & (\mathsf D-\mathsf C \mathsf A^{-1} \mathsf B)^{-1}
\end{pmatrix} \notag \\
&\times
\begin{pmatrix}
\mathsf I & -\mathsf B \mathsf D^{-1} \\
-\mathsf C \mathsf A^{-1} & \mathsf I
\end{pmatrix}\!, \label{frob_inv_formula}
\end{align}
system \eqref{global_lin_sys1_two_particles} can be readily recast to the following equivalent form in which unknown $\Tilde{\mathbf G}_i$ are independently resolved:
\begin{equation}
\label{global_lin_sys1_two_particles_explicit_G}
\!\!\!\begin{pmatrix}
\!\Tilde{\mathbf G}_1 \!\!\\
\!\Tilde{\mathbf G}_2 \!\!
\end{pmatrix}
\!\!=\!\!\begin{pmatrix}
\!(\mathsf I \!-\! \mathsf A_1^{-1} \mathsf B_{1 2} \mathsf A_2^{-1} \mathsf B_{2 1}\!)^{-1} \mathsf A_1^{-1} (\mathbf S_1 \!-\! \mathsf B_{1 2} \mathsf A_2^{-1}\mathbf S_2\!)\!\! \\
\!(\mathsf I \!-\! \mathsf A_2^{-1} \mathsf B_{2 1} \mathsf A_1^{-1} \mathsf B_{1 2}\!)^{-1} \mathsf A_2^{-1} (\mathbf S_2 \!-\! \mathsf B_{2 1} \mathsf A_1^{-1}\mathbf S_1\!)\!\!
\end{pmatrix}\!.\!
\end{equation}
Relation \eqref{global_lin_sys1_two_particles_explicit_G}, although obtained in a different way (and also with the re-expansion coefficients differently determined -- see \eqref{two_particles_contraction_formula0} below), was the starting cornerstone on which the 2-sphere (with $z$-aligned centers) analytical theory of work \cite{our_jcp} has rested -- in this sense, the new formalism (i.e.~developed in the present study) encompasses that of \cite{our_jcp}, and thus the relations for the potential/energy, derived in work \cite{our_jcp}, can also be reproduced here (i.e.~using the present formalism and also taking into account \eqref{two_particles_contraction_formula0} derived below). However, the reverse statement is not true, since relation \eqref{global_lin_sys1_two_particles_explicit_G} (derived at $N=2$) does not allow generalization to many-body ($N>2$) cases (as well as there is no explicit formulas, similar to Frobenius formula \eqref{frob_inv_formula}, for inverting arbitrary-size $N\times N$ block matrices) -- thus, the formalism developed in \cite{our_jcp} cannot handle many-bodyness (apart from the sequential application of Wigner rotations with all the attendant disadvantages of such an approach and its actual inability to obtain meaningful analytical expressions -- see discussion in Sec.~\ref{spherical_Fourier_coeffs_subsec}). Concerning the re-expansion coefficients derived in \cite{our_jcp}, the following link between them and identity \eqref{Yu3_reexp} can be established (for instance, we put $i=1$ now, since the complementary expression for $i=2$ can be derived in a completely similar way) taking into account that in \cite{our_jcp} it was assumed that the spheres' centers are $z$-aligned ($\mathbf x_1=(0,0,0)$, $\mathbf x_2=(0,0,R)$,~$R>0$):
\begin{equation}
\label{two_particles_contraction_formula0}
\begin{gathered}
 \sqrt{\frac{\pi(2 n+1)}{2(2 l+1)}} \frac{I_{n+1/2}(\Tilde r_1)}{\sqrt{\Tilde r_1}} (-1)^m \sum_{l_2=|n-l|}^{n+l} C_{n 0 l_2 0}^{l 0} C_{n m l_2 0}^{l m} (2 l_2+1) \\ 
 \times \frac{K_{l_2+1/2}(\Tilde R)}{\sqrt{\Tilde R}} = \sqrt{\frac{(2 l+1) (l-\left| m \right|)! (n+ \left| m \right|)!}{(2 n+1) (l+\left| m \right|)! (n- \left| m \right|)!}} b_{n \left| m \right| l}(\Tilde r_1, \Tilde R),
\end{gathered}
\end{equation}
where $\Tilde R \mathrel{:=}\kappa R$,  $b_{n m l}(\Tilde r,\Tilde R)$ is the expression derived in~\cite{our_jcp}:
\begin{equation}
\label{b_nml_Theta_jcp}
b_{n m l}(\Tilde r,\Tilde R) = \frac{\pi}{2}\frac{I_{n+1/2}(\Tilde r)}{\sqrt{\Tilde r}}\frac{e^{-\Tilde R}}{\Tilde R^{m+1}}\Theta_{n m l}(\Tilde R)
\end{equation}
with $\Theta_{n m l}(\Tilde R)$ being a polynomial in $\Tilde R^{-1}$ of degree $n+l-m$ having an explicit representation
\begin{align*}
&\Theta_{n m l}(\Tilde R) = \frac{(2 n+1) (n-m)! (l+m)!}{2^m (n+m)! (l-m)!} \sum_{p=0}^{n-m} \sum_{s=0}^l \frac{1}{2^{p+s} \Tilde R^{s+p}} \\ 
& \quad \times \frac{(l+s)! (s+m+p)! (n+m+p)! }{s! (l-s)! (s+m)! p! (n-m-p)! (m+p)!} \\
& = \frac{(2 n+1) (n-m)! (l+m)!}{2^m (n+m)! (l-m)!} \sum_{s=0}^{n+l-m} \Biggl(\: \sum_{p=\max(0,\, s+m-n)}^{\min(l,\, s)} \frac{(l+p)!}{(l-p)!} \\ 
& \times \frac{(n+m+s-p)!}{p! (n-m-s+p)! (s-p)! (p+m)! (m+s-p)!} \! \Biggr)\frac{(m+s)!}{2^s \Tilde R^s} .
\end{align*}
\emph{Proof of \eqref{two_particles_contraction_formula0}:} Using \cite[Eq.~(65)]{our_jpcb} we can then arrive at the relation $\sum\limits_{L,M}\frac{K_{L+1/2}(\Tilde r_2)}{\sqrt{\Tilde r_2}}(-1)^{L+M} Y_L^M(\Hat{\mathbf r}_2) G_{L M,2} = \sum\limits_{n,m}\Bigl(\,\sum\limits_{l=|m|}^{+\infty} \Breve b_{n m l}(\Tilde r_1,\Tilde R) G_{l m,2} \Bigr)Y_n^m(\Hat{\mathbf r}_1)$, where $\Breve b_{n m l}(\Tilde r_1,\Tilde R)$ is determined by $\Breve b_{n m l}(\Tilde r_1,\Tilde R) \mathrel{:=} \sqrt{\frac{(2 l+1) (l-\left| m \right|)! (n+ \left| m \right|)!}{(2 n+1) (l+\left| m \right|)! (n- \left| m \right|)!}} b_{n \left| m \right| l}(\Tilde r_1, \Tilde R)$ (factor $(-1)^{L+M}$ has appeared in the left-hand side because papers \cite{our_jpcb,our_jcp} operated with the complementary angle $\pi-\theta_2$ instead of $\theta_2$). On the other hand, from re-expansion \eqref{Yu3_reexp} we can obtain $\sqrt{2/\pi}\sum\limits_{L,M}\frac{K_{L+1/2}(\Tilde r_2)}{\sqrt{\Tilde r_2}} Y_L^M(\Hat{\mathbf r}_2) (-1)^{L+M} G_{L M,2} = \sqrt{\pi/2}\sum\limits_{n,m} \frac{I_{n+1/2}(\Tilde r_1)}{\sqrt{\Tilde r_1}}\Bigl(\,\sum\limits_{L,M} \mathcal H_{n m}^{L M}(\mathbf{R}_{1 2}) (-1)^{L+M} G_{L M,2} \Bigr)Y_n^m(\Hat{\mathbf r}_1)$, thence comparing the above expressions we can then deduce the following relation: 
\begin{equation}
\label{rel_bnml_H_temp0}
\begin{aligned}
& \sqrt{2/\pi}\sum\nolimits_{l=|m|}^{+\infty} \Breve b_{n m l}(\Tilde r_1,\Tilde R) G_{l m,2} \\
&\quad = i_n(\Tilde r_1)\sum\nolimits_{L,M} \mathcal H_{n m}^{L M}(\mathbf{R}_{1 2}) (-1)^{L+M} G_{L M,2} \, ;
\end{aligned}
\end{equation}
now, taking into account that $\mathbf{R}_{1 2} = R \Hat{\mathbf z}$ and therefore $Y_{l_2}^{m_2}(\Hat{\mathbf R}_{1 2}) = \sqrt{\frac{2 l_2+1}{4\pi}}\delta_{m_2,0}$ ($\delta_{\cdot,\cdot}$ is a Kronecker delta), and the fact that $C_{n m l_2 0}^{L M}=0$ if $M\ne m$ (see footnote~\ref{footnote_ClebshGordan_properties}), after algebraic transformations we finally arrive at~\eqref{two_particles_contraction_formula0}.
\subsection{Interaction energy in the system of two monopolar spheres}
\label{two_spheres_kappa_nonzero_appendix}
\noindent
Let us consider two monopolar spheres with free charges $q_1$ and $q_2$, and (for convenience of calculations, as the resulting energy expressions are rigid-motions invariant) centers $\mathbf x_1=(0,0,0)$, $\mathbf x_2=(0,0,R)$ ($R>0$). Let us define the dimensionless factors $\vartheta_i \mathrel{:=} \frac{e^{\Tilde a_i}}{1+\Tilde a_i}$ and
\begin{align*}
\mathfrak e^i_n & \mathrel{:=} \frac{(\varepsilon_i-\varepsilon_\text{sol}) n i_n(\Tilde a_i)/\Tilde a_i - \varepsilon_\text{sol} i_{n+1}(\Tilde a_i)}{\alpha_n(\Tilde a_i,\varepsilon_i)} \\ 
&\qquad = \frac{(\varepsilon_i-\varepsilon_\text{sol}) n i_n(\Tilde a_i) - \varepsilon_\text{sol} \Tilde a_i i_{n+1}(\Tilde a_i)}{(\varepsilon_i-\varepsilon_\text{sol}) n k_n(\Tilde a_i) + \varepsilon_\text{sol} \Tilde a_i k_{n+1}(\Tilde a_i)} .
\end{align*}
Note that $\alpha_n(\Tilde a_i,\varepsilon_i)>0$ (see \eqref{alpha_is_positive_ineq}). In this AS situation under consideration, relation \eqref{L_l_elementwise} and operator $\mathbb K = \bigl(\begin{smallmatrix}
0 & \mathsf K_{1 2} \\
\mathsf K_{2 1} & 0
\end{smallmatrix}\bigr)$ (see \eqref{global_lin_sys1}) boil down to the following element-wise identities:
\begin{subequations}
\label{L_K_N_2_kappa_nonzero}
\begin{gather}
(\mathsf K_{i j})_{n 0, l 0} = (-1)^l \mathfrak e^i_n \frac{\Upsilon_{l,j}}{\Upsilon_{n,i}}\sqrt{\frac{2 l+1}{2 n+1}}\frac{e^{-\Tilde R}}{\Tilde R} \Theta_{n 0 l}(\Tilde R) ,\label{K_N_2_kappa_nonzero} \\
L_{0 0,i}^{(\ell)}\bigr|_{N=2} = \sum_{l=0}^{+\infty} (-1)^l \vartheta_i \sqrt{2 l+1} \frac{e^{-\Tilde R}}{\Tilde R} \Theta_{0 0 l}(\Tilde R) \Upsilon_{l,j} \cdot \Tilde G_{l 0,j}^{(\ell-1)}\bigr|_{N=2} \label{L_ell_N_2_kappa_nonzero}
\end{gather}
\end{subequations}
as $i=1$ and $j=2$; the corresponding complementary identities at $i=2$ and $j=1$ are the same but without factor $(-1)^l$ in \eqref{L_ell_N_2_kappa_nonzero}, and with $(-1)^n$ instead of $(-1)^l$ in \eqref{K_N_2_kappa_nonzero} (as immediately follows from~\eqref{symmetry_H_nmLM_3}). Here $\Theta_{n 0 l}(\Tilde R)$ is a polynomial in $\Tilde R^{-1}$ with positive coefficients defined after~\eqref{b_nml_Theta_jcp}. Relations \eqref{L_K_N_2_kappa_nonzero} readily follow from definitions of the corresponding quantities (see \eqref{global_lin_sys1}, \eqref{L_l_elementwise}, \eqref{various_matrix_definitions_0}, \eqref{alpha_beta_definitions}), identities of Sec.~\ref{list_of_some_H_nmLM_values_}, $Y_{l}^{0}(\Hat{\mathbf R}_{1 2}) = \sqrt{\frac{2 l+1}{4\pi}}$, $(-1)^l \sqrt{\frac{2 l+1}{2 n+1}} \frac{e^{-\Tilde R}}{\Tilde R} \Theta_{n 0 l}(\Tilde R) = \mathcal H_{n 0}^{l 0}(\mathbf{R}_{1 2})$ (see \eqref{b_nml_Theta_jcp}, \eqref{rel_bnml_H_temp0}), and equality $\frac{e^{-\Tilde R}}{\Tilde R} \Theta_{0 0 l}(\Tilde R) = k_l(\Tilde R)$ (this can be seen by directly calculating the polynomial~$\Theta_{0 0 l}(\Tilde R)$).

Now taking into account relations \eqref{K_N_2_kappa_nonzero}, $\Tilde G_{n 0,i}^{(\ell)}\bigr|_{N=2} = - \sum\limits_{l=0}^{+\infty}(\mathsf K_{i j})_{n 0, l 0} \Tilde G_{l 0,j}^{(\ell-1)}\bigr|_{N=2}$ (see \eqref{G_componentwise0}), and the starting values $\Tilde G^{(0)}_{n 0,i}\bigr|_{N=2} = \frac{q_i \delta_{n,0}}{\sqrt{4\pi}\varepsilon_0\varepsilon_\text{sol}(1+\Tilde a_i)}$ (see Sec.~\ref{Appendix_screened_potential_coefficients_point_kappa_nonzero}) as $i=\overline{1, 2}$, $j=3-i$, we then readily arrive at the following expressions for $\Tilde G^{(\ell)}_{n 0,i}\bigr|_{N=2}$:
\begin{align*}
&\Tilde G^{(1)}_{n 0,i}\bigr|_{N=2} = \frac{-q_j \vartheta_j \kappa}{\sqrt{4\pi}\varepsilon_0\varepsilon_\text{sol}}\frac{\mathfrak e^i_n \Upsilon_{n,i}^{-1}}{\sqrt{2 n+1}} \frac{e^{-\Tilde R}}{\Tilde R} \Theta_{n 0 0}(\Tilde R), \\
& \Tilde G^{(2)}_{n 0,i}\bigr|_{N=2} = \frac{q_i \vartheta_i \kappa}{\sqrt{4\pi}\varepsilon_0\varepsilon_\text{sol}}\frac{\mathfrak e^i_n \Upsilon_{n,i}^{-1}}{\sqrt{2 n+1}} \Bigl(\frac{e^{-\Tilde R}}{\Tilde R}\Bigr)^2 \sum_{l_1=0}^{+\infty} \mathfrak e^j_{l_1} \\ 
&\qquad \times \Theta_{n 0 l_1}(\Tilde R) \Theta_{l_1 0 0}(\Tilde R), \\
& \Tilde G^{(3)}_{n 0,i}\bigr|_{N=2} = \frac{-q_j \vartheta_j \kappa}{\sqrt{4\pi}\varepsilon_0\varepsilon_\text{sol}}\frac{\mathfrak e^i_n \Upsilon_{n,i}^{-1}}{\sqrt{2 n+1}} \Bigl(\frac{e^{-\Tilde R}}{\Tilde R}\Bigr)^3 \sum_{l_1=0}^{+\infty} \sum_{l_2=0}^{+\infty} \mathfrak e^j_{l_1} \mathfrak e^i_{l_2} \\
&\qquad \times \Theta_{n 0 l_1}(\Tilde R) \Theta_{l_1 0 l_2}(\Tilde R)\Theta_{l_2 0 0}(\Tilde R),\\
& \Tilde G^{(4)}_{n 0,i}\bigr|_{N=2} = \frac{q_i \vartheta_i \kappa}{\sqrt{4\pi}\varepsilon_0\varepsilon_\text{sol}}\frac{\mathfrak e^i_n \Upsilon_{n,i}^{-1}}{\sqrt{2 n+1}} \Bigl(\frac{e^{-\Tilde R}}{\Tilde R}\Bigr)^4 \sum_{l_1=0}^{+\infty} \sum_{l_2=0}^{+\infty} \sum_{l_3=0}^{+\infty} \mathfrak e^j_{l_1} \mathfrak e^i_{l_2} \mathfrak e^j_{l_3} \\ 
&\qquad \times \Theta_{n 0 l_1}(\Tilde R) \Theta_{l_1 0 l_2}(\Tilde R) \Theta_{l_2 0 l_3}(\Tilde R) \Theta_{l_3 0 0}(\Tilde R),\\
& \Tilde G^{(5)}_{n 0,i}\bigr|_{N=2} = \frac{-q_j \vartheta_j \kappa}{\sqrt{4\pi}\varepsilon_0\varepsilon_\text{sol}}\frac{\mathfrak e^i_n \Upsilon_{n,i}^{-1}}{\sqrt{2 n+1}} \Bigl(\frac{e^{-\Tilde R}}{\Tilde R}\Bigr)^5 \sum_{l_1=0}^{+\infty} \sum_{l_2=0}^{+\infty} \sum_{l_3=0}^{+\infty} \sum_{l_4=0}^{+\infty} \mathfrak e^j_{l_1} \\
&\qquad \times \mathfrak e^i_{l_2} \mathfrak e^j_{l_3} \mathfrak e^i_{l_4} \Theta_{n 0 l_1}(\Tilde R) \Theta_{l_1 0 l_2}(\Tilde R)\Theta_{l_2 0 l_3}(\Tilde R)\Theta_{l_3 0 l_4}(\Tilde R)\Theta_{l_4 0 0}(\Tilde R),\\
&\ldots,
\end{align*}
where $i=1$ and $j=2$ (the corresponding equalities determining $\Tilde G^{(\ell)}_{n 0,i}\bigr|_{N=2}$ at $i=2$ and $j=1$ are the same but with the prepended factor~$(-1)^n$). Now leveraging these expressions with \eqref{L_ell_N_2_kappa_nonzero}, \eqref{energy_N_monopoles_ell_screened_} and the symmetry property \cite[Eq.~(41)]{our_jcp} $\Theta_{n 0 l}(\Tilde R) = \frac{2 n+1}{2 l+1} \Theta_{l 0 n}(\Tilde R)$, we arrive at the following sequence of exact relations determining addends $\mathcal E^{(\ell)}$ of \eqref{energy_expansion_components_interaction} at $N=2$:
\begin{align*}
& \mathcal E^{(1)}\bigr|_{N=2} = \frac{q_i q_j \vartheta_i \vartheta_j \kappa}{4\pi\varepsilon_0\varepsilon_\text{sol}}\frac{e^{-\Tilde R}}{\Tilde R}, \\
& \mathcal E^{(2)}\bigr|_{N=2} = \frac{-\kappa}{8\pi\varepsilon_0\varepsilon_\text{sol}}\Bigl(\frac{e^{-\Tilde R}}{\Tilde R}\Bigr)^2 \sum_{l_1=0}^{+\infty}(q_i^2 \vartheta_i^2 \mathfrak e^j_{l_1} + q_j^2 \vartheta_j^2 \mathfrak e^i_{l_1}) \\ 
&\quad \times \Theta_{0 0 l_1}(\Tilde R)\Theta_{l_1 0 0}(\Tilde R) ,\\
& \mathcal E^{(3)}\bigr|_{N=2} = \frac{q_i q_j \vartheta_i \vartheta_j \kappa}{4\pi\varepsilon_0\varepsilon_\text{sol}}\Bigl(\frac{e^{-\Tilde R}}{\Tilde R}\Bigr)^3 \sum_{l_1=0}^{+\infty} \sum_{l_2=0}^{+\infty} \mathfrak e^i_{l_1} \mathfrak e^j_{l_2} \Theta_{0 0 l_1}(\Tilde R) \\
&\quad \times \Theta_{l_1 0 l_2}(\Tilde R) \Theta_{l_2 0 0}(\Tilde R) , \\
& \mathcal E^{(4)}\bigr|_{N=2} = \frac{-\kappa}{8\pi\varepsilon_0\varepsilon_\text{sol}}\Bigl(\frac{e^{-\Tilde R}}{\Tilde R}\Bigr)^4 \sum_{l_1=0}^{+\infty} \sum_{l_2=0}^{+\infty} \sum_{l_3=0}^{+\infty} (q_i^2 \vartheta_i^2 \mathfrak e^j_{l_1}\mathfrak e^i_{l_2}\mathfrak e^j_{l_3} \\ 
&\quad + q_j^2 \vartheta_j^2 \mathfrak e^i_{l_1} \mathfrak e^j_{l_2} \mathfrak e^i_{l_3}) \Theta_{0 0 l_1}(\Tilde R) \Theta_{l_1 0 l_2}(\Tilde R) \Theta_{l_2 0 l_3}(\Tilde R) \Theta_{l_3 0 0}(\Tilde R) , \\
& \mathcal E^{(5)}\bigr|_{N=2} = \frac{q_i q_j \vartheta_i \vartheta_j \kappa}{4\pi\varepsilon_0\varepsilon_\text{sol}}\Bigl(\frac{e^{-\Tilde R}}{\Tilde R}\Bigr)^5 \sum_{l_1=0}^{+\infty} \sum_{l_2=0}^{+\infty} \sum_{l_3=0}^{+\infty} \sum_{l_4=0}^{+\infty} \mathfrak e^i_{l_1} \mathfrak e^j_{l_2} \mathfrak e^i_{l_3} \mathfrak e^j_{l_4} \\ 
&\quad \times \Theta_{0 0 l_1}(\Tilde R) \Theta_{l_1 0 l_2}(\Tilde R) \Theta_{l_2 0 l_3}(\Tilde R) \Theta_{l_3 0 l_4}(\Tilde R) \Theta_{l_4 0 0}(\Tilde R), \\
& \mathcal E^{(6)}\bigr|_{N=2} = \frac{-\kappa}{8\pi\varepsilon_0\varepsilon_\text{sol}}\Bigl(\frac{e^{-\Tilde R}}{\Tilde R}\Bigr)^6 \sum_{l_1=0}^{+\infty} \sum_{l_2=0}^{+\infty} \sum_{l_3=0}^{+\infty} \sum_{l_4=0}^{+\infty} \sum_{l_5=0}^{+\infty} (q_i^2 \vartheta_i^2 \mathfrak e^j_{l_1}\mathfrak e^i_{l_2} \\
&\quad \times \mathfrak e^j_{l_3}\mathfrak e^i_{l_4}\mathfrak e^j_{l_5} + q_j^2 \vartheta_j^2 \mathfrak e^i_{l_1} \mathfrak e^j_{l_2} \mathfrak e^i_{l_3} \mathfrak e^j_{l_4} \mathfrak e^i_{l_5}) \Theta_{0 0 l_1}(\Tilde R) \Theta_{l_1 0 l_2}(\Tilde R) \Theta_{l_2 0 l_3}(\Tilde R) \\
&\quad\times\Theta_{l_3 0 l_4}(\Tilde R) \Theta_{l_4 0 l_5}(\Tilde R) \Theta_{l_5 0 0}(\Tilde R), \\
&\ldots
\end{align*}
where $i\in\{1, 2\}$, $j=3-i$. As follows from the appearance of the above addends, the energy series \eqref{energy_expansion_components_interaction} can exhibit an oscillatory convergent behavior (nevertheless, the results of \textcolor{red}{\cite{supplem_pre_math}} ensure its absolute convergence).
\begin{remark}
\label{appendix_lca_isnt_possible}
Note that $\mathfrak e^i_n < 0$ $\forall n\ge0$ if $\varepsilon_i \le \varepsilon_\text{sol}$ (as immediately follows from \eqref{alpha_is_positive_ineq} and \eqref{in_kn_are_positive_}). Thus, if $\varepsilon_1 \le \varepsilon_\text{sol}$ and $\varepsilon_2 \le \varepsilon_\text{sol}$, then also taking into account that $\Theta_{n m l}(\Tilde R)$ is a polynomial in $1/\Tilde R$ with positive coefficients, we readily conclude from the $\mathcal E^{(\ell\ge1)}\bigr|_{N=2}(R)$ expressions derived above that all screened energy addends $\mathcal E^{(\ell)}\bigr|_{N=2}(R)$ with \emph{odd} $\ell$ behave qualitatively similarly to the conventional DLVO addend $\mathcal E^{(1)}\bigr|_{N=2}(R)$ (i.e.~they monotonically decrease / increase as $R$ grows if $q_i q_j>0$ / $q_i q_j<0$), and in a like manner all addends $\mathcal E^{(\ell)}\bigr|_{N=2}(R)$ with \emph{even} $\ell$ behave qualitatively similarly to addend $\mathcal E^{(2)}\bigr|_{N=2}(R)$ (i.e.~they monotonically decrease irrelevantly to the sign of $q_i q_j$ as $R$ grows) which is repulsive and constitutes the leading correction to the DLVO term. Accordingly, taking $-\frac{\partial\mathcal E^\text{Int}(R)}{\partial R}$ in turn allows one to determine the electrostatic force exerted on an $R$-centered particle.
\par

When $\varepsilon_1 > \varepsilon_\text{sol}$ and $\varepsilon_2 > \varepsilon_\text{sol}$, the variety of possible types of interaction becomes more involved --- indeed, it is always $\mathfrak e^i_0 < 0$ independently of the dielectric constants and the inequality $\mathfrak e^i_n < 0$ may still hold for some starting $n=\overline{1, 2, \ldots}$ for which the difference $\varepsilon_i-\varepsilon_\text{sol}>0$ is small enough (so that the positive value of the first term $(\varepsilon_i-\varepsilon_\text{sol}) n i_n(\Tilde a_i)$ does not surpass the negative value of the second one, i.e.~$-\varepsilon_\text{sol}\Tilde a_i i_{n+1}(\Tilde a_i)$, taking into account that the $i_n(x)$ values of any fixed $x>0$ rapidly decrease with increasing $n$). This implies that e.g.~the $\mathcal E^{(\ell)}\bigr|_{N=2}(R)$-terms with even $\ell$ are not necessarily attractive (as would be in the case of the $\kappa=0$ limit, see Sec.~\ref{two_spheres_kappa0_appendix} below) but may also be of repulsive character, and a non-trivial crossover of competing terms may arise. Hence, both effects of like-charge attraction and opposite-charge repulsion may occur as $\varepsilon_1, \varepsilon_2 > \varepsilon_\text{sol}$ and $\kappa>0$, a situation that is impossible when~$\kappa=0$.
\end{remark}
\subsection{Interaction energy in the system of two monopolar spheres in the~$\kappa\to0$ limit}
\label{two_spheres_kappa0_appendix}
\noindent
Again, let us consider two monopolar spheres with free charges $q_1$ and $q_2$, and centers $\mathbf x_1=(0,0,0)$, $\mathbf x_2=(0,0,R)$ ($R>0$). Although the values of $\mathcal E^{(\ell)}\bigr|_{\kappa\to0,\; N=2}$ (i.e.~$\mathcal E^{(\ell)}\bigr|_{N=2}$ in the $\kappa\to0$ limit) can be derived from the previously found values of $\mathcal E^{(\ell)}\bigr|_{N=2}$, see Appendix~\ref{two_spheres_kappa_nonzero_appendix}, it is insightful to look at the behavior of operator $\mathbb K$ and relations coupling the potential coefficients in the $\kappa\to0$ limit directly. Namely, in this AS situation, in the $\kappa\to0$ limit, relation \eqref{L_l_elementwise} and operator $\mathbb K = \bigl(\begin{smallmatrix}
0 & \mathsf K_{1 2} \\
\mathsf K_{2 1} & 0
\end{smallmatrix}\bigr)$ (see \eqref{global_lin_sys1}) boil down to the following element-wise identities:
\begin{subequations}
\label{L_K_N_2_kappa_0}
\begin{gather}
L_{0 0,i}^{(\ell)}\bigr|_{\kappa\to0,\; N=2} = \sum_{l=0}^{+\infty}\frac{(-1)^l}{\sqrt{2 l+1}}\frac{a_j^l}{R^{l+1}} \cdot \Tilde G_{l 0,j}^{(\ell-1)}\bigr|_{\kappa\to0,\; N=2} ,\label{L_ell_N_2_kappa_0}\\
\!\!\!\! (\mathsf K_{i j}|_{\kappa\to0})_{n 0, l 0} \!=\! (-1)^l \frac{n(\varepsilon_i-\varepsilon_\text{sol}) a_i^{n+1} a_j^l}{n\varepsilon_i+(n+1)\varepsilon_\text{sol}}\!\frac{\sqrt{2 n\!+\!1}}{\sqrt{2 l\!+\!1}}\!\frac{(n+l)!}{n! l! R^{n+l+1}} \label{K_N_2_kappa_0}
\end{gather}
\end{subequations}
as $i=1$ and $j=2$; the corresponding complementary identities at $i=2$ and $j=1$ are the same but without factor $(-1)^l$ in \eqref{L_ell_N_2_kappa_0}, and with $(-1)^n$ instead of $(-1)^l$ in \eqref{K_N_2_kappa_0} as readily follows from \eqref{symmetry_H_nmLM_3}. The proof of relations \eqref{L_K_N_2_kappa_0} is given below.

Now taking into account relations \eqref{K_N_2_kappa_0}, $\Tilde G_{n 0,i}^{(\ell)}\bigr|_{\kappa\to0,\; N=2} = - \sum\limits_{l=0}^{+\infty}(\mathsf K_{i j}|_{\kappa\to0})_{n 0, l 0} \Tilde G_{l 0,j}^{(\ell-1)}\bigr|_{\kappa\to0,\; N=2}$ (see \eqref{G_componentwise0}), and the starting values $\Tilde G^{(0)}_{n 0,i}\bigr|_{\kappa\to0,\;N=2} = \frac{q_i \delta_{n,0}}{\sqrt{4\pi}\varepsilon_0\varepsilon_\text{sol}}$ (see Sec.~\ref{Appendix_screened_potential_coefficients_point_kappa_nonzero}) as $i=\overline{1, 2}$, $j=3-i$, we then readily arrive at the following expressions for $\Tilde G^{(\ell)}_{n 0,i}\bigr|_{\kappa\to0,\;N=2}$ (i.e.~in the limit $\kappa\to0$):
\begin{align*}
&\Tilde G^{(1)}_{n 0,i}\bigr|_{\kappa\to0,\;N=2} = \frac{-q_j \sqrt{2 n+1}}{\sqrt{4\pi}\varepsilon_0\varepsilon_\text{sol}}\Bigl(\frac{R}{a_i}\Bigr)^n \Hat{\mathfrak e}^i_n ,\\
& \Tilde G^{(2)}_{n 0,i}\bigr|_{\kappa\to0,\;N=2} = \frac{q_i \sqrt{2 n+1}}{\sqrt{4\pi}\varepsilon_0\varepsilon_\text{sol}} \Bigl(\frac{R}{a_i}\Bigr)^n \Hat{\mathfrak e}^i_n \sum_{l_1=1}^{+\infty} \Hat{\mathfrak e}^j_{l_1} \frac{(n+l_1)!}{n! l_1!} , \\
& \Tilde G^{(3)}_{n 0,i}\bigr|_{\kappa\to0,\;N=2} = \frac{-q_j \sqrt{2 n+1}}{\sqrt{4\pi}\varepsilon_0\varepsilon_\text{sol}} \Bigl(\frac{R}{a_i}\Bigr)^n \Hat{\mathfrak e}^i_n \sum_{l_1=1}^{+\infty}\sum_{l_2=1}^{+\infty} \Hat{\mathfrak e}^j_{l_1} \Hat{\mathfrak e}^i_{l_2} \frac{(n+l_1)!}{n! l_1!} \\ 
&\qquad \times\frac{(l_1+l_2)!}{l_1! l_2!} , \\
&\Tilde G^{(4)}_{n 0,i}\bigr|_{\kappa\to0,\;N=2} = \frac{q_i \sqrt{2 n+1}}{\sqrt{4\pi}\varepsilon_0\varepsilon_\text{sol}} \Bigl(\frac{R}{a_i}\Bigr)^n \Hat{\mathfrak e}^i_n \sum_{l_1=1}^{+\infty}\sum_{l_2=1}^{+\infty}\sum_{l_3=1}^{+\infty} \Hat{\mathfrak e}^j_{l_1} \Hat{\mathfrak e}^i_{l_2} \Hat{\mathfrak e}^j_{l_3} \\
&\qquad \times \frac{(n+l_1)!}{n! l_1!}\frac{(l_1+l_2)!}{l_1! l_2!}\frac{(l_2+l_3)!}{l_2! l_3!} ,\\
&\Tilde G^{(5)}_{n 0,i}\bigr|_{\kappa\to0,\;N=2} = \frac{-q_j \sqrt{2 n+1}}{\sqrt{4\pi}\varepsilon_0\varepsilon_\text{sol}} \Bigl(\frac{R}{a_i}\Bigr)^n \Hat{\mathfrak e}^i_n \sum_{l_1=1}^{+\infty}\sum_{l_2=1}^{+\infty}\sum_{l_3=1}^{+\infty}\sum_{l_4=1}^{+\infty} \Hat{\mathfrak e}^j_{l_1} \Hat{\mathfrak e}^i_{l_2} \\ 
&\qquad \times \Hat{\mathfrak e}^j_{l_3} \Hat{\mathfrak e}^i_{l_4}\frac{(n+l_1)!}{n! l_1!}\frac{(l_1+l_2)!}{l_1! l_2!}\frac{(l_2+l_3)!}{l_2! l_3!}\frac{(l_3+l_4)!}{l_3! l_4!} ,\\
& \ldots ,
\end{align*}
where $i=1$, $j=2$, and the dimensionless factor $$\Hat{\mathfrak e}^i_l \mathrel{:=} \frac{l(\varepsilon_i-\varepsilon_\text{sol})}{l\varepsilon_i+(l+1)\varepsilon_\text{sol}}\Bigl(\frac{a_i}{R}\Bigr)^{2 l+1}$$ (one may observe that it gives the conventionally defined Clausius-Mossotti factor \cite{Derb2} weighted by $(a_i/R)^3$ at $l=1$); the corresponding equalities for $\Tilde G^{(\ell)}_{n 0,i}\bigr|_{\kappa\to0,\;N=2}$ in the case of a complementary pair $i=2$ and $j=1$ are the same but with the prepended factor~$(-1)^n$. 

Now leveraging the above $\Tilde G^{(\ell)}_{n 0,i}\bigr|_{\kappa\to0,\;N=2}$ values with \eqref{L_ell_N_2_kappa_0} and \eqref{energy_N_monopoles_ell_screened_} we immediately arrive at the following expressions for addends of the interaction energy expansion~\eqref{energy_expansion_components_interaction} in the limit $\kappa\to0$:
\begin{align*}
& \mathcal E^{(1)}\bigr|_{\kappa\to0, \, N=2} \!=\! \frac{q_i q_j}{4\pi\varepsilon_0\varepsilon_\text{sol}R} ,\\
& \mathcal E^{(2)}\bigr|_{\kappa\to0, \, N=2} \!=\! \frac{-1}{8\pi\varepsilon_0\varepsilon_\text{sol} R}\sum_{l_1=1}^{+\infty} (q_i^2 \Hat{\mathfrak e}^j_{l_1} + q_j^2 \Hat{\mathfrak e}^i_{l_1}) ,\\
& \mathcal E^{(3)}\bigr|_{\kappa\to0, \, N=2} \!=\! \frac{q_i q_j}{4\pi\varepsilon_0\varepsilon_\text{sol} R}\sum_{l_1=1}^{+\infty}\sum_{l_2=1}^{+\infty}\Hat{\mathfrak e}^i_{l_1} \Hat{\mathfrak e}^j_{l_2}\frac{(l_1+l_2)!}{l_1! l_2!} ,\\
& \mathcal E^{(4)}\bigr|_{\kappa\to0, \, N=2} \!=\! \frac{-1}{8\pi\varepsilon_0\varepsilon_\text{sol} R}\sum_{l_1=1}^{+\infty}\sum_{l_2=1}^{+\infty}\sum_{l_3=1}^{+\infty} (q_i^2 \Hat{\mathfrak e}^j_{l_1}\Hat{\mathfrak e}^i_{l_2}\Hat{\mathfrak e}^j_{l_3} \\ 
& \qquad + q_j^2 \Hat{\mathfrak e}^i_{l_1}\Hat{\mathfrak e}^j_{l_2}\Hat{\mathfrak e}^i_{l_3}) \frac{(l_1+l_2)!}{l_1! l_2!} \frac{(l_2+l_3)!}{l_2! l_3!} ,\\
& \mathcal E^{(5)}\bigr|_{\kappa\to0, \, N=2} \!=\! \frac{q_i q_j}{4\pi\varepsilon_0\varepsilon_\text{sol} R}\sum_{l_1=1}^{+\infty}\sum_{l_2=1}^{+\infty}\sum_{l_3=1}^{+\infty}\sum_{l_4=1}^{+\infty}\Hat{\mathfrak e}^i_{l_1} \Hat{\mathfrak e}^j_{l_2}\Hat{\mathfrak e}^i_{l_3} \Hat{\mathfrak e}^j_{l_4} \\ 
& \qquad \times \frac{(l_1+l_2)!}{l_1! l_2!}\frac{(l_2+l_3)!}{l_2! l_3!}\frac{(l_3+l_4)!}{l_3! l_4!} ,\\
& \mathcal E^{(6)}\bigr|_{\kappa\to0, \, N=2} \!=\! \frac{-1}{8\pi\varepsilon_0\varepsilon_\text{sol} R}\!\sum_{l_1=1}^{+\infty}\sum_{l_2=1}^{+\infty}\sum_{l_3=1}^{+\infty}\sum_{l_4=1}^{+\infty}\sum_{l_5=1}^{+\infty}\!(q_i^2 \Hat{\mathfrak e}^j_{l_1}\Hat{\mathfrak e}^i_{l_2}\Hat{\mathfrak e}^j_{l_3}\Hat{\mathfrak e}^i_{l_4}\Hat{\mathfrak e}^j_{l_5} \\ 
&\qquad + q_j^2 \Hat{\mathfrak e}^i_{l_1}\Hat{\mathfrak e}^j_{l_2}\Hat{\mathfrak e}^i_{l_3} \Hat{\mathfrak e}^j_{l_4}\Hat{\mathfrak e}^i_{l_5}) \frac{(l_1+l_2)!}{l_1! l_2!}\!\frac{(l_2+l_3)!}{l_2! l_3!}\!\frac{(l_3+l_4)!}{l_3! l_4!}\!\frac{(l_4+l_5)!}{l_4! l_5!} ,\\
&\ldots,
\end{align*}
where $i\in\{1, 2\}$, $j=3-i$. From the above expressions it is easy to discern the general scheme for constructing the $\mathcal E^{(\ell)}\bigr|_{\kappa\to0, \, N=2}$ quantities at even and odd $\ell$ values (also explicitly reflecting iterative ``cascade-type'' character of propagation of polarization effects, as was recently asserted numerically in \cite{Lindgren_jctc2025}). It can also be readily concluded from the above energy expressions that $$\mathcal E^{(\ell)}\bigr|_{\kappa\to0, \, N=2} = O(R^{-(3\ell-2)})$$ as $R\to+\infty$, $\ell\ge1$ (hence, $\mathcal E^{(2)}\bigr|_{\kappa\to0, \, N=2}$ starts with the term $\propto R^{-4}$ as we have already observed in \eqref{E2_point_charges_N_spheres_s_kappa_zero_N_2}, $\mathcal E^{(3)}\bigr|_{\kappa\to0, \, N=2}$ starts with the term $\propto R^{-7}$, etc.). Besides, as one can observe from the above energy relations, they also suggest a method for directly explicitly calculating arbitrary coefficients $\mathfrak C_i$ of \eqref{Energy_correction_to_Coulombic_kappa_zero}, so that a practical algorithm for calculating $\mathfrak C_i$ then readily follows.
\begin{remark}
\label{appendix_lca_isnt_possible_kappa_zero}
Likewise to what was noticed in the case of nonzero $\kappa$ (see Remark~\ref{appendix_lca_isnt_possible}), the definition of $\Hat{\mathfrak e}^i_l$ and explicit expressions of $\mathcal E^{(\ell\ge1)}\bigr|_{\kappa\to0, \, N=2}(R)$
found above tell us that when dielectric constants $\varepsilon_1$ and $\varepsilon_2$ lie on the same side of $\varepsilon_\text{sol}$ (i.e.~either both $\varepsilon_1<\varepsilon_\text{sol}$ and $\varepsilon_2<\varepsilon_\text{sol}$, or $\varepsilon_1>\varepsilon_\text{sol}$ and $\varepsilon_2>\varepsilon_\text{sol}$), then $\mathcal E^{(\ell)}\bigr|_{\kappa\to0, \, N=2}(R)$ with \emph{odd} $\ell$ behave qualitatively similarly to the Coulombic term $\mathcal E^{(1)}\bigr|_{\kappa\to0, \, N=2}(R)$, and in a like manner $\mathcal E^{(\ell)}\bigr|_{\kappa\to0, \, N=2}(R)$ with \emph{even} $\ell$ behave qualitatively similarly to $\mathcal E^{(2)}\bigr|_{\kappa\to0, \, N=2}(R)$ which provides the leading correction to the Coulombic energy.
\end{remark}

\begin{remark}
\label{Remark_conducting_spheres_in_touch}
Formally setting the spheres to be equal-sized ($a_1=a_2=a$), conducting ($\varepsilon_1,\varepsilon_2 = +\infty$) and touching ($2 a = R$), we derive that the total energy expansion $\sum_{\ell=0}^{+\infty}\mathcal E^{(\ell)}\bigr|_{\kappa\to0, \, N=2}$ (see \eqref{energy_expansion_components_abs_gen}) with the above addends $\mathcal E^{(\ell)}\bigr|_{\kappa\to0, \, N=2}$ still produces a convergent series and in fact it converges to a well-known result \cite{Lekner2012,QKW2016,LianQin2018,LianQin2022}: $$\mathcal E = \frac{(q_i+q_j)^2}{16\pi\varepsilon_0\varepsilon_\text{sol} a \ln 2} .$$ Indeed, in this case $\Hat{\mathfrak e}^i_l = 2^{-2 l -1}$ ($l\ge1$), then taking into account the binomial expansion $(1-x)^{-l-1} = \sum_{n=0}^{+\infty}\frac{(l+n)!}{l! n!} x^n$ and the fact that once the spheres touch their initial charges will be equalized (each sphere acquires $(q_i+q_j)/2$), the relation \eqref{energy_expansion_components_abs_gen} generates the series having the following first few terms: $\frac{(q_i+q_j)^2}{16\pi\varepsilon_0\varepsilon_\text{sol} a}\Bigl(1+\frac{1}{2}-\frac{1}{12}+\frac{1}{24}-\frac{19}{720}+\frac{3}{160}-\frac{863}{60480}+\cdots\Bigr)$; the series in braces converges to~$\frac{1}{\ln 2}$ (noting the Maclaurin series \cite{GradRyzh} of $\frac{x}{\ln(1+x)}$ converging at $x=1$).
\end{remark}

\medskip

\emph{Proof of relations \eqref{L_K_N_2_kappa_0}:} 

Relation \eqref{L_ell_N_2_kappa_0} readily follows from \eqref{L_l_elementwise} by employing there relations $\mathcal H_{0 0}^{l 0}(R \Hat{\mathbf z}) = \left|\text{see Sec.~\ref{list_of_some_H_nmLM_values_}}\right| = \sqrt{4\pi}(-1)^l k_l(\Tilde R) Y_l^0(R \Hat{\mathbf z}) = (-1)^l k_l(\Tilde R) \sqrt{2 l+1}$, $\alpha_0(\Tilde a_i,\varepsilon_i) = \varepsilon_\text{sol}\frac{\Tilde a_i+1}{\Tilde a_i^2 e^{\Tilde a_i}}$ (see Appendix~\ref{appendix_properties_asympt_alpha_beta}), and asymptotics~\eqref{small_Bessel_i_k}.

Relation \eqref{K_N_2_kappa_0} follows from the very definition of $\mathsf K_{i j} = \mathsf A_i^{-1}\mathsf B_{i j}$ (see \eqref{global_lin_sys1}, \eqref{various_matrix_definitions_0}, \eqref{alpha_beta_definitions}) by employing relation $(-1)^l \sqrt{\frac{2 l+1}{2 n+1}} b_{n 0 l}(\Tilde r_i, \Tilde R) = \frac{\pi}{2}\frac{I_{n+1/2}(\Tilde r_i)}{\sqrt{\Tilde r_i}} \mathcal H_{n 0}^{l 0}(\mathbf{R}_{i j})$ (see \eqref{b_nml_Theta_jcp}, \eqref{rel_bnml_H_temp0}), asymptotics $b_{n 0 l}(\Tilde r, \Tilde R) \sim \sqrt{\frac{\pi}{2}}(2 l-1)!!\frac{(n+l)!}{n! l!}\frac{r^n}{R^{n+l+1}\kappa^{l+1}}$ as $\kappa\to0$ (see \cite[Eq.~(B7)]{our_jcp}), \eqref{alpha_asymptotic_small_arg_0}, and~\eqref{small_Bessel_i_k}.

\section{Janus particles (derivation of relations of Section~\ref{subsection_janus_particles})}
\label{appendix_Janus_proofs}
\subsection{Derivation of \eqref{Janus_sigma_nm_i_canonical}}
\label{appendix_Janus_proofs_canonical}
\noindent
From \eqref{sigma_nm_i_def} one trivially obtains the relation $\sigma_{0 0,i}^\text{f} = \sqrt{\pi}\Bigl(\int_0^{\theta_{i,1}} \varsigma_{i,1} \sin\theta_i d \theta_i + \int_{\pi-\theta_{i,2}}^\pi \varsigma_{i,2} \sin\theta_i d \theta_i\Bigr) = \sqrt{\pi}\bigl( \varsigma_{i,1}(1-\cos\theta_{i,1}) + \varsigma_{i,2}(1-\cos\theta_{i,2}) \bigr) = \frac{q_{i,1}+q_{i,2}}{\sqrt{4\pi} a_i^2}$. Next, $\sigma_{n m,i}^\text{f} = 0$ when $m\ne0$ (due to the fact that the charge distribution under consideration is AS, see Fig.~\ref{Janus_sigma_figure_canonical}),~and
\begin{align*}
& \sigma_{n 0,i}^\text{f} = a_i^{-2} \oint_{\partial\Omega_i} \sigma_i^\text{f}(\Hat{\mathbf s}_i) \sqrt{\frac{2 n+1}{4\pi}} P_n(\cos\theta_i) d s_i = \left|\mu_i \mathrel{:=} \cos\theta_i\right| \\
& = \sqrt{(2 n+1)\pi}\biggl(\! \varsigma_{i,1}\int\limits_{\cos\theta_{i,1}}^1 P_n(\mu_i) d\mu_i + \varsigma_{i,2}\int\limits_{-1}^{-\cos\theta_{i,2}} P_n(\mu_i) d\mu_i \!\biggr) \\
&\quad = \biggl| \int_{-1}^{-\cos\theta_{i,2}} P_n(\mu_i) d\mu_i =  (-1)^n\int_{\cos\theta_{i,2}}^1 P_n(\mu_i) d\mu_i \, ;\\
&\quad \text{ and }\ \int_{\cos\theta}^1 P_n(x) d x = \left|\text{see \cite[Eqs. (7.111), (8.753)]{GradRyzh}}\right| \\ 
&\quad = \frac{1-\cos^2\theta}{n(n+1)} \frac{d P_n(\cos\theta)}{d\cos\theta} = \frac{P_{n-1}(\cos\theta)-P_{n+1}(\cos\theta)}{2 n+1} \biggr| \\
& = \text{relation \eqref{Janus_sigma_nm_i_canonical} at $n>0$, $m=0$} .
\end{align*}
\subsection{Derivation of \eqref{Janus_sigma_nm_i_rotated}}
\label{appendix_Janus_proofs_general}
\noindent
It is well-known (see \cite[Chapter~12~B]{HirschCurBird}) that transformations of spherical harmonics $Y_l^\cdot$ under rotations can be expressed via Wigner D-matrices $D^l \mathrel{:=} \{D^l_{m n}(\omega)\}$, where a set $\omega=\{\alpha,\beta,\gamma\}$ of Euler angles describes the orientation of the new (rotated) coordinate system (with its spherical angles $(\Bar\theta, \Bar\varphi)$) with respect to the old coordinate system (with its spherical angles $(\theta, \varphi)$), $D^l_{m n}(\omega)$ are Wigner D-functions. In our case (see Fig.~\ref{Janus_sigma_figure_rotated}) it is enough to utilize the precession and nutation angles $\alpha=\varphi_i^r$ and $\beta=\theta_i^r$ to give the orientation of the new charge symmetry axis $\mathbf z_i^\prime$, while the intrinsic rotation angle $\gamma$ is $0$ (as further rotations around $\mathbf z_i^\prime$ are senseless owing to AS of the charge distributions considered). Then, from \cite[Eq.~(12.B-11)]{HirschCurBird} we obtain (via matrix product of a row-vector with~$D^l$)
\begin{equation*}
[Y_l^{-l}(\theta, \varphi),\ldots,Y_l^{l}(\theta, \varphi)] = [Y_l^{-l}(\Bar\theta, \Bar\varphi),\ldots,Y_l^{l}(\Bar\theta, \Bar\varphi)] D^l 
\end{equation*}
or, taking into account that $D^l$ is a unitary square matrix of dimension $2 l+1$ (thus, the inversion of $D^l$ equals to its conjugate transpose),
\begin{equation*}
Y_l^n(\Bar\theta, \Bar\varphi) = \sum\nolimits_{m=-l}^l Y_l^m(\theta, \varphi) D_{n m}^l(\omega)^\star .
\end{equation*}
Now taking into account this identity and relation $D_{0 m}^l(\omega) = \sqrt{4\pi/(2 l+1)}(-1)^m Y_l^m(\beta,\alpha)$ (see \cite[Eq.~(12.B-22)]{HirschCurBird}; note also that \cite{HirschCurBird} uses spherical harmonics $Y_n^m$ that differ from our \eqref{Ynm_definition} by Condon-Shortley factors $(-1)^m$), we transform charge density $\sigma_i^\text{f}$ that is AS with respect to~$\mathbf z_i^\prime$:
\begin{align*}
& \sigma_i^\text{f} = \sum_{l=0}^{+\infty} \sigma_{l 0,i}^\text{f, can} Y_l^0(\Bar\theta, \Bar\varphi) = \sum_{l=0}^{+\infty} \sigma_{l 0,i}^\text{f, can} \sum_{m=-l}^l (-1)^m Y_l^m(\theta, \varphi) \\
& \times D_{0 m}^l(\omega)^\star = \sum_{n,m}\sqrt{\frac{4\pi}{2 n+1}}(-1)^m \sigma_{n 0,i}^\text{f, can} Y_n^{-m}(\beta,\alpha)Y_n^m(\theta, \varphi),
\end{align*}
from which relation \eqref{Janus_sigma_nm_i_rotated} immediately follows.
\subsection{Derivation of \eqref{Janus_Born_general_expr}}
\label{appendix_Janus_proofs_Born}
\noindent
Although surface charge's Fourier coefficients \eqref{Janus_sigma_nm_i_rotated} in general depend on index $m$ and angles $(\theta_i^r, \varphi_i^r)$, note a (quite expected from the physical point of view) rotational invariance of the $0$-screened (``Born") energy $\mathcal E^{(0)}$ in \eqref{Janus_Born_general_expr} (so that each addend in the sum $\sum_{i=1}^N$ of \eqref{Janus_Born_general_expr} represents the solvation energy of the corresponding particle $\Omega_i$, independently of all other particles and as if this particle did not experience any rotation). Indeed, from the addition theorem for spherical harmonics (see \eqref{Ynm_addition_theorem}) $\frac{4\pi}{2 n+1}\sum_{m=-n}^n \left|Y_n^m(\theta_i^r, \varphi_i^r)\right|^2 = P_n(1) = 1$, then
\begin{align*}
& \mathcal E^{(0)} = \frac{1}{2}\sum_{i=1}^N a_i^2 \sum_{n,m} \sigma_{n m,i}^{\text{f} \ \star} \Tilde a_i^n L_{n m,i}^{(0)} = \left|\text{use~\eqref{L_nmi_single_screened_components_L0}}\right| \\
& = \frac{1}{2}\sum_{i=1}^N a_i^2 \sum_{n,m} \sigma_{n m,i}^{\text{f} \ \star} \Tilde a_i^n \frac{k_n(\Tilde a_i) \sigma_{n m,i}^\text{f}}{\varepsilon_0 \kappa \Tilde a_i^n \alpha_n(\Tilde a_i,\varepsilon_i)} = \left|\text{use~\eqref{Janus_sigma_nm_i_rotated}}\right| \\
& = \frac{1}{2 \varepsilon_0 \kappa} \sum_{i=1}^N a_i^2 \sum_{n=0}^{+\infty}\frac{k_n(\Tilde a_i) (\sigma_{n 0,i}^\text{f, can})^2}{\alpha_n(\Tilde a_i,\varepsilon_i)} \frac{4\pi}{2 n+1}\sum_{m=-n}^n \left|Y_n^m(\theta_i^r, \varphi_i^r)\right|^2 \\
& = \text{relation~\eqref{Janus_Born_general_expr}} .
\end{align*}
Let us note that in the case of uniform free charge distribution over sphere, i.e.~$\varsigma_{i,1} = \varsigma_{i,2}$ and $\theta_{i,1}+\theta_{i,2}=\pi$, the right-hand side of relation \eqref{Janus_sigma_nm_i_rotated} boils down (simply take into account the parity $P_n^m(-x) = (-1)^{n+m} P_n^m(x)$) to that of \eqref{uniform_surf_charge_distr_}, and \eqref{Janus_Born_all_} then yields the conventional solvation energy~\eqref{Born_energy_N_spheres_central_surf_unif}.
\subsection{Double-screened energy for Janus particles}
\label{appendix_ell_2_screened_proof}
\noindent
From \eqref{L_nmi_single_screened_components}, \eqref{energy_expansion_components_abs_surface} and \eqref{Janus_sigma_nm_i_rotated}, denoting also $\Hat\beta_L(\Tilde a_j,\varepsilon_j) \mathrel{:=} (\varepsilon_j-\varepsilon_\text{sol}) L \Tilde a_j^{-1} i_L(\Tilde a_j) -\varepsilon_\text{sol}i_{L+1}(\Tilde a_j)$, we derive for the monopolar component of~$\mathcal E^{(2)}$:
\begin{align*}
& \mathcal E^{(2)}_\text{mon,mon} = \frac{1}{2}\sum_{i=1}^N a_i^2 \sigma_{0 0,i}^{\text{f} \ \star} L_{0 0,i}^{(2)} = \frac{1}{2\sqrt{4\pi}}\sum_{i=1}^N \frac{q_i}{\Tilde a_i^2}\sum_{j=1,\,j\ne i}^N \: \sum_{L,M} \frac{-\varepsilon_\text{sol}}{\varepsilon_0\kappa} \\
& \times\!\frac{\mathcal H_{0 0}^{L M}(\mathbf{R}_{i j})}{\alpha_0(\Tilde a_i,\varepsilon_i)}\!\sum_{p=1,\, p\ne j}^N \! \frac{\beta_{L M, 0 0}(\Tilde a_j,\varepsilon_j,\mathbf R_{j p})}{\alpha_L(\Tilde a_j,\varepsilon_j) \alpha_0(\Tilde a_p,\varepsilon_p)}\!\frac{q_p}{\sqrt{4\pi} a_p^2} = \bigl|\text{use~\eqref{alpha_beta_definitions}}\\
&\text{and Secs. \ref{list_of_some_H_nmLM_values_}, \ref{appendix_properties_asympt_alpha_beta}}\bigr| = \frac{-\kappa}{2\varepsilon_0\varepsilon_\text{sol}}\sum_{i=1}^N \frac{q_i e^{\Tilde a_i}}{1+\Tilde a_i}\sum_{j=1,\,j\ne i}^N \: \sum_{L,M} (-1)^L \\
& \times k_L(\Tilde R_{i j}) \frac{\Hat\beta_L(\Tilde a_j,\varepsilon_j)}{\alpha_L(\Tilde a_j,\varepsilon_j)} \sum_{p=1,\, p\ne j}^N \frac{q_p e^{\Tilde a_p}}{1+\Tilde a_p} k_L(\Tilde R_{j p}) Y_L^M(\Hat{\mathbf R}_{i j}) Y_L^M(\Hat{\mathbf R}_{j p})^\star \\
& = \bigl|\text{use~\eqref{Ynm_addition_theorem}}\bigr| = \frac{-\kappa}{8\pi\varepsilon_0\varepsilon_\text{sol}}\sum_{i=1}^N \frac{q_i e^{\Tilde a_i}}{1+\Tilde a_i}\sum_{j=1,\, j\ne i}^N \, \sum_{p=1,\, p\ne j}^N \frac{q_p e^{\Tilde a_p}}{1+\Tilde a_p} \\
&\times\sum_{L=0}^{+\infty} \frac{(2 L+1) \Hat\beta_L(\Tilde a_j,\varepsilon_j)}{\alpha_L(\Tilde a_j,\varepsilon_j)} k_L(\Tilde R_{i j}) k_L(\Tilde R_{j p}) P_L(\cos\gamma_{j i,j p}) ,
\end{align*}
which actually coincides with~\eqref{energy_N_monopoles_2_screened_legendre_full}; note that only the $n=m=0$ coefficient of \eqref{Janus_sigma_nm_i_rotated} was elaborated here (as the next ones, i.e.~those with $n=1$, produce dipole $\mathbf p_i$ defined in Sec.~\ref{subsection_janus_particles}).

\section{On the properties and asymptotics of quantities~\eqref{coeffs_HlmLM_definition} and~\eqref{alpha_beta_definitions}}
\label{appendix_properties_asympt_alpha_beta_H_nmLM}
\subsection{Properties of quantities~\eqref{coeffs_HlmLM_definition}}
\label{appendix_properties_asympt_H_nmLM}
\noindent
We will establish (see Appendix~\ref{appendix_properties_asympt_H_nmLM_proof}) that the following symmetries for $\mathcal H_{n m}^{L M}(\mathbf{R}_{i j})$ (with arbitrary indices $0\le |m|\le n$, $0\le |M|\le L$) hold:
\begin{subequations}
\label{symmetry_H_nmLM_all}
\begin{align}
& \mathcal H_{n, -m}^{L, -M}(\mathbf{R}_{i j}) = (-1)^{m+M}\mathcal H_{n m}^{L M}(\mathbf{R}_{i j})^\star ,\label{symmetry_H_nmLM_1} \\
& \mathcal H_{n m}^{L M}(\mathbf{R}_{i j}) = (-1)^{n+L} \mathcal H_{L M}^{n m}(\mathbf{R}_{i j})^\star,\label{symmetry_H_nmLM_2} \\
& \mathcal H_{n m}^{L M}(\mathbf{R}_{j i}) = (-1)^{n+L}\mathcal H_{n m}^{L M}(\mathbf{R}_{i j}).\label{symmetry_H_nmLM_3}
\end{align}
\end{subequations}
These symmetries are actively used in our analytical calculations throughout this work. They are also useful in numerical implementations of the analytical formalism proposed in the current study allowing one to crucially minimize the amount of calculations of re-expansion coefficients when forming the key matrix $\mathbb K$ (see Sec.~\ref{spherical_Fourier_coeffs_subsec}), as well as to store it in memory more economically. 
\subsubsection{Proof of symmetry relations~\eqref{symmetry_H_nmLM_all}}
\label{appendix_properties_asympt_H_nmLM_proof}
\emph{Proof of \eqref{symmetry_H_nmLM_1}.} Using the relation between Clebsch-Gordan coefficients and Wigner 3-$j$ symbols $C_{j_1 m_1 j_2 m_2}^{j m} = (-1)^{j_1-j_2+m} \sqrt{2 j+1} \left(\begin{smallmatrix}
j_1 & j_2 & j\\m_1 & m_2 & -m\end{smallmatrix}\right)$ (see e.g.~\cite[Chap.~8]{Varsh}) and the symmetry $\left(\begin{smallmatrix}j_1 & j_2 & j\\m_1 & m_2 & -m\end{smallmatrix} \right) = (-1)^{j_1+j_2+j} \left(\begin{smallmatrix}j_1 & j_2 & j\\-m_1 & -m_2 & m\end{smallmatrix} \right)$ (see \cite[\S~8.4.2]{Varsh}) we have $C_{j_1 m_1 j_2 m_2}^{j m} = (-1)^{j_1+j_2+j} C_{j_1, -m_1, j_2, -m_2}^{j, -m}$; then taking into account that $C_{n 0 l_2 0}^{L 0} = 0$ if $L+n+l_2$ is odd we obtain $C_{n 0 l_2 0}^{L 0} C_{n m l_2, M-m}^{L M} = C_{n 0 l_2 0}^{L 0} C_{n, -m, l_2, m-M}^{L, -M}$. Now employing the last relation and the elementary identity $Y_n^{-m}(\Hat{\mathbf R}_{i j}) = (-1)^m Y_n^m(\Hat{\mathbf R}_{i j})^\star$ we arrive~at
\begin{align*}
& \mathcal H_{n, -m}^{L, -M}(\mathbf{R}_{i j}) = \left|\text{see footnote~\ref{footnote_ClebshGordan_properties}}\right| = (-1)^L \sum\nolimits_{l_2} C_{n,-m,l_2,m-M}^{L,-M} \\
&\quad\times C_{n 0 l_2 0}^{L 0} \sqrt{\tfrac{4\pi (2 l_1+1) (2 l_2+1)}{2 L+1}} k_{l_2}(\Tilde R_{i j}) Y_{l_2}^{m-M}(\Hat{\mathbf R}_{i j}) \\ 
& = (-1)^L \sum\nolimits_{l_2} C_{n 0 l_2 0}^{L 0} C_{n m l_2, M-m}^{L M} \sqrt{\tfrac{4\pi (2 l_1+1) (2 l_2+1)}{2 L+1}} k_{l_2}(\Tilde R_{i j}) \\
&\quad\times (-1)^{M-m} Y_{l_2}^{M-m}(\Hat{\mathbf R}_{i j})^\star = (-1)^{m+M}\mathcal H_{n m}^{L M}(\mathbf{R}_{i j})^\star 
\end{align*}
(let us note that, besides spherical functions $Y$, other quantities are real there).

\medskip

\emph{Proof of~\eqref{symmetry_H_nmLM_2}.} One has
\begin{align*}
& \mathcal H_{n m}^{L M}(\mathbf{R}_{i j}) = \left|\text{use~\cite[Eq.~(D10)]{Yu3}}\right| \\
& = (-1)^L \sum\nolimits_{l_2,m_2} H_{L M l_2 m_2}^{n m} k_{l_2}(\Tilde R_{i j}) Y_{l_2}^{m_2}(\Hat{\mathbf R}_{i j})^\star \\
& = (-1)^{n+L} \mathcal H_{L M}^{n m}(\mathbf{R}_{i j})^\star .
\end{align*}

\medskip

\emph{Proof of~\eqref{symmetry_H_nmLM_3}.} Using the elementary identity $Y_n^m(-\Hat{\mathbf R}_{i j}) = (-1)^n Y_n^m(\Hat{\mathbf R}_{i j})$ and footnote~\ref{footnote_ClebshGordan_properties} we immediately arrive~at
\begin{align*}
\mathcal H_{n m}^{L M}(\mathbf{R}_{j i}) & = (-1)^L\sum\nolimits_{l_2,m_2} H_{n m l_2 m_2}^{L M} k_{l_2}(\Tilde R_{i j}) (-1)^{l_2} Y_{l_2}^{m_2}(\Hat{\mathbf R}_{i j}) \\
& = (-1)^{n+L} \mathcal H_{n m}^{L M}(\mathbf{R}_{i j}).
\end{align*}
\subsubsection{Some handy $\mathcal H_{n m}^{L M}(\mathbf{R}_{i j})$ expressions}
\label{list_of_some_H_nmLM_values_}
\noindent
Let us list some values of $\mathcal H_{n m}^{L M}(\mathbf{R}_{i j})$ just for the convenience of their further use:
\begin{align*}
\mathcal H_{n m}^{0 0}(\mathbf{R}_{i j}) & = (-1)^n \mathcal H_{0 0}^{n m}(\mathbf{R}_{i j})^\star  = \sqrt{4\pi} k_n(\Tilde R_{i j}) Y_n^m(\Hat{\mathbf R}_{i j})^\star, \\
\mathcal H_{1 1}^{1 1}(\mathbf{R}_{i j}) &= \mathcal H_{1, -1}^{1, -1}(\mathbf{R}_{i j}) = k_2(\Tilde R_{i j})\frac{3\cos^2\!\theta_{i j}-1}{2} - k_0(\Tilde R_{i j}), \\
\mathcal H_{1 0}^{1 0}(\mathbf{R}_{i j}) & = -k_2(\Tilde R_{i j})(3\cos^2\!\theta_{i j}-1) - k_0(\Tilde R_{i j}), \\
\mathcal H_{1, -1}^{1 0}(\mathbf{R}_{i j}) & = \mathcal H_{1 0}^{1, -1}(\mathbf{R}_{i j})^\star = -\mathcal H_{1 0}^{1 1}(\mathbf{R}_{i j}) = -\mathcal H^{1 0}_{1 1}(\mathbf{R}_{i j})^\star \\
& = -\frac{3}{\sqrt{2}}k_2(\Tilde R_{i j})\sin\theta_{i j} \cos\theta_{i j} e^{\imath \varphi_{i j}}, \\
\mathcal H_{1, -1}^{1 1}(\mathbf{R}_{i j}) & = \mathcal H_{1 1}^{1, -1}(\mathbf{R}_{i j})^\star = \frac{3}{2}k_2(\Tilde R_{i j})\sin^2\!\theta_{i j} e^{2 \imath \varphi_{i j}} 
\end{align*}
($\theta_{i j}$ and $\varphi_{i j}$ are spherical angles of $\Hat{\mathbf R}_{i j}$).
\subsection{Properties and asymptotics of quantities~\eqref{alpha_beta_definitions}}
\label{appendix_properties_asympt_alpha_beta}
\noindent
From \eqref{i_n_k_n_modified_Bessel} one immediately arrives at the expanded forms $\alpha_0(\Tilde a_i,\varepsilon_i) = \varepsilon_\text{sol}\frac{\Tilde a_i+1}{\Tilde a_i^2 e^{\Tilde a_i}}$, $\alpha_1(\Tilde a_i,\varepsilon_i) = \frac{(\varepsilon_i+2\varepsilon_\text{sol}) (1+\Tilde a_i) + \varepsilon_\text{sol}\Tilde a_i^2}{\Tilde a_i^3 e^{\Tilde a_i}}$, $\alpha_2(\Tilde a_i,\varepsilon_i) = \frac{(2\varepsilon_i+3\varepsilon_\text{sol})(\Tilde a_i^2+3\Tilde a_i+3)+\varepsilon_\text{sol}\Tilde a_i^2(1+\Tilde a_i)}{\Tilde a_i^4 e^{\Tilde a_i}}$, etc. It is easy to conclude that
\begin{equation}
\label{alpha_is_positive_ineq}
\alpha_n(\Tilde a_i,\varepsilon_i)>0
\end{equation}
(for arbitrary $n\ge 0$, $\Tilde a_i>0$, $\varepsilon_i>0$, $\varepsilon_\text{sol}>0$) -- indeed, using recurrence \eqref{in_kn_recurrences0} one gets $$\alpha_n(\Tilde a_i,\varepsilon_i) = \bigl(\varepsilon_\text{sol} k_{n-1}(\Tilde a_i) + (n\varepsilon_i+(n+1)\varepsilon_\text{sol}) k_n(\Tilde a_i)/\Tilde a_i \bigr)>0$$ for general~$n$. From the last equality for $\alpha_n(\Tilde a_i,\varepsilon_i)$ and \eqref{i_n_k_n_modified_Bessel} it also follows that representation $\alpha_n(\Tilde a_i,\varepsilon_i) = \frac{\mathcal P_{n+1}(\Tilde a_i)}{\Tilde a_i^{n+2} e^{\Tilde a_i}}$ holds, where $\mathcal P_{n+1}(\Tilde a_i)$ is a polynomial of degree $n+1$ in $\Tilde a_i$ with all positive coefficients and with the free term $\mathcal P_{n+1}(0) = (n\varepsilon_i+(n+1)\varepsilon_\text{sol}) (2 n-1)!!$ (note that $(-1)!! = 1$ is always assumed \cite{GradRyzh}); in particular, this leads to a rapid decay of the $\alpha_n(\Tilde a_i,\varepsilon_i)^{-1}$ values (with increasing $n$) appearing in screening-ranged expansions of the potential coefficients constructed in Sec.~\ref{general_expansions_pot_en}. From the above representation of $\alpha_n(\Tilde a_i,\varepsilon_i)$ and \eqref{small_Bessel_i_k} it also follows that asymptotically for small~$\Tilde a_i\to0$
\begin{equation}
\label{alpha_asymptotic_small_arg_0}
\alpha_n(\Tilde a_i,\varepsilon_i)^{-1} \sim \frac{\Tilde a_i^{n+2}}{(n\varepsilon_i+(n+1)\varepsilon_\text{sol}) (2 n-1)!!} .
\end{equation}

For small $a_i\to0$ we can derive from asymptotics \eqref{small_Bessel_i_k} that 
\begin{subequations}
\label{asymptotics_Kij_small_spheres_radii}
\begin{align}
\frac{\beta_{n m, L M}(\Tilde a_i,\varepsilon_i,\mathbf R_{i j})}{\alpha_n(\Tilde a_i,\varepsilon_i)} &= O(\Tilde a_i^{\max(3,\, 2 n+1)}) \label{asymptotics_Kij_small_spheres_radii_1}
\intertext{(or $O(\Tilde a_i^{2 n+3})$ if incidentally $\varepsilon_i = \varepsilon_\text{sol}$) and} 
\frac{\beta_{n m, L M}(\Tilde a_i,\varepsilon_i,\mathbf R_{i j})\Upsilon_{L,j}}{\alpha_n(\Tilde a_i,\varepsilon_i)\Upsilon_{n,i}} &= O(\Tilde a_i^{\max(3,\, 2 n+1)-n}) \label{asymptotics_Kij_small_spheres_radii_2}
\end{align}
\end{subequations}
(note that $\forall n\ge0$ one has $(\max(3,\, 2 n+1)-n)\ge2$).

For large $R_{i j}\to+\infty$, since multiplier $e^{-\Tilde R_{i j}}/\Tilde R_{i j}$ is present in any $k_{l_2}(\Tilde R_{i j})$ of~\eqref{Yu3_reexp}-\eqref{coeffs_HlmLM_definition},  we immediately obtain asymptotics
\begin{subequations}
\label{asymptotics_Kij_large_Rij}
\begin{align}
& \frac{\beta_{n m, L M}(\Tilde a_i,\varepsilon_i,\mathbf R_{i j})}{\alpha_n(\Tilde a_i,\varepsilon_i)} = O(e^{-\Tilde R_{i j}}/\Tilde R_{i j}) , \label{asymptotics_Kij_large_Rij_1} \\
& \frac{\beta_{n m, L M}(\Tilde a_i,\varepsilon_i,\mathbf R_{i j})\Upsilon_{L,j}}{\alpha_n(\Tilde a_i,\varepsilon_i)\Upsilon_{n,i}} = O(e^{-\Tilde R_{i j}}/\Tilde R_{i j}) . \label{asymptotics_Kij_large_Rij_2}
\end{align}
\end{subequations}

Using \eqref{wronsky_in_kn} we can prove the following useful relations (see~\eqref{alpha_beta_definitions} for the definitions of $\alpha_n$ and $\beta_{n m,L M}$):
\begin{subequations}
\label{wronsky_in_kn_corollary}
\begin{align}
&\begin{aligned}
&\left(n i_n(\!\Tilde a_i\!) \!+\! \Tilde a_i i_{n+1}(\!\Tilde a_i\!)\!\right)\!\mathcal H_{n m}^{L M}(\!\mathbf R_{i j}\!) \! - \! \frac{n k_n(\!\Tilde a_i\!)\! - \!\Tilde a_i k_{n+1}(\!\Tilde a_i\!)}{\alpha_n(\Tilde a_i,\varepsilon_i)} \\
&\quad \times \beta_{n m, L M}(\Tilde a_i,\varepsilon_i,\mathbf R_{i j}) = \frac{n\varepsilon_i}{\Tilde a_i^2 \alpha_n(\Tilde a_i,\varepsilon_i)}\mathcal H_{n m}^{L M}(\mathbf R_{i j}) ,
\end{aligned} \label{wronsky_in_kn_corollary_d} \\
&\begin{aligned}
&i_n(\Tilde a_i) \mathcal H_{n m}^{L M}(\mathbf R_{i j}) - \frac{k_n(\Tilde a_i)}{\alpha_n(\Tilde a_i,\varepsilon_i)}\beta_{n m, L M}(\Tilde a_i,\varepsilon_i,\mathbf R_{i j}) \\
&\quad = \frac{\varepsilon_\text{sol}}{\Tilde a_i^2 \alpha_n(\Tilde a_i,\varepsilon_i)}\mathcal H_{n m}^{L M}(\mathbf R_{i j}), 
\end{aligned} \label{wronsky_in_kn_corollary_f}
\end{align}
\end{subequations}
which will be used further for simplifying various expressions.

\section{Selected facts on modified Bessel functions and spherical~harmonics}
\label{appendix_bessel_functions_summary}
\noindent
Since we extensively use spherical and modified Bessel functions in calculations, let us first collect some elementary facts on them just for the convenience of further referencing.

Modified Bessel functions of the first and second kinds of semi-integer order $n+1/2$, respectively $I_{n+1/2} ( \cdot )$ and $K_{n+1/2} ( \cdot )$, have the exact analytic representations \cite{Wat, GradRyzh} $K_{n+1/2}(x) = \frac{\sqrt{\pi}}{\sqrt{2 x}} e^{-x}\sum_{l=0}^n\frac{(n+l)!}{l! (n-l)! (2 x)^l}$ and $I_{n+1/2}(x) = \frac{1}{\sqrt{2\pi x}}\Bigl(e^x\sum_{l=0}^n\frac{(-1)^l (n+l)!}{l! (n-l)! (2 x)^l} + (-1)^{n+1} e^{-x} \sum_{l=0}^n \frac{(n+l)!}{l! (n-l)! (2 x)^l}\Bigr)$ as $n\ge0$. For further convenience, we also recall the first few functions: $K_{1/2}(x)=\sqrt{\frac{\pi}{2}}\frac{e^{-x}}{\sqrt{x}}$,  $I_{1/2}(x)=\sqrt{\frac{2}{\pi}}\frac{\sinh x}{\sqrt{x}}$,  $K_{3/2}(x)=\sqrt{\frac{\pi}{2}}\frac{e^{-x}(x+1)}{x\sqrt{x}}$,  $I_{3/2}(x)=\sqrt{\frac{2}{\pi}}\frac{x \cosh x - \sinh x }{x\sqrt{x}}$. Then spherical modified Bessel functions of the first and second kinds, respectively $i_n(x)$ and $k_n(x)$, are defined~as
\begin{equation}
\label{i_n_k_n_modified_Bessel}
i_n(x) \mathrel{:=} \sqrt{\frac{\pi}{2}}\frac{I_{n+1/2}(x)}{\sqrt{x}}, \qquad k_n(x) \mathrel{:=} \sqrt{\frac{2}{\pi}}\frac{K_{n+1/2}(x)}{\sqrt{x}}.
\end{equation}

There are recurrences~\cite[Eq.~(8.486)]{GradRyzh} 
\begin{equation}
\label{in_kn_recurrences0}
\begin{aligned}
& x k_{n+1}(x) = x k_{n-1}(x) + (2 n+1)k_n(x), \\ 
& x i_{n+1}(x) = x i_{n-1}(x) - (2 n+1)i_n(x).
\end{aligned}
\end{equation}
In addition, for their derivatives there are the expressions
\begin{equation}
\label{diff_modifiedBessel}
\begin{aligned}
& \frac{d}{d x}i_n(x) = \frac{n}{x} i_n(x) + i_{n+1}(x), \\  
& \frac{d}{d x}k_n(x) = \frac{n}{x} k_n(x) - k_{n+1}(x).
\end{aligned}
\end{equation}

For small $x\to0^+$ one has~\cite{Wat}
\begin{equation}
\label{small_Bessel_i_k}
\begin{aligned}
i_n(x) & = \frac{x^n}{(2 n+1)!!} + \frac{x^{n+2}}{2 (2 n+3)!!} + O(x^{n+4}) \\ 
& \sim \frac{x^n}{(2 n+1)!!} \quad \text{and} \quad  k_n(x) \sim \frac{(2 n-1)!!}{x^{n+1}},
\end{aligned}
\end{equation}
while for large $x\to+\infty$ one has $k_n(x) \sim \frac{e^{-x}}{x}$ (notation ``$f(x)\sim g(x)$ as $x\to y$'' here means that $f(x)$ behaves asymptotically like $g(x)$ as $x\to y$, $\lim_{x\to y}(f(x)/g(x))=1$). 

Note also (see \cite{Wat}) that for arbitrary $x>0$ and $n\ge0$
\begin{equation}
\label{in_kn_are_positive_}
i_n(x)>0, \qquad k_n(x)>0.
\end{equation}

Finally, composing the Wronskian determinant $W[I_{n+1/2}(x), K_{n+1/2}(x)]$ (see \cite[Eq.~(8.474)]{GradRyzh}) we get a useful relation
\begin{equation}
\label{wronsky_in_kn}
i_n(x) k_{n+1}(x) + i_{n+1}(x) k_n(x) = x^{-2}.
\end{equation}

Throughout the work we use the standard spherical (complex-valued) harmonics defined as~\cite{Jack} 
\begin{equation}
\label{Ynm_definition}
Y_n^m(\Hat{\mathbf r}_i) = \sqrt{\tfrac{(2 n+1)(n-m)!}{4\pi(n+m)!}} P_n^m(\cos\theta_i) e^{\imath m\varphi_i} ,
\end{equation}
where $P_n^m(x)$ denotes the conventional associated Legendre polynomial, $P_n^m(x) = \frac{(-1)^m}{2^n n!}(1-x^2)^{m/2}\frac{d^{n+m}}{d x^{n+m}}(x^2-1)^n$, and $\imath$ is a complex unit. When simplifying expressions, we sometimes find the following standard addition theorem \cite{Jack} useful:
\begin{equation}
\label{Ynm_addition_theorem}
\frac{4\pi}{2 n+1}\sum_{m=-n}^n Y_n^m(\Hat{\mathbf r}_i) Y_n^m(\Hat{\mathbf r}_j)^\star = P_n(\cos\gamma_{i j}),
\end{equation}
where $\gamma_{i j}$ is the angle between $\Hat{\mathbf r}_i$ and $\Hat{\mathbf r}_j$ (one has $\cos\gamma_{i j} = \cos\theta_i \cos\theta_j + \sin\theta_i \sin\theta_j \cos(\varphi_i-\varphi_j) = \Hat{\mathbf r}_i\cdot\Hat{\mathbf r}_j$, as per unit vectors $\Hat{\mathbf r}_i$, $\Hat{\mathbf r}_j$), and $P_n(\cdot) = P_n^0(\cdot)$ is the (standard) Legendre polynomial. Legendre polynomials can also be defined via generating function
\begin{equation}
\label{leg_pol_gen_func}
1/\sqrt{1-2 x t +t^2} = \sum\nolimits_{n=0}^{+\infty} t^n P_n(x) ,
\end{equation}
$|x|\le1$, $|t|<1$. One has the asymptotics as $x\to1^-$ (see \cite[Eq.~(8.812)]{GradRyzh})
\begin{equation}
\label{small_theta_asymptotic_Leg}
P_n(x) = 1-\frac{n(n+1)}{2}(1-x) + O((1-x)^2) .
\end{equation}

\bibliography{LinPaper}

\end{document}